\documentclass[11pt]{article}

\usepackage{jheppub} 

\usepackage{amsmath,amssymb,epsfig,amsfonts}
\usepackage{graphicx,subfigure}
\usepackage{epsf}
\usepackage{makeidx}
\usepackage[all]{xy}
\usepackage{slashed}
\usepackage{verbatim} 
\usepackage{float}
\restylefloat{table}
\usepackage[all]{xy}
\usepackage{pdflscape}
\usepackage{tikz}
\usepackage{array}
\usepackage{youngtab}
\usepackage{multirow}
\usepackage{color}
\usepackage{ulem}
\usepackage{cleveref}

\usepackage{tensor}

\makeatletter

\newcommand{\be}{\begin{equation}}
\newcommand{\ee}{\end{equation}}
\newcommand{\ba}{\begin{aligned}}
\newcommand{\ea}{\end{aligned}}

\newcommand{\AdS}{\mathrm{AdS}}
\newcommand{\CFT}{\mathrm{CFT}}

\newcommand{\cIIBL}{c^{\IIB}_{L}}
\newcommand{\cIIBR}{c^{\IIB}_{R}}
\newcommand{\cIIBsugra}{c_{\text{sugra}}^{\IIB}}

\newcommand{\cML}{c^{11}_{L}}
\newcommand{\cMR}{c^{11}_{R}}
\newcommand{\cMsugra}{c_{\text{sugra}}^{11}}

\newcommand{\Del}{A}

\newcommand{\n}{q_N}
\newcommand{\Ntt}{q_N}
\newcommand{\Mti}{q_M}

\newcommand\F{\mathbb{F}}

\newcommand\D{\mathcal{D}}

\newcommand\diff{\mathrm{d}}

\newcommand{\E}{\mathbb{E}}
\newcommand{\dd}{\mathrm{d}}
\newcommand{\me}{\mathrm{e}}
\newcommand{\ii}{\mathrm{i}}

\newcommand{\vol}{\mathrm{vol}}

\newcommand{\Imag}{\mathrm{Im}\, }
\newcommand{\Real}{\mathrm{Re}\, }

\newlength{\sswidth}

\newcommand{\M}{\mathcal{M}}

\newcommand{\mb}{m_{B}}

\newcommand{\lb}{\left(}
\newcommand{\rb}{\right)}

\newcommand{\tr}{\mathrm{Tr}}

\newcommand{\IIB}{\text{IIB}}

\newcommand{\Tra}{\text{T}}
\newcommand{\sff}{\slashed{F}^{(2)}}
\newcommand{\ff}{F^{(2)}}

\newcommand{\alphai}{\alpha}

\renewcommand{\P}{\mathbb{P}}

\newcommand{\bea}{\begin{eqnarray}}
\newcommand{\eea}{\end{eqnarray}}

\newcommand{\R}{{\mathbb R}}
\newcommand{\Z}{{\mathbb Z}}

\def\Re{\mathop{\mathrm{Re}}\nolimits}

\def\Tr{\mathop{\mathrm{Tr}}\nolimits}
\def\tr{\mathop{\mathrm{Tr}}\nolimits}

\def\half{{\frac{1}{2}}}

\def\unit{{1\kern-.65ex {\rm l}}}
\def\1{{1\kern-.65ex {\rm l}}}

\def\ls{{\ell_{\rm s}}}
\def\lp{{\ell_{\rm p}}}
\def\gs{}

\newcount\hour \newcount\minute
\hour=\time \divide \hour by 60
\minute=\time
\def\now{%
\ifnum \hour<13
  \ifnum \hour=0 \advance \hour by 12 \number\hour:\else \number\hour:\fi%
     \ifnum \minute<10 0\fi%
     \number\minute%
\ A.M.%
\else \advance \hour by -12 \number\hour:%
  \ifnum \minute<10 0\fi%
  \number\minute%
  \ P.M.%
\fi%
}

\makeatother

\begin{document}

\baselineskip=18pt  
\numberwithin{equation}{section}  


%
%


\thispagestyle{empty}

\vspace*{-2cm} 
\begin{flushright}
{\tt KCL-MTH-17-02}\\
\end{flushright}

\vspace*{1cm} 
\begin{center}
{\huge F-theory and $\AdS_3/\CFT_2$}\\

 \vspace*{1.7cm}
 
{Christopher Couzens$^1$, \, Craig Lawrie$^2$, \,  Dario Martelli$^1$,  \\
\smallskip

Sakura Sch\"afer-Nameki$^3$,  \, and\,   Jin-Mann Wong$^1$}\\

 \vspace*{.8cm} 
{\it $^1$ Department of Mathematics, King's College London, \\
  The Strand, London, WC2R 2LS,  UK}\\
  \smallskip
{\it $^2$ Institut f\"ur Theoretische Physik, Ruprecht-Karls-Universit\"at,\\
 Philosophenweg 19, 69120 Heidelberg, Germany }\\
\smallskip
{\it $^3$ Mathematical Institute, University of Oxford \\
Woodstock Road, Oxford, OX2 6GG, UK}\\
\bigskip

{\tt {kcl.ac.uk}: christopher.couzens, dario.martelli}\\
{\tt {gmail:  craig.lawrie1729, sakura.schafer.nameki, jinmannwong}}
\vspace*{0.8cm}
\end{center}
\vspace*{.5cm}

\noindent
We construct supersymmetric AdS$_3$ solutions in F-theory, that is Type IIB
supergravity with varying axio-dilaton, which are holographically dual to 2d
$\mathcal{N}=(0,4)$ superconformal field theories with small superconformal algebra.  In F-theory these  arise from
D3-branes wrapped  on curves in the base of an elliptically fibered Calabi--Yau
threefold $Y_3$ and correspond to self-dual 
strings in the 6d $\mathcal{N}=(1,0)$ theory obtained from F-theory on $Y_3$. The non-trivial fibration over the wrapped curves implies a 
varying coupling of the $\mathcal{N}=4$ Super-Yang--Mills theory on the
D3-branes.  We compute the holographic central charges  and show that these
agree with the field theory and with the anomalies of self-dual strings in 6d. We complement our
analysis with a discussion of the  dual M-theory solutions and a comparison of the central charges.

\newpage

\tableofcontents


\section{Introduction}
\label{intro}

F-theory \cite{Vafa:1996xn,Morrison:1996na,Morrison:1996pp} has a firm standing as a framework for constructing Type IIB Minkowski vacua in even dimensions, which preserve minimal supersymmetry. The main focus thus far in utilising F-theory has been on the construction and classification of Type IIB vacua with varying axio-dilaton $\tau$, as well as $(p,q)$ 7-branes, which are naturally encoded in the singularities of $\tau$. The canonical setup to construct such vacua is the compactification on elliptic Calabi--Yau varieties $Y_d$ of complex dimension $d$ with base $B_{d-1}$, where the complex structure of the elliptic fiber models the axio-dilaton. 

In the current paper we will deviate from this and consider supergravity
solutions with AdS vacua of Type IIB supergravity, where we allow the
axio-dilaton $\tau$ to vary consistently with the $SL(2, \mathbb{Z})$ duality transformations. 
In this sense we are constructing F-theory solutions. Our
main focus will be supersymmetric solutions with AdS factors, which
allow for a holographic interpretation. The motivation to study these
backgrounds arises from various points of view. It is in general an
interesting question to characterise systematically such Type IIB solutions,
as a method for exploring superconformal field theories (SCFTs). Moreover, recently\footnote{D3-branes wrapped
  on curves in compact Calabi--Yau threefolds were studied much earlier
  \cite{Bershadsky:1995vm}, however in those setups the coupling of the
D3-brane remains constant.} various D3-brane configurations in F-theory have
been shown to give rise to 2d  SCFTs
\cite{Haghighat:2013gba, Haghighat:2014vxa, DelZotto:2016pvm,
Haghighat:2015ega,Lawrie:2016axq}. These constructions are based on D3-branes
wrapped on a complex curve $C$ in the base $B_{d-1}$ of the elliptic
fibration. Our goal here is to construct holographic duals to such 2d SCFTs.
In this paper we are interested in the case where the curve $C$ is deformable, which is in contrast to the case of the strings in 6d non-Higgsable clusters (NHC) \cite{Morrison:2012np}, where the curve is rigid.

The main novelty in our configuration is that we consider a background, where the complexified coupling $\tau$ of the $\mathcal{N}=4$ Super-Yang Mills (SYM) theory on the D3-brane varies over the internal compact dimensions. 
This requires a particular topological twist of the $\mathcal{N}=4$ SYM theory, which is known as the topological duality twist which accounts for the additional varying coupling constant, and was introduced for $U(1)$ gauge groups in \cite{Martucci:2014ema} and generalised to non-abelian groups in \cite{Assel:2016wcr}. {For the case of a single D3-brane wrapping a curve above which the coupling varies, as we consider here}, the 2d SCFTs were studied in \cite{Haghighat:2015ega, Lawrie:2016axq} and the central charges for these abelian theories were obtained therein. 

Concretely, we will consider F-theory on elliptically fibered Calabi--Yau
threefolds $Y_3$ with base $B$, where the D3-branes wrap a curve
$C \subset B$, and an $\mathbb{R}^{1,1}$ factor, {giving rise to strings with
2d $(0,4)$ supersymmetric theories on their worldvolume}. We will see that the
holographic dual AdS$_3$ solutions have a non-trivial axio-dilaton profile and
the geometry of the solution is of the form 
\be
\mathrm{AdS}_3 \times S^3 \times B  \,,
\ee
with $\tau$ varying holomorphically over $B$, which is a K\"ahler surface, and a specific form of
five-form flux\footnote{Geometries of this type were discussed in
  \cite{Haghighat:2015ega} from the point of view of effective six-dimensional
  supergravity \cite{Bonetti:2011mw}. However, to the best of our knowledge
  the problem of lifting these to supersymmetric solutions of ten dimensional
  supergravity was not addressed. Our findings demonstrate that only a
  restricted class of 6d vacua discussed in  \cite{Haghighat:2015ega} can be
embedded consistently into F-theory.}. Constraints on $B$ follow from the
existence of a minimal elliptic Calabi--Yau fibration above it, which imply that  $B$
is either $\mathbb{P}^2$, a Hirzebruch surface $\mathbb{F}_m$, blow-ups
thereof, or an Enriques surface \cite{MR1109635,MR1217381}.  As an F-theoretic
solution this could formally be said to correspond to a background of the form
AdS$_3 \times S^3 \times Y_3$. The fibration will necessarily have
fibers which degenerate and are thus singular,
\emph{i.e.} there are codimension one loci in the base, above which $\tau$ will have log-singularities. 
Physically these correspond to the presence of $(p,q)$
7-branes, which are sourced by the singularity in $\tau$.  The specific type
of 7-brane is canonically determined by the singularity type, using the
Kodaira-N\'eron classification of singular fibers \cite{Kodaira, Neron}. The
base is a smooth K\"ahler surface, however the metric induced from the
Calabi--Yau fibration above it has singularities due to the presence of
7-branes. This makes the direct Type IIB analysis somewhat more delicate. We
will show that these  solutions exist, preserve the required supersymmetry,
and we will be able to determine some properties of the dual SCFTs, despite
the singular nature of the Type IIB solutions.

In view of the subtleties with regards to the singularities of the metric on $B$, we will corroborate these results by considering the M/F-dual solutions in 11d supergravity, which can be obtained by T-duality and uplift to M-theory, where they become solutions of the type
\be
\mathrm{AdS}_3 \times S^2 \times Y_3 \, \qquad\mathrm{with}\qquad  \mathbb{E}_\tau \hookrightarrow Y_3  \rightarrow B\,.
\ee
These  fall into the classification of supersymmetric AdS$_3$ solutions
presented in  \cite{Colgain:2010wb,Kelekci:2016uqv}. Once lifted to 11d supergravity,
the singularities of the six-dimensional metric can be resolved, yielding a
smooth Ricci-flat K\"ahler metric on the compact Calabi--Yau variety, which is the resolution of $Y_3$.
This allows us to compute reliably the holographic central charges, including
certain sub-leading corrections.

More generally, we will show that in the Type IIB solution one can perform 
a quotient of $S^3$ by a discrete subgroup $\Gamma =\Gamma_{ADE} \subset SU(2)_L \subset SU(2)_L\times SU(2)_R$ retaining the full $(0,4)$ supersymmetry.  The resulting
  \begin{equation}
    \mathrm{AdS}_3 \times S^3/\Gamma \times B \,,
  \end{equation}
solution  is the most general  F-theory geometry (with vanishing three-form
fluxes) preserving $(0,4)$ supersymmetry.
We will concentrate on the A series\footnote{The D and E series result in three-dimensional spaces with non-abelian fundamental groups, which do not posses continuous isometries.}, where $\Gamma=\Z_M$, with the quotient acting on the Hopf fiber of $S^3$. In this case,
the solution can be interpreted as the near-horizon limit of D3-branes wrapped on a curve  $C \subset B$ transverse
to an $M$-centered\footnote{In the geometry, the $M$ centers coincide, which results in a $\mathbb{Z}_M$ singularity.} Taub-NUT space, corresponding to the presence of $M$ Kaluza-Klein (KK)
monopoles. Of course, for $M=1$ the near-horizon limit of the solutions
corresponding to one KK or no KK-monopoles are
indistinguishable. However, we will see that the explicit dual 11d
supergravity solution can be argued to correspond to the configuration with
\emph{one KK-monopole} in Type IIB.
 
{\emph{A priori}} the M/F-dual setup to the D3-branes wrapped on $C\times
\mathbb{R}^{1,1}$ are M5-branes that in addition wrap the elliptic fiber above
the curve $C$, \emph{i.e.} the M5-branes wrap an elliptic surface, which we
will denote by $\widehat{C}$, times $\mathbb{R}^{1,1}$. This is closely
related to the MSW-string \cite{Maldacena:1997de, Vafa:1997gr, Minasian:1999qn}, although here
$\widehat{C}$ is \emph{not} an ample divisor of $Y_3$, and  as a result the
supergravity solution for this configuration does not exist. However, upon
explicit T-duality and lift to 11d, we will find that our F-theory
solution can be interpreted in terms of M5-branes wrapped on an ample divisor
that is a linear combination of  $N$ times $\widehat{C}$  and $M$ times the section of the
fibration, $B$. The latter arise precisely from dualizing $M$ KK-monopoles in
Type IIB\footnote{Aspects of these backgrounds were studied in \cite{Bena:2006qm} from a different point of view.}. The spectrum of the wrapped M5-brane on $\widehat{C}\times \mathbb{R}^{1,1}$ and its relation to a single D3-brane wrapped on $C \times \mathbb{R}^{1,1}$ with the topological duality twist, was analysed in \cite{Lawrie:2016axq}.

The entire class of F-theory solutions $\mathrm{AdS}_3 \times S^3/\Z_M \times B$, and their 11d supergravity duals $\mathrm{AdS}_3 \times S^2 \times Y_3$, is expected to be holographically dual to 2d $(0,4)$ SCFTs.
The details of the SCFTs will depend on a choice of  K\"ahler manifold $B$ and the elliptic fibration above it, which specifies the 7-brane configuration, as well as the choice of holomorphic curve $C\subset B$ that the D3-branes wrap.  
Therefore, in this paper we will not attempt to construct explicit 2d field theories 
that should  flow to the SCFTs in the IR, beyond the discussion in \cite{Lawrie:2016axq}. Nevertheless, some generic
 features of these SCFTs can be inferred from the supergravity solutions and may be checked by using different methods, mainly based on anomaly inflow on branes as well as anomalies of strings in 6d SCFTs.
 For $M>1$ the supergravity solutions
 include an $SO(2,2)\times SU(2)_R$ isometry, with the Killing spinors transforming each in the representation ${\bf 2}$ of $SU(2)_R$.  This can therefore be identified as the R-symmetry of a \emph{small}  superconformal algebra \cite{Ademollo:1975an}. For $M=1$ the F-theory solution (but not the 11d supergravity dual) has an enhanced $SO(4)\simeq SU(2)_L\times SU(2)_R$ 
 isometry. Since the Killing spinors do not transform under $SU(2)_L$ this corresponds to an emergent flavour symmetry in the theory arising from D3-branes, with no KK-monopoles.  In the abelian case, this symmetry was shown to be realised in terms of a current algebra  \cite{Haghighat:2015ega}.
 
Information on  the field theories arising from a stack of D3-branes only and
those involving in addition $M$ KK-monopoles can be obtained using complementary
approaches.  The former give rise to self-dual strings in 6d and we can
determine the central charges $c_L,c_R$ from their anomaly polynomial. We can also reproduce these central charges in the M-theory dual setup involving a stack of M5-branes wrapped on the (non-ample) divisor $\widehat C$, by evaluating the relevant  anomaly polynomial 
\cite{Harvey:1998bx}. However, as remarked above, this setup does not admit a supergravity dual, and there will be additional subtleties that we shall discuss in detail. 
Conversely, for the theories arising in the presence of $M$ KK-monopoles, the anomaly polynomial for the 6d strings is not known, so in this paper we will not be able to compute the central charges microscopically on the F-theory side in these cases. For the M-theory configurations however, we can determine them from the M5-brane anomaly polynomial, and successfully compare the results with the gravity duals.

We arrive at the F-theory solutions  described  above by performing a systematic analysis of the supersymmetry equations in Type IIB supergravity. We adopt the method of G-structures, that has been  employed to investigate supersymmetric solutions with AdS factors in various settings -- see \cite{Martelli:2003ki,Gauntlett:2004zh,Gauntlett:2005ww,Gabella:2012rc,Apruzzi:2013yva,Apruzzi:2014qva,Apruzzi:2015zna}    for a representative  list of references. In this paper we study warped AdS$_3$ solutions of Type IIB supergravity, allowing for a general five-form flux and arbitrary axio-dilaton, while setting the three-form fluxes to zero. Here we will  concentrate on solutions preserving $(0,4)$  supersymmetry, where our analysis shows that the class of solutions is essentially unique. Generalisations of this setup will be discussed elsewhere \cite{toappear}.  

Examples of supersymmetric AdS$_3$ solutions that can be thought of as the near-horizon limit of brane intersections  have been known for a long time \cite{Boonstra:1998yu}, although in many cases the dual SCFTs have remained somewhat 
elusive \cite{Tong:2014yna}. The  conditions for AdS$_3$ solutions preserving at least $(0,2)$ supersymmetry,  arising purely from wrapped D3-branes with constant coupling, were spelled out in  \cite{Kim:2005ez}. Following this work, a number of explicit solutions have been found  
in \cite{Gauntlett:2006qw,Gauntlett:2006ns,Gauntlett:2007ph}, which also determined some properties of the putative dual 2d $(0,2)$ SCFTs. The authors of \cite{Donos:2008ug} presented generalisations to include a particular form of three-form fluxes.
Finally, new interesting examples of solutions to the equations in  \cite{Kim:2005ez} were found in \cite{Benini:2013cda,Bobev:2014jva,Benini:2015bwz}, where the precise dual $(0,2)$ SCFTs were identified and non-trivial checks of the central charges
were performed. In particular,  the central charges of $(0,2)$ SCFTs can be determined by the method of $c$-extremisation \cite{Benini:2012cz}. In all these 
examples, however, the axio-dilaton is constant. In the present paper we will initiate a line of investigation in which this can vary non-trivially over the internal manifold,  setting the stage for studying holography in the context of F-theory. 

 We should remark on other supergravity solutions with holographic duals, where non-trivial profiles of the axio-dilaton have appeared -- however none of which include a varying axio-dilaton with the full $SL(2, \mathbb{Z})$ monodromy, which we incorporate in this paper. $\AdS$-duals with particular constant, but not necessarily perturbative, values of the axio-dilaton, which correspond to F-theory at constant coupling were studied  in \cite{Aharony:1998xz, Ahn:1998tv, Aharony:2007dj}.

 There are also  holographic setups with D3- and D7-branes, where the latter play the role of flavour symmetries in the field theory dual, see \emph{e.g.} \cite{Erdmenger:2007cm} for a review. 
Typically these correspond to configurations of D3- and D7-branes sharing four flat space-time directions, corresponding to non-conformal four-dimensional field theories. 
 When the backreaction of the D7-branes is included, the supergravity solutions do not have an AdS factor.
 Another closely related setup involving D3- and D7-branes was discussed in
 \cite{Buchbinder:2007ar, Harvey:2008zz, Harvey:2007ab}. These are
 configurations where, as in the present paper,  the branes share two
 dimensions. Here the D3-branes are placed into the supergravity background
 sourced by the D7-branes, however the Type IIB solution does not possess $SO(2,2)$ isometry, and therefore it is not holographically dual to a 2d SCFT.
 This is distinct from the setup that we consider, in that it corresponds to a 4d gauge theory in the presence of 2d defects.

Recently,  AdS$_6$ solutions dual to 5d SCFTs were constructed in Type IIB supergravity, which have a non-trivial $\tau$ profile that allows for poles in $\tau$, but does not include any 
$SL(2, \mathbb{Z})$ monodromy \cite{DHoker:2016ujz, DHoker:2017mds}. Furthermore there is the class of holographic duals to Janus configurations, \cite{Bak:2003jk, DHoker:2007zhm, Clark:2004sb,  Bachas:2013vza} where the gauge coupling varies along a real line, which was later generalised to the  $\theta$-angle varying along the 1d line  \cite{Gaiotto:2008sd}. In contrast, in our configurations, the complexified coupling $\tau$ varies holomorphically along {the base of the fibration, which is a complex surface in the present case,}  giving rise to an elliptic fibration with general $SL(2, \mathbb{Z})$ monodromy.

The rest of the paper is organised as follows. Section \ref{sec:Fsugra} contains a short overview of F-theory and elliptic Calabi--Yau manifolds. In Section 
\ref{marypoppins} we derive the Type IIB supergravity solutions using G-structure methods. In Section \ref{sec:FTheoryCC} we present the computation of the holographic central charges for the F-theory solutions. In Section \ref{sec:FMduality} we discuss the duality to 11d supergravity making contact with \cite{Maldacena:1997de}.  In Section \ref{sec:Mthccs}
 we present the computation of the holographic central charges for the M-theory solutions. Section \ref{peppapig} contains the microscopic computations of the central charges of the dual SCFTs via anomaly inflow for 6d self-dual strings and MSW-strings. We discuss our results in Section \ref{writesomethingsensible}. We include  several appendices containing supergravity computations as well as 
 summarising mathematical properties of the geometries used in the main part of the paper. 
Moreover, in appendix \ref{app:GfluxSol} we present a new $\AdS_3$ solution of Type IIB supergravity, including three-form fluxes, which is thereby outside the analysis in the rest of the paper.


\section{F-theory and Wrapped D3-branes}
\label{sec:Fsugra}

\subsection{F-theory}

We begin with some remarks regarding F-theory and solutions of Type IIB supergravity. 
Consider a Calabi--Yau manifold, $Y$, which is elliptically fibered over a compact K\"ahler base $B$, which we abbreviate as
$\mathbb{E}_\tau \hookrightarrow Y \rightarrow B$. 
If there are no singular fibers, the fibration is in fact topologically 
trivial, \emph{i.e.} $Y = \mathbb{E}_\tau \times B$, implying that the base
$B$ is itself a Calabi--Yau variety, and also that $\tau$ does not vary over $B$. Here we will be interested in non-trivial fibrations, which
necessarily include so-called Kodaira singular fibers \cite{Kodaira,
Neron}\footnote{As a cautious remark, despite the name, these are the resolved
fibers above singular loci of the fibration.}. We 
consider an elliptic fibration with a section, \emph{i.e.} a map $B\rightarrow
\mathbb{E}_\tau$, which implies the existence of a Weierstrass form of the
elliptic fibration
\be\label{Weier}
y^2 = x^3 + fx w^4+ g w^6 \,,
\ee
where $f, g$ are sections of $K_B^{-4}$ and $K_B^{-6}$, respectively, and thereby depend on the base $B$. Here $K_B$ denotes the canonical class of the base.
The hypersurface equation (\ref{Weier}) is written in an ambient space, which is the fourfold obtained as the total space of the projectivization of the sum of line bundles over $B$ 
\be
X_4= \mathbb{P} (\mathcal{O}_B \oplus (\mathcal{O}_B \oplus K_B^{-1})^2 \oplus (\mathcal{O}_B \oplus K_B^{-1})^3)\,,
\ee
where $\mathcal{O}_B$ is the structure sheaf of $B$. 
The coordinates $[w, x, y]$ satisfy standard projective relations in
{$\mathbb{P}^{1,2,3}$} and their class is given by $[w] = \sigma$, $[x] = 2
(\sigma + c_1(B))$ and $[y] = 3(\sigma + c_1(B))$. In particular, the class of
the section of the elliptic fibration is denoted here by $\sigma$. For a more
in depth review of elliptic fibrations and their geometry and more
specifically the intersection theory used in the following, we refer the reader to \emph{e.g.} \cite{Denef:2008wq, Weigand:2010wm, Marsano:2011hv}.
For each point in the base, this equation defines an elliptic curve, whose complex structure can be determined via the $j$-function, which in turn depends on $f$ and $g$. 
Singularities in the elliptic fibration are characterised by the vanishing of the discriminant $\Delta$ of the Weierstrass equation, {\it i.e.} 
\be
\Delta = 4 f^3 + 27 g^2 =0 \,,
\ee
which define complex codimension one loci in $B$. The type of singular fibers that can occur
were classified by Kodaira--N\'eron, and are characterised in terms of the order
of vanishing of $(f, g, \Delta)$ along the discriminant locus. More concretely,
the different singularities are characterised in terms of the fibers that are
obtained upon resolution\footnote{More precisely, these are K\"ahler
deformations, which retain the Calabi--Yau property, so-called crepant resolutions.}. The resulting Kodaira
singular fibers are collections of rational curves ($\mathbb{P}^1$s) with
precise intersection patterns and multiplicities, given \emph{e.g.} in terms of affine Dynkin
diagrams of ADE Lie algebras. The simplest Kodaira fiber is $I_1$, which has 
\be
I_1: \qquad \hbox{ord} (f,g, \Delta) = (0,0,1) \,.
\ee
The fiber is a node, \emph{i.e.} a $\mathbb{P}^1$ with a self-intersection.
This fiber does not induce a singularity in the total space of the fibration and thus
no resolution is necessary, as one can check directly. 
In F-theory, singularities of the axio-dilaton $\tau$, \emph{i.e.} vanishing of the discriminant $\Delta$,  determine the loci of the 7-branes, and the type of singularity characterises the $(p,q)$ charges of the 7-branes under the $SL(2, \mathbb{Z})$ self-duality of Type IIB. The $I_1$ singular fiber corresponds to a single D7-brane. The worldvolume of the 7-branes is $\Delta \times \mathbb{R}^{1,d}$. 
The effective theory of F-theory on $Y\times \mathbb{R}^{1,d}$ is given in terms of the gauge theories on the 7-branes coupled to $\mathcal{N}=1$ supergravity in $d+1$ dimensions. 

For a given Weierstrass model, the complex structure $\tau$ of the elliiptic curve can be extracted from
\be
j(\tau) = - 1728 {(4 f)^3 \over \Delta}\,,
\ee
where $j(\tau)$ is the Jacobi $j$-function. By expanding 
\be
j(\tau) = {1\over q} + 744 +  \cdots\,,
\ee
where $q= \me^{2 \pi \ii \tau}$. 
Using the asymptotic expansion along the loci where $\Delta=0$
one can extract the local behaviour of $\tau$.  For example, 
the axio-dilaton close to a 7-brane wrapping the local divisor $z=0$ in the base $B$ has the profile 
\be\label{tauLog}
\tau = {1\over 2\pi \ii} \log z + \cdots \,,
\ee
which has a singularity at $z=0$, and undergoes a monodromy $\tau \rightarrow \tau +1$ around this locus. More general $(p,q)$ 7-branes will have $\gamma \in SL(2, \mathbb{Z})$ monodromy. 
For elliptic K3s, this was derived in \cite{Greene:1989ya}.
Specifically this singular behaviour of the axio-dilaton  implies that the metric on the base will  have singularities. 

In the present context we are interested in solutions to the effective theory. Naively we would define the F-theory supergravity solutions on $Y$ in terms of Type IIB supergravity, on $B$ including $\tau$ which varies over $B$. However when the elliptic fibration has singularities as in (\ref{tauLog}), the metric that is induced on the base $B$ is expected to be singular. In the case of K3 surfaces, this can be made explicit for non-compact 
\cite{Greene:1989ya} and for compact K3s \cite{GrossWilson, MSDbraneBook}, who also give a precise measure for the divergence of the curvature scalar close to the singular fibers. Thus a supergravity approach seems at first sight to be somewhat questionable. 

In the absence of a first principle formulation of F-theory, the effective
action of a compactification on an elliptically fibered manifold $Y$ is
defined in terms of M/F-duality: F-theory on $Y \times S^1$ defines the same
effective theory as M-theory on $Y$. This will be the approach that we will
use to define the F-theory solution: there is a dual M-theory solution on an
elliptically fibered manifold $Y$, which is smooth and Calabi--Yau, with a
smooth Ricci-flat K\"ahler metric. Reducing to Type IIA, and performing a
T-duality results in a Type IIB solution with varying $\tau$. The effective
theory is obtained solely by considerations in the 11d supergravity
compactification. {Whenever we allow for not only $I_1$ but enhanced singular
fibers, the computations are done in the resolved geometry, where the
singularities are blown up retaining the Calabi--Yau property of $Y_3$.}

\subsection{Elliptic Fibrations}
\label{subsec:EllFib}

{In the following it will be useful to have some geometric basics about elliptic fibrations in place for studying F-theory solutions. Here our main interest is in elliptic threefolds, but much can be generalised to other dimensions.
We consider Calabi--Yau threefolds $Y_3$, which are elliptically fibered over a base $B$, which is a complex surface. Denote the projection map $\pi: Y_3 \rightarrow B$. Furthermore we assume there is a section, which as explained earlier implies the existence of a Weierstrass model. 
 
It will be very important in the following to determine the possible divisors
(4d submanifolds) in such a geometry, which is  the content of the
Shioda-Tate-Wazir theorem \cite{Wazir}, which  implies that the divisors of an
elliptic Calabi--Yau threefold $Y_3$ with a section, fall into the following three classes: }
\begin{enumerate}
\item \underline{Section}: This is the divisor obtained by the image of the
  base $B$ in $Y_3$. We denote it simply by $B$. The dual $(1,1)$-form will be
  denoted by $\omega_0$. \footnote{An elliptic fibration can have more
  rational sections, in which case there are additional divisors and $(1,1)$
forms, which generate the Mordell-Weil group of the fibration. As this will not play any role here, we refrain from discussing these further.}
\item \underline{Pull-back of curves in the base $B$}: For every
  effective curve $C_\alpha \in H_2 (B)$ we have a divisor in $Y_3$ given
  by $\pi^*(C_\alpha)$. We will refer to these as $\widehat{C}_\alpha \equiv
  \pi^* (C_\alpha)$, and denote the dual $(1,1)$-forms by $\omega_\alpha$. 
\item \underline{Resolution/Cartan divisors}: These divisors occur whenever
  there is a singularity in the Weierstrass model of the elliptic
  fibration, and they are given in terms of rational curves, that are obtained from the resolution of the singularities, fibered over a curve in the base (which are components of the discriminant). In the literature these are often referred to as Cartan divisors, as they are (in many cases) labeled by the simple roots of the Lie algebra associated to the Kodaira singular fiber. The Cartan divisors will be denoted by $D_i$, and the dual $(1,1)$-forms by $\omega_i$. 
\end{enumerate}
For most part of the paper we will consider smooth Weierstrass fibrations, \emph{i.e.} there are no Cartan divisors. However this can be easily generalised and we will comment on this throughout the paper. 
{Divisors are dual to $(1,1)$-forms, and the Shioda-Tate-Wazir theorem thus implies that the K\"ahler form of the Calabi--Yau can be expanded as}
\be \label{KahlerFormExp}
J_{Y_3}= k_0 \omega_0 +\sum_\alpha k_\alpha  \omega_\alpha + \sum_i k_i \omega_i \,.
\ee
We will require that the K\"ahler class of the base
\be
J_B = \sum_\alpha k_\alpha  \omega_\alpha \,,
\ee
is dual inside $B$ to a curve $C$, implying that $k_\alpha \in \mathbb{Z}^+$.  
This means that $J_B$ is in fact the K\"ahler class associated to the Hodge metric on $B$ \cite{MR0068871}.
However, we do not require any such integrality for $k_0$. 

The non-trivial triple intersections of the basis $\omega_I = \{\omega_0, \omega_\alpha, \omega_i\}$ in the Calabi--Yau
\be
C_{IJK}= D_{I} \cdot_{Y_3} D_J \cdot_{Y_3} D_K = \int_{Y_{3}}\omega_{I}\wedge\omega_{J}\wedge\omega_{K}\,,
\ee 
can be evaluated in terms of data of the base $B$ as follows\footnote{Whenever we write $\cdot$ without any subscript in the following, this will denotethe intersection in $B$, unless otherwise stated.}
\be \label{eqn:IntersectionNumbers}\ba 
C_{000} &= \int_B c_1(B)^2= 10- h^{1,1}(B)  \cr 
C_{00 \alpha} &= - c_1(B) \cdot C_\alpha  \cr 
C_{0 \alpha \beta} &= Q_{\alpha \beta}  \cr 
C_{\alpha ij} &=  -\mathcal{C}_{ij} Q_{\alpha\beta} C^{\beta}  \,,
\ea \ee
where $Q_{\alpha \beta}$ is the intersection form on $B$ and $\mathcal{C}_{ij}$ the Cartan matrix of the gauge algebra $\mathfrak{g}$ associated to the singularity. The triple intersection $C_{ijk}$ were determined in \cite{Intriligator:1997pq, Grimm:2011fx, Hayashi:2014kca} and depend on codimension two singularities, which are labeled by representations of $\mathfrak{g}$. 

In deriving these intersection numbers we have made use of the intersection relation in $Y_3$\footnote{{This follows from the fact that $w, x, y$ cannot vanish at the same time, and thus $[w] \cdot [x] \cdot [y] =0$ as intersections in $X_4$, or noting that the class of the hypersurface (\ref{Weier}) is $[y^2]$, this becomes $\sigma \cdot (\sigma + c_1(B)) \cdot [y^2] =0$.}  }
 \be \label{IntersectionRel}
\sigma \cdot_{Y_3} (\sigma + c_1(B)) =0\,.
\ee
We will also need to compute intersections with $c_2(Y_3)$. The total Chern class for the Calabi--Yau can be written as 
\be
{{c}(Y_3) = (1+ [w]) (1+ [x]) (1+ [y])  {{c}(B) \over (1+ [y^2])} } \,,
\ee 
where ${c}(B)$ is the total Chern class of the base and the denominator corresponds to the class of the hypersurface equation (\ref{Weier}). Expanding this to second order we obtain\footnote{Here we used the relation (\ref{IntersectionRel}), which holds on $Y_3$.}
\be\label{c2cy}
c_2 (Y_3) = c_2(B) + 11 c_1(B)^2 + 12 \omega_0 \wedge c_1(B)  \,,
\ee
which allows the computation of integrals over $c_2(Y_3)$ using the intersection numbers in \eqref{eqn:IntersectionNumbers}.

Finally, we will often consider curves $C \subset B$, and it will be useful to recall the adjunction formula
\be 
K_C = (K_{B} + C)|_C\,,
\ee
where $K_C$ and $K_{B}$ are the canonical classes of $C$ and $B$, respectively. For a genus $g$ curve this implies 
\be  \label{adjunction}
2(g - 1)  = C \cdot C - c_1(B)\cdot C\,.
\ee
Throughout our considerations we will assume the base $B$ to be smooth as
a variety, albeit the induced metric on $B$ will have singularities that we will discuss in some detail later on.


\subsection{D3-branes in F-theory and 2d $(0,4)$ SCFTs}
\label{sec:FieldTheory}
 
The supergravity solutions that will be studied in this paper are dual to 2d $\mathcal{N} = (0,4)$ SCFTs arising from D3-branes wrapped on a curve, $C$, in the base of an elliptic Calabi--Yau threefold, in an F-theory compactification to 6d. The field content for these 2d SCFTs was worked out  in \cite{Lawrie:2016axq}, where in particular the abelian zero mode spectrum and the left and right central charges were computed. The zero mode spectrum in terms of $(0,4)$ multiplets was found to be
\be \begin{array}{c|c|c}
(0,4) \text{ multiplet} & \text{Multiplicity} & (c_R, c_L) \\ \hline
\text{Hyper} & \half C\cdot C + \half c_1(B)\cdot C  & (6,4)\\
\text{ Twisted Hyper} & 1   & (6,4) \\
\text{ Fermi} & \half C\cdot C - \half c_1(B) \cdot C + 1 &  (0,2)
\end{array} \,.\ee
 In addition, one has half-Fermi multiplets arising from 3-7 strings, which contribute $c_{37} = 8 c_1(B) \cdot C$ to the left-moving central charge. The left and right central charges are computed by summing the contributions from each multiplet with their appropriate multiplicity, and are given by \cite{Haghighat:2015ega,Lawrie:2016axq}
\be \label{CentralChargeAbelian} 
\ba 
c_R &=3 C\cdot C + 3 c_1(B) \cdot C +6 \, ,\\
c_L &= 3 C \cdot C + 9 c_1(B) \cdot C + 6\,.
\ea \ee
Notice that upon using the adjunction formula (\ref{adjunction}), the right central charge may be rewritten as
\be  \label{spectrumtimes6}
c_R = 6\left(g + c_1(B)\cdot C \right)\, ,
\ee
which is manifestly a multiple of 6, as expected generically for $(0,4)$ SCFTs with small superconformal algebra \cite{Schwimmer:1986mf}.
Under M/F-duality, this is equivalent to M5-branes wrapped on the elliptic surface $\widehat{C} = \pi^*(C)$ in the Calabi--Yau threefold. The 2d spectrum obtained from a single M5-brane wrapped on an elliptic surface was also determined in \cite{Lawrie:2016axq} as 
\be \begin{array}{c|c|c}
(0,4) \text{ multiplet} & \text{Multiplicity}  & (c_R, c_L) \\ \hline
\text{Hyper} & \half C\cdot C + \half c_1(B)\cdot C +1 & (6,4)\\
\text{ Fermi} & \half C\cdot C - \half c_1(B) \cdot C + 1 &  (0,2) \\
\text{Half-Fermi} & 8c_1(B) \cdot C  & (0,1) \\
\end{array} \,.\ee
 Here, the half-Fermi multiplets arise directly from the reduction of the 6d $\mathcal{N} = (2,0)$ tensor multiplet. This spectrum matches that of the D3-brane wrapped on $C$ and therefore the left and right central charges are also given by \eqref{CentralChargeAbelian}. 
These central charges are computed for a single D3-brane, {\it i.e.} $N=1$. In the following we will compute these holographically for general $N$. 


\section{AdS$_3$ Solutions in F-theory dual to 2d $(0,4)$ Theories}

\label{marypoppins}

In this section we switch gears and turn to an explicit analysis of the supersymmetry equations of the ten-dimensional Type IIB supergravity. We will derive the AdS$_3$ solutions of interest starting from
 a very general ansatz and requiring the existence of $(0,4)$ supersymmetry in the boundary SCFTs. The methods we use are by now completely standard, but we nevertheless include some details here, for the benefit of readers that may not be familiar with these.


\subsection{Type IIB Killing Spinor Equations}

To find supersymmetric solutions we first determine the constraints implied by
the existence of certain Killing spinors in Type IIB. We follow the Type IIB
supergravity conventions presented in \cite{Gauntlett:2005ww}, which we
briefly summarise here. The equations of motion and supersymmetry
transformations for bosonic configurations were originally found in
\cite{Schwarz:1983qr,Howe:1983sra}. The Neveu--Schwarz Neveu--Schwarz (NS-NS)
sector of Type IIB supergravity consists of the metric $g$, a real scalar
$\phi$, called the dilaton, and a real two-form potential, $B^{(2)}$. The Ramond
Ramond (RR) sector includes a real scalar potential $C^{(0)}$, a real two-form
potential $C^{(2)}$, and a real four-form potential, $C^{(4)}$, with self dual-field strength. It is convenient to combine these fields into complex ones: we combine the scalars as 
\be
\tau =   \tau_{1}+\ii \tau_{2}\equiv C^{(0)}+\ii \me^{-\phi}\,,
\ee
which we refer to as the axio-dilaton, and the three-form field strengths as
\be
G\equiv\ii \me^{\phi/2}(\tau \dd B^{(2)} -\dd C^{(2)})\,.
\ee
We are interested in bosonic solutions and therefore we set the fermionic content of the theory to zero. We still wish to preserve supersymmetry, therefore the supersymmetry variations must vanish identically. The bosonic variations, being proportional to fermionic fields vanish trivially, whilst for the fermionic variations
\bea
\delta\psi_{M}&=& \mathcal{D}_{M}\epsilon-\frac{1}{96}\left(\tensor{\Gamma}{_{M}^{P_{1}..P_{3}}}G_{P_{1}..P_{3}}-9\Gamma^{P_{1}P_{2}}G_{MP_{1}P_{2}}\right)\epsilon^{c}
+\frac{\ii}{192}\Gamma^{P_{1}...P_{4}}F_{MP_{1}...P_{4}}\epsilon \label{susy1}\,,\\   
\delta \lambda&=& \ii \Gamma^{M}P_{M}\epsilon^{c}+\frac{\ii}{24}\Gamma^{P_{1}..P_{3}}G_{P_{1}...P_{3}}\epsilon \label{susy2} \,,
\eea
we must choose the bosonic fields such that they vanish. These are the Killing spinor equations of Type IIB supergravity.  The supersymmetry parameter $\epsilon$ is a Weyl spinor satisfying the projection condition $\Gamma_{11}\epsilon=-\epsilon$.

The covariant derivative $\mathcal{D}$ is with respect to both Lorentz transformations and local $U(1)$ transformations,
\be
\mathcal{D}_{\mu}=\nabla_{\mu}-\ii q Q_{\mu} \,,
\ee
where $Q$ is the gauge field for the $U(1)$ transformations. It takes the form
\be
Q=-\frac{1}{2 \tau_{2}}\dd \tau_{1} \,,
\ee
where $\tau=\tau_{1}+\ii \tau_{2}$ is the axio-dilaton and $q$ is the charge
under the $U(1)$.  The Killing spinors have $U(1)$ charge $1/2$, $P$ has
charge $2$, and $G$ has charge $1$.
The field $P$, appearing in the dilatino equation, is constructed from the axio-dilaton as
\be
P=\frac{\ii}{2 \tau_{2}}\dd \tau\,.
\ee
The equations of motion consist of the Einstein equation
\be
R_{MN}=2P_{(M}P^{*}_{N)}+\frac{1}{96}F_{MP_{1}..P_{4}}\tensor{F}{_{N}^{P_{1}..P_{4}}}+\frac{1}{8}\lb
2\tensor{G}{_{(M}^{P_{1}P_{2}}}G^{*}_{N)P_{1}P_{2}}-\frac{1}{6}g_{MN}G^{P^{1}..P_{3}}G^{*}_{P_{1}..P_{3}}\rb
\,, \label{Einsteineq}
\ee
and the equations of motion for the fluxes
\bea
\mathcal{D}*G&=&P\wedge *G^{*}-\ii F\wedge G\,,\\
\mathcal{D}*P&=&-\frac{1}{4}G\wedge *G\,.
\eea
These are supplemented by the Bianchi identities
\bea
\mathcal{D}P&=&0\,,\\
\mathcal{D}G&=&-P\wedge G^{*}\,,\\
\dd F&=&\frac{\ii}{2}G\wedge G^{*}\,,
\eea
and the self-duality constraint
\be
F=*F\,.
\ee

\subsection{AdS$_{3}$ Ansatz}

In this paper we consider the most general class of bosonic Type IIB supergravity solutions with $SO(2,2)$ symmetry and vanishing three-form fluxes. We take the 10d metric in Einstein frame to be a warped product of the form
\be
\dd s^{2}=\me^{2 \Del}\lb \dd s^{2}(\text{AdS}_{3})+\dd s^{2}(\M_{7})\rb\,,
\ee
where $\dd s^{2}($AdS$_{3})$ is the metric on AdS$_{3}$, with Ricci tensor $R_{ab}=-2 m^{2}g_{ab}$, and  $\dd s^{2}(\M_{7})$ is the metric on an arbitrary internal 7d manifold $\M_{7}$. To preserve the $SO(2,2)$ symmetry of AdS$_3$ we impose $\Del\in\Omega^{(0)}(\M_{7},\mathbb{R}),~P\in \Omega^{(1)}(\M_{7},\mathbb{C})$ and $~\tau\in \Omega^{(0)}(\M_{7},\mathbb{C})$. In this paper we will not consider solutions with non-trivial three-form fluxes, thus our fluxes take the form
\be
F=(1+*)\text{dvol}(\text{AdS}_{3})\wedge F^{(2)} \,,\qquad G = 0\,,
\ee
with $F^{(2)}\in\Omega^{(2)}(\M_{7},\mathbb{R})$.  The Bianchi identity for the five-form flux gives two equations for $\ff$, that read
\bea
\dd F^{(2)}=0 \,,\qquad 
\dd\hat{*}_{7}F^{(2)} =0 \,,
\eea
where $\hat{*}_{7}$ is the Hodge star on the unwarped metric $\dd s^{2}(\M_{7})$. 

We now use the spinor ansatz developed in appendix \ref{spinconv} and apply the results of appendix \ref{Wsimp} for Killing spinors of $\AdS_3$. We consider the Killing spinor ansatz
\be
\epsilon=\psi_{1}\otimes \me^{\Del/2} \xi_{1}\otimes \theta+\psi_{2}\otimes
\me^{\Del/2}\xi_{2}\otimes \theta\label{KSansatz} \,,
\ee
where $\psi_{i}$ are Majorana Killing spinors on AdS$_{3}$ and satisfy 
\be
\nabla_{a}\psi_{i}=\frac{\alphai_{i} m}{2}\rho_{a}\psi_{i}\label{AdSKSeq}\,.
\ee
The constants $\alphai_{i}=\pm 1$ are the eigenvalues of the matrix $W$ discussed in
appendix \ref{Wsimp}. We assume that the $\psi_{i}$ are independent and that
the $\xi_{i}$ are Dirac spinors. With this ansatz we can preserve
$\mathcal{N}=\{(4,0),~(2,2),~(0,4)\}$, depending on the signs in
(\ref{AdSKSeq}), in the dual SCFT, as discussed in appendix \ref{Wsimp}. Each Majorana AdS$_{3}$ Killing spinor will contribute a single superconformal supercharge and a single Poincar\'e supercharge. The two independent Dirac spinors will then imply that we preserve four superconformal supercharges and four Poincar\'e supercharges and hence $\mathcal{N}=\{(4,0),~(2,2),~(0,4)\}$ depending on the signs in (\ref{AdSKSeq}).

Decomposing the 10d supersymmetry equations, (\ref{susy1}) and (\ref{susy2}), by using the above ansatz (\ref{KSansatz}) we obtain the equations
\bea
\gamma^{\mu}P_{\mu}\xi_{j}^{c}&=&0\label{dilalgg=0}\,,\\[2mm]
\lb \frac{1}{2} \partial_{\mu}\Del\gamma^{\mu}-\frac{\ii \alphai_{j}m}{2}+\frac{ \me^{-4\Del}}{8}\slashed{F}^{(2)}\rb \xi_{j}&=&0\,,\label{gravalgg=0}\\
\lb\D_{\mu} +\frac{\ii \alphai_{j}m}{2}\gamma_{\mu}-\frac{\me^{-4 \Del}}{8}F^{(2)}_{\nu_{1}\nu_{2}}\tensor{\gamma}{_{\mu}^{\nu_{1}\nu_{2}}}\rb\xi_{j}&=&0~ \label{gravdifg=0},
\eea
for each of the two spinors $\xi_{i}$. Note that we can derive some immediate consequences of the algebraic condition (\ref{dilalgg=0}), which implies
\bea
P^{2}\bar{\xi}_{j}\xi_{i}=0\,.
\eea
In particular for $i=j$ we have that $\bar{\xi}_{i}\xi_{i}\not=0$ and therefore we see that necessarily 
\be
P^{2}=0  \,.
\label{upsy}
\ee
Together with the equation of motion $\diff* P =0$, this implies that $\tau$ is harmonic 
\be
\Box \, \tau =0 \,.
\label{desy}
\ee
We are specifically interested in solutions where $\tau$ varies over the
compact part of the space. In a compact space without boundary the only
 harmonic
functions are constant \cite{MR2451566}.  This was noted in  \cite{Gauntlett:2005ww} in the
context of supersymmetric AdS$_5$ solutions and it implies that we must allow
for singularities in $\tau$, as anticipated from general F-theory
considerations; in particular we will allow for log singularities as in (\ref{tauLog}).

\subsection{Constraints for 2d $(0,4)$ Supersymmetry}

 In the rest of the paper we specialise to the case where the dual 2d SCFTs  have chiral supersymmetry. In particular, we choose the convention where the supercharges are right-moving \emph{i.e.} $\mathcal{N} = (0,4)$ supersymmetry, which  implies we take $\alpha_1 = \alpha_2 = 1$. 

To classify the solutions we shall analyse the G-structure defined by the Killing spinors of the solution. This is a standard technique for finding the necessary and sufficient conditions imposed by supersymmetry  \cite{Gauntlett:2002sc}. Let the Killing spinors of the solution have isotropy group G, this then defines a canonical G-structure. One may construct tensors from these Killing spinors as spinor bilinears, it follows that these tensors are then G-invariant.  These G-invariant tensors will satisfy a number of algebraic relations  depending on the particular G-structure defined. One then analyses the information obtained from the Killing spinor equations by computing the differential and algebraic conditions they impose on the G-invariant tensors, these define the so called ``intrinsic torsion'' of the G-structure. Finally one computes the integrability conditions of the Killing spinor equations and the torsion conditions of the G-invariant tensors, and determines which equations of motion and Bianchi identities are automatically  satisfied, imposing the remaining conditions. This classification procedure provides the most general local form of the solution given in terms of the information contained in the G-structure. 

We find that the most general solutions in this class admit an $SU(2)$ structure. In seven-dimensions an $SU(2)$ structure implies the existence of three independent one-forms orthogonal to a four-dimensional foliation with $SU(2)$ structure. This is specified by a real two-form of maximal rank and a complex two-form satisfying the $SU(2)$ structure relations which we give later. We define the G-invariant tensors obtained from the Killing spinors in appendix \ref{app:TorsionConditions}. To compute the algebraic relations imposed by the $SU(2)$  structure we shall introduce an orthonormal frame using the gamma matrices defined in appendix \ref{spinconv}. One may recover these results by making use of Fierz identities. 

In the following we summarise the results, with more detailed provided in appendix \ref{app:TorsionConditions}, where in particular the torsion conditions for general $\alpha_i$ are written down. Here we specialise to the relevant case $\alpha_1 = \alpha_2 = 1$.

From (\ref{xicixij=0}) and (\ref{dsijg=0}) we obtain the following conditions on the scalar bilinears
\bea
S_{11}=S_{22}&=&1\,,\\
A_{11}=A_{22}=A_{12}=S_{12}&=&0\,.
\eea
From (\ref{killg=0}) we see that there are three independent Killing vectors.
Imposing that the Killing vectors lie along a subspace defined by the vielbeins $e^{5},~e^{6},~e^{7}$, consistent with an $SU(2)$ structure, is equivalent to imposing the projection condition
\be
\gamma_{1234}\xi_{i}=-\xi_{i}\,.
\ee
In addition we have the freedom to choose $K_{11}$ to be parallel to $e^{7}$.
In this frame the independent one-forms and two-forms are given by \footnote{For notational simplification, we shall make use of the following shorthand notation for the wedge of multiple vielbein $e^{a_1}\wedge ..\wedge e^{a_n}=e^{a_1..a_n}$.}
\bea
K_{11}&=&-K_{22}=e^{7}\,,\\
K_{12}&=&e^5-\ii e^6\,,\\
B_{12}&=&0 \,,\\
U_{11} &=&-\ii (e^{12}+e^{34}-e^{56})\,,  \\
U_{22}&=&-\ii (e^{12}+e^{34}+e^{56})\,,\\
V_{11}&=&V_{22}=0  \,,\\
V_{12}&=&-(e^{1}-\ii e^{2})\wedge (e^{3}-\ii e^{4})\,.
\eea
The remaining forms may be expressed in terms of the forms defined above as
\be
\ba
U_{12}=&K_{11}\wedge K_{12}\,,\\
X_{11}=&U_{11}\wedge K_{11}\,,\\
X_{22}=&U_{22}\wedge K_{22}\,,\\
X_{12}=&U_{11}\wedge K_{12}=U_{22}\wedge K_{12}\,,\\
Y_{11}=&V_{12}\wedge K_{12}^{*}\,,\\
Y_{22}=&-V_{12}\wedge K_{12}\,,\\
Y_{12}=&-V_{12}\wedge K_{11}=V_{12}\wedge K_{22}\,.
\ea\ee


\subsection{F-theory AdS$_3$ Solution}
\label{Fads3:section}

After introducing the frame, it is now possible to reduce the differential conditions and determine the final Type IIB supergravity solution. The remaining conditions are 
\bea
\me^{-4 \Del}\dd \lb \me^{4 \Del}K_{jj}\rb &=&-2\ii m  U_{jj}-\me^{-4 \Del}\ff\,,\label{computeF}\\
\me^{-4 \Del}\dd \lb \me^{4 \Del}K_{12}\rb&=&-2\ii m  U_{12}\,,\label{K12clear}\\
\dd \lb \me^{4 \Del}U_{ij}\rb&=&0\,,\label{Uijclear}\\
\D\lb \me^{6 \Del}V_{12}\rb &=&0\label{V12clear} \,.
\eea
First we determine $\ff$. The frame computation implies $K_{11}=-K_{22}=e^{7}$, and inserting this into (\ref{computeF}) we find
\bea
\ff&=&-\ii  m \me^{4 \Del}(U_{11}+U_{22})= - 2  m \me^{4 \Del}\lb e^{12}+e^{34}\rb\,.\label{fexplicit}
\eea
Notice that (\ref{Uijclear}) implies that \eqref{fexplicit} satisfies the Bianchi identity for $\ff$.  From this explicit expression we may also compute $*\ff$ and show that it satisfies its equation of motion. Observe from the algebraic equations (\ref{algg=01})-(\ref{algg=04}) we have the relation
\be \label{RicciFRel}
\partial_{\mu}\Del=-\frac{\me^{-4\Del}}{4}F_{\mu\nu}\bar{\xi}_{1}\gamma^{\nu}\xi_{1}\,,
\ee
which together with \eqref{fexplicit} implies that the warp factor is constant
\be
\dd \Del=0 \,.
\ee
Next we can determine the Killing vectors. 
From (\ref{killg=0}) we see that there are three independent Killing vectors of the \emph{full} solution whose dual one-forms are
\be
\ba
K_{11}&=-K_{22}=e^{7}\,,\\
\Real[K_{12}]&=e^{5}\,,\\
\Imag[K_{12}]&=-e^{6}\,.
\ea\ee
From the torsion conditions (\ref{computeF}) the dual one-forms to these Killing vectors satisfy the differential conditions
\bea
\dd  e^{5}&=&2m e^{67}\,,\label{S3cond1}\\
\dd  e^{6} &=& 2m e^{75}\,,\label{S3cond2}\\
\dd e^{7} &=& 2m e^{56}\,,\label{S3cond3}
\eea 
which is a warped form of the equations obeyed by the $SU(2)$  invariant
one-forms \footnote{The Maurer--Cartan left-invariant one-forms in the coordinates we are using are
$$
\sigma_{1,L}=-\sin \psi \dd \theta +\cos \psi \sin \theta \dd \varphi\,,~~\sigma_{2,L}=\cos \psi \dd \theta +\sin \psi \sin \theta \dd \varphi\,,~~\sigma_{3,L}=\dd \psi +\cos \theta \dd \varphi\,.
$$
The coordinates have periods
$\psi\in[0,4\pi],~\varphi \in [0,2 \pi],~\theta \in [0,\pi]$\,.These satisfy
$\dd \sigma_{i,L}=\frac{1}{2}\epsilon_{ijk}\sigma_{j,L}\wedge \sigma_{k,L}$.
The right-invariant one-forms we shall take are
$$
\sigma_{1,R}=\sin \varphi \dd \theta-\cos \varphi\sin \theta \dd
\psi\,,~~\sigma_{2,R}=\cos \varphi \dd \theta+\sin \varphi \sin \theta \dd
\psi\,,~~ \sigma_{3,R}=\dd \varphi +\cos \theta \dd \psi \,.
$$ 
These satisfy $\dd \sigma_{i,R}=-\frac{1}{2}\epsilon_{ijk}\sigma_{j,R}\sigma_{k,R}$. Of course we could take the right-invariant one-forms to solve the same equation as the left-invariant one-forms however in the present case we have the desirable property $\sigma_{1,L}\wedge \sigma_{2,L}\wedge \sigma_{3,L}=\sigma_{1,R}\wedge \sigma_{2,R}\wedge \sigma_{3,R}$.}. 

On the 4d subspace $B$ we have an $SU(2)$ structure with the K\"ahler-form given by
\be
J_B=\frac{\ii}{2}\lb U_{11}+U_{22}\rb = e^{12}+e^{34} \,,
\ee
with corresponding holomorphic two-form satisfying  $\half \Omega_B \wedge \bar\Omega_B = J_B \wedge J_B = 2 \text{dvol}(B)$, which takes the form
\be
\Omega_B =- V^*_{12}=\lb e^{1}+\ii e^{2}\rb \wedge \lb e^{3}+\ii e^{4}\rb \,.
\ee
With these definitions the remaining torsion conditions become
\bea
\dd J_B&=&0\,,\label{dJB=0}\\
\bar{D} \Omega_B&=&0\,.\label{Domegag=0}
\eea
Furthermore, from (\ref{Pkilled1}) and (\ref{Pkilled2}) it follows that $P$ may only have components along $B$ and using (\ref{dilalgg=0}) we find
\be
\tensor{J}{_{B }_{m}^{n}}P_{n}=\ii P_{m}\,,
\label{iggle}
\ee
and hence $P$ is holomorphic. 
The form of $P$ then implies that $\tau$ is holomorphic and therefore $\tau$ satisfies
\be
\dd \tau \wedge \Omega_B=0 \,.
\label{piggle}
\ee 

Notice that  (\ref{iggle}) implies the necessary conditions (\ref{upsy}) and (\ref{desy}). From (\ref{Domegag=0}) we may identify $-Q$ as the canonical Ricci-form potential on the K\"ahler manifold $B$ and hence we have
\be
\mathfrak{R}+ \dd Q =0 ~\label{dQg=0},
\ee
where $\mathfrak{R}$ is the Ricci-form on $B$. 
Making use of the identity,
\be
\tensor{\mathfrak{R}}{^{m_{1}}_{m_{2}}}=\tensor{J}{^{m_{1}}_{n}}\tensor{R}{^{n}_{m_2}}\,,
\ee
the condition in \eqref{dQg=0} may be expressed as
\be
R_{mn}^{Y_3} \equiv  R_{mn}- \frac{1}{2 \tau_{2}^{2}}\lb \partial_{m}\tau_{1}\partial_{n}\tau_{1}+\partial_{m}\tau_{2}\partial_{n}\tau_{2}\rb =0 \,.\label{Riccicondition}
\ee
This equation relates the Ricci tensor of the base to the variation of $\tau$ over $B$. In particular, this is the Ricci flatness condition for the metric of an elliptically fibered Calabi--Yau threefold $Y_3$ valid away from the singularities in the fiber. 

Since $\tau$ is holomorphic, away from loci where the fiber degenerates, the metric for the elliptically fibered Calabi--Yau can be written as
\be
\dd s^{2}(Y_3)=\frac{1}{\tau_{2}}\lb (\dd x +\tau_{1}\dd y)^{2}+\tau_{2}^{2}\dd y^{2}\rb +\dd s^{2}(B)\,.\label{singularcy}
\ee
Indeed, imposing that this metric is Ricci flat implies that the Ricci tensor on $B$ satisfies (\ref{Riccicondition}).  As was noted in \cite{Greene:1989ya} this local metric is singular over the discriminant locus of the elliptic fibration.

To exhibit the Calabi--Yau condition, we construct the K\"ahler form and holomorphic three-form of the Calabi--Yau threefold from the corresponding quantities of the base, which define an $SU(3)$ structure. Let the vielbein on the fibration be
\be
e^1=\frac{1}{\sqrt{\tau_2}}(\dd x +\tau_1 \dd y)\,,~~e^2=\sqrt{\tau_{2}} \dd y\,,\label{EFframe}
\ee
then
\be
J_{Y_3}=e^{12}+J_{B} \,,\qquad \Omega_{Y_3}=(e^1+\ii e^2)\wedge \Omega_B\,.
\ee
With this frame on the elliptic fibration and giving indices $3,4,5,6$ to the base, some relevant components of the spin connection which are useful later on are
\bea \label{SpinConnections}
\tensor{\omega}{^{1}_{2}} = Q, \qquad 
\tensor{\omega}{^{3}_{4}}+\tensor{\omega}{^{5}_{6}}=-Q\,,\qquad 
\tensor{\omega}{^{4}_{5}}+\tensor{\omega}{^{3}_{6}}=0\,,\qquad 
\tensor{\omega}{^{3}_{5}}-\tensor{\omega}{^{4}_{6}}= 0\,.
\eea
The Calabi--Yau condition in terms of this $SU(3)$ structure is equivalent to
\be
\dd J_{Y_3}=0\,,~~\quad \dd \Omega_{Y_3}=0\,.
\ee
Upon using $\dd J_B=0$ it follows trivially that $\dd J_{Y_3}=0$.  Consider instead $\dd \Omega_{Y_3}=0$; we have
\bea
\dd \Omega_{Y_3}&=&\lb-\frac{1}{2 \tau_{2}}\dd \tau_{2}-\ii Q\rb \wedge \Omega_{Y_3} +\frac{1}{\tau_{2}}\dd \tau \wedge \dd y \wedge \Omega_{B}\nonumber\\
&=&\frac{\ii}{2 \tau_{2}}\dd \tau\wedge \Omega_{Y_3}+\frac{1}{\tau_{2}}\dd \tau \wedge \dd y \wedge \Omega_{B}\,,
\eea
where we have used (\ref{Domegag=0}) in the first line. Upon using the holomorphicity of $\tau$ and that it depends only on the base coordinates this is identically zero. This shows that (\ref{Domegag=0}) and therefore also (\ref{dQg=0}) and (\ref{Riccicondition}) are equivalent to $B$ being the base of an elliptically fibered Calabi--Yau threefold. 

The Type IIB equations of motion follow immediately from supersymmetry. The equation of motion for $F$, $\diff F=0$, follows immediately from (\ref{dJB=0}). Moreover one may rewrite (\ref{Riccicondition}) as
\be
R_{mn}=2 P_{(m}P^{*}_{n)}\, ,
\ee 
which is precisely the form in which $\tau$ appears in the Einstein equation (\ref{Einsteineq}). With the explicit form of $F$ it is easy to show that indeed the Einstein equation is also satisfied.

In summary, the solution
\be
\ba
& \mathbb{E}_\tau \cr 
& \downarrow \cr 
\mathrm{AdS}_3 \times S^3 \times \ & B
\ea\ee
where the elliptic fibration over $B$ gives rise to a Calabi--Yau threefold,  is thus given by\footnote{Of course, the one-forms $\sigma_i$ can be taken to be either $\sigma_{i,L}$ or $\sigma_{i,R}$.}
\begin{align}
\dd s^{2}&=\me^{2 \Del}\lb\dd s^{2}(\text{AdS}_{3})+\frac{1}{4 m^{2}}\lb \sigma_{1}^{2}+\sigma_{2}^{2}+\sigma_{3}^{2}\rb +{\frac{1}{\mb^{2}}}\dd s^{2}(B)\rb\,,\label{IIBmetric} \\
F&=-(1+*) \frac{2m \me^{4 \Del}}{{\mb^2}}  J_B\wedge \dd\vol(\mathrm{AdS}_{3})\, ,
\label{IIBF5} \\
P&=\frac{\ii}{2 \tau_{2}}\dd \tau\,,
\end{align}
where $J_B$ is the K\"ahler form on the base of the elliptically fibered Calabi--Yau threefold, and $\tau$ varies holomorphically over $B$. Here, $1/\mb$ is the length scale associated to the base $B$. 

The possible base manifolds of an elliptically fibered Calabi--Yau threefold
were determined in \cite{MR1109635,MR1217381} and found to be one of the
following: $\mathbb{P}^2$, Hirzebruch surfaces $\mathbb{F}_m$, blow-ups
thereof, and Enriques surfaces. In the case where the elliptic fibration is trivial then the base itself must be a Calabi--Yau two-fold, which is either a K3 surface or $T^4$. This is precisely the solution obtained in \cite{Kim:2005ez} which results in $(4,4)$ supersymmetry, and is the dual to the classic D1-D5 system \cite{Maldacena:1997re}.

At this point it is perhaps timely  to recall that our description is valid away from the singular loci of $\tau$ \footnote{This manifests itself \emph{e.g.} by noting that since  $ c_1(B)=2 \pi \mathfrak{R}$ we would have $c_1(B)\wedge c_1(B)=0$, contradicting the global property that 
$$
\int_{B}c_1(B)\wedge c_1(B) = 10-h^{1,1}(B)\,.
$$
}.
As explained earlier, we will allow for singularities in $\tau$, given by for instance by
(\ref{tauLog}), which have a characterisation in terms of Kodaira singular fibers. The Ricci-flatness condition then takes the form
\be
K_B = - \sum_i a_i D_i \,,
\ee
where $D_i$ are the Cartan divisors of the resolution of the singularity and $a_i$ depend on the Kodaira type of the singular fiber \cite{Morrison:1996na, Morrison:1996pp}.

For the case of an elliptically fibered K3 surface with 24 $I_1$ singularities, a semi-Ricci-flat metric was constructed in \cite{GrossWilson}. The metric in the neighborhood of each $I_1$ fiber is given by the Ooguri-Vafa metric \cite{Ooguri:1996me}. The semi-flat metric was constructed by gluing the Ooguri-Vafa metric to the metric constructed in \cite{Greene:1989ya} around the 24 points where the fiber becomes singular. It was shown in \cite{GrossWilson} that in the limit $\text{vol}(\mathbb{E}_\tau) \rightarrow 0 $ the semi-flat metric reduced to a singular metric on $\mathbb{P}^1$, the base of the elliptic K3, where the singularities are exactly at the points where the fiber is singular. 
In \cite{Aharony:1998xz, Ahn:1998tv} the metric in \cite{Greene:1989ya} was used to give some estimate of the curvature singularity, and it was argued that in the large $N$ limit, the gravity approximation can still be trusted. 
One expects in higher dimensions that the metric on the base is also singular in the F-theory limit. 
However, as we shall discuss in section \ref{sec:FTheoryCC}, one is still able to compute quantities of the dual CFT using this solution. It would be interesting to estimate the curvature singularities in these higher-dimensional cases, to support these findings. 

In the next subsection we shall describe a supersymmetry preserving $\Z_M$ quotient of these solutions. 
This will be important for identifying the superconformal R-symmetry of the dual $(0,4)$ SCFT in the IR, and furthermore will be a key ingredient in performing the duality to 11d supergravity in section \ref{sec:FMduality}.

\subsection{Lens Space Solution}\label{Lens}
\label{LensSpace}

Manifest in the solution is an $S^{3}$ which has isometry group
$SO(4)\simeq SU(2)_L\times SU(2)_R$, a subgroup of which realises the R-symmetry of 
the dual SCFT geometrically. The Killing spinors transform non-trivially under
the R-symmetry but are singlets under flavour symmetries. We shall find that
the Killing spinors of this solution are only charged under one of the
$SU(2)$s, which identifies the small $\mathcal{N}=(0,4)$ superconformal R-symmetry.
Furthermore, by inspection of the Killing spinors it is apparent that one can extend the solution found above by quotienting the $S^3$ by 
a discrete group $\Gamma\subset SU(2)_L$ and still preserve the same amount of supersymmetry. This
generalises the solution described in section \ref{Fads3:section} to the class
\be 
\mathrm{AdS}_3 \times S^3/\Gamma \times B\,.
\ee
We will focus on the case that $\Gamma=\Z_M$, where the quotient has the effect of changing the period of $\psi$, the coordinate of the Hopf fiber, so that $\psi \sim  \psi+4\pi/M$ rather than being $4\pi$ periodic. We shall show that the Killing spinors we obtain are $SU(2)_{L}$ singlets, and in particular independent of $\psi$, therefore quotienting by $\Z_M$ does not break any supersymmetry. 

It suffices to compute the Killing spinors in Einstein frame as this will not affect the above analysis. Moreover, as we have taken the Killing spinors to be a direct product as in (\ref{KSansatz}) we need only consider solving the seven-dimensional Killing spinor equations (\ref{dilalgg=0})-(\ref{gravdifg=0}). The Killing spinor equation obtained by restricting (\ref{gravdifg=0}) to the base of the elliptically fibered Calabi--Yau is
\be
\nabla_{m}\xi-\frac{\ii}{2}Q_{m}\xi=0\,.
\ee
This follows by restricting the covariantly constant Killing
spinor equation of the elliptically fibered Calabi--Yau to the base by using
the results for the spin connection in (\ref{SpinConnections}). Equivalently,
one can notice that this is precisely the canonical spin$^c$ Killing spinor
equation on a K\"ahler manifold where $-Q$ is the Ricci one-form potential, as
shown in the previous subsection\footnote{Recall that this is a local
equation as $Q$ and the metric on $B$ are singular.}. One may take the
Killing spinor on the base of an elliptically fibered Calabi--Yau manifold to
be constant if one imposes suitable projection conditions. Using the relations
for the spin-connection of an elliptically fibered Calabi--Yau, as computed in
(\ref{SpinConnections}), one finds that the projection conditions are
\be
\gamma^{34}\xi=\gamma^{56}\xi=-\ii \xi\,,\label{projection}
\ee
where the indices are flat.
In conclusion, to solve the Killing spinor equation on the base we need only to consider a constant spinor satisfying the projection conditions (\ref{projection}). 
Note that (\ref{gravalgg=0}) is automatically satisfied thanks to (\ref{projection}) and (\ref{fexplicit}). Moreover, holomorphicity of $\tau$ and (\ref{projection}) imply 
 that (\ref{dilalgg=0}) is also satisfied. One therefore needs only solve (\ref{gravdifg=0}) for the $S^{3}$ indices.

One may use the explicit form of the flux (\ref{fexplicit}) to reduce (\ref{gravdifg=0}) on the $S^{3}$ to
\be
\nabla_{\hat{a}}\xi=\frac{\ii m}{2}\gamma_{\hat{a}}\xi\,.
\ee
With the vielbein
\be
e^{1}_{(S^{3})}=-\frac{1}{2m}\sigma_{1,R}\,,~~ e^{2}_{(S^{3})}=-\frac{1}{2m}\sigma_{2,R}\,,~~ e^{3}_{(S^{3})}=-\frac{1}{2m}\sigma_{3,R}\,,
\ee
where $\sigma_{i,R}$ are right-invariant one-forms, one finds that the constant spinor solves this final set of conditions. The Killing spinor is therefore a constant spinor subject to the projection conditions (\ref{projection}), and therefore has four real components consistent with preserving $(0,4)$ supersymmetry. As the solution is constant in $\psi$, there is no ambiguity in the definition of the spinor if we quotient the $S^{3}$ by $\Z_{M}$. We may therefore replace the $S^{3}$ factor in the solution by the Lens spaces $S^{3}/\Z_{M}$ without breaking supersymmetry and still satisfying all equations of motion and Bianchi identities. We shall give a physical interpretation of this quotient in section \ref{sec:KKM}.

Having computed the Killing spinors we may now determine the R-symmetry. On the $S^{3}$ there are six Killing vectors corresponding to the six generators of $SO(4)\simeq SU(2)_{L}\times SU(2)_{R}$. These are the three dual to the left-invariant one-forms 
\be\ba
k_{(1)}&=\partial_{\psi}\,,~k_{(2)}=-\cos\psi\cot\theta \partial_{\psi}-\sin \psi \partial_{\theta}+\frac{\cos\psi}{\sin\theta}\partial_{\varphi}\,, \cr 
k_{(3)}&=-\sin \psi \cot\theta \partial_{\psi}+\cos\psi \partial_{\theta}+\frac{\sin\psi}{\sin\theta}\partial_{\varphi}\,,
\ea\ee
and  the three dual to the right-invariant one-forms
\be
k_{(4)}=\frac{\sin\varphi}{\sin\theta}\partial_{\psi}+\cos\varphi\partial_{\theta}-\cot\theta
\sin \varphi \partial_{\varphi}\,,~~k_{(5)}=-\frac{\cos\varphi}{\sin
\theta}\partial_{\psi}+\sin \varphi\partial_{\theta}+\cot \theta \cos\varphi
\partial_{\varphi}\,,~~k_{(6)}=\partial_{\varphi} \,,
\ee
with each set satisfying the $SU(2)$ Lie algebra. The spinorial Lie-derivative along a Killing direction, $K$, is defined to be 
\be
\mathcal{L}_{K}\epsilon=\lb K^{\mu}\nabla_{\mu}+\frac{1}{8}(\dd K)_{\nu_1 \nu_2}\gamma^{\nu_1 \nu_2}\rb \epsilon \,.
\ee
In order to ascertain along which directions the Killing spinor is charged one computes the spinorial Lie derivative along these directions. We find that the Killing spinor is invariant under the left-invariant Killing vectors and charged under the right-invariant Killing vectors. This implies that we can take the quotient by $\Gamma\subset SU(2)_L$, preserving the same amount of supersymmetry. Moreover, as discussed above this means that 
 we can identify $SU(2)_R$ with the $SU(2)_r$ R-symmetry of the dual SCFT. We note that the spinorial Lie derivative is frame independent (subject to preserving the same orientation, which is correlated with the choice of $SU(2)$ under which the Killing spinors are charged) and therefore this result is non-ambiguous.

It is a well known fact in the literature that performing a T-duality along a Killing direction with vanishing spinorial Lie-derivative for the Killing spinor along the Killing vector leads to a Killing spinor in the dual solution. It is clear from the results above that one may dualize along the Hopf fiber without breaking supersymmetry, 
which will be used later on to determine the dual M-theory solution.

\subsection{Flux Quantisation}

To complete the solution, we need to ensure that the five-form field strength, $F$, is properly quantized through all the integral five-cycles in the 7d manifold transverse to AdS$_{3}$. We impose that\footnote{In the following we will set $g_s =1$.}
\be
n(M_{\alpha})=\frac{1}{(2 \pi l_{s})^{4}\gs}\int_{M_{\alpha}}F \in \mathbb{Z}
\ee
for all $M_{\alpha}\in H_{5}(\M_{7},\Z)$. The five cycles which contribute are of the form $S^{3} \times \mathcal{C}$, where $\mathcal{C}$ is any two-cycle in the  base $B$ of the Calabi--Yau. We therefore find \footnote{We have defined $\dd \vol(S^{3}/\Z_{M})=\sigma_{1,R}\wedge \sigma_{2,R}\wedge \sigma_{3,R}$ which gives $\vol(S^{3}/\Z_{M})=\frac{16 \pi^{2}}{M}$. Notice that this is not the volume form of the unit radius Lens space $S^{3}/\Z_{M}$.}
\be \ba 
n(M_{\alpha})&=-\frac{\me^{4 \Del}\vol(S^{3}/\Z_M)}{(2 \pi l_{s})^{4}\gs4 m^{2}{\mb^{2}}}\int_{Y_3}\omega_0 \wedge \pi^*J_B \wedge \omega_{\alpha} \\
&=-\frac{\me^{4 \Del}\vol(S^{3}/\Z_M)}{(2 \pi l_{s})^{4}\gs4 m^{2}{\mb^2}}\int_{C_\alpha}J_B \,	,
\ea \ee
where the $C_{\alpha}$ form a basis of cycles in $H_{2}(B,\Z)$. 

The possible bases $B$ for an elliptic Calabi--Yau threefold, as listed earlier, are projective, and therefore also Hodge manifolds \cite{MR0068871}, and moreover they admit an integral K\"ahler form. As $J_B$ is dual to a curve, we in fact have that $B$ is not only a Hodge manifold, but we in fact pick the Hodge metric on it. This implies that we can take
\be
\int_{C_\alpha}J_B= k_{\alpha} \  \in \ \mathbb{Z}^+ \,.
\label{betadef}
\ee
 Using \eqref{betadef} we find that $n(M_\alpha)$ are integer if we impose
\be
N  =\frac{\me^{4 \Del} \vol(S^{3}/\Z_M)}{4 m^{2}{\mb^2}(2 \pi l_{s})^{4}\gs} \in \Z \, .
\ee

\subsection{Brane Solutions and the Interpretation of the Quotient}
\label{sec:KKM}

In this subsection we shall give an interpretation of the $\Z_M$ quotient performed in section \ref{LensSpace}. To do so we shall construct smeared brane solutions whose near-horizon geometry is\footnote{For simplicity  we set the warp factor, $\Del$, to 0 in this section.}
\be
\dd s^{2}=\dd s^{2}(\text{AdS}_3)+\frac{1}{4m^{2}}\dd s^{2}(S^{3}/\Z_M)+\frac{1}{{\mb^{2}}}\dd s^{2}(B)\,.
\ee
We shall need to combine various D3-brane solutions, employing the harmonic function rule (see \cite{Peet:2000hn} for a review). 

We shall use this strategy to obtain a UV completion of the AdS$_3$ solution that we have in Type IIB in the near-horizon limit, which we refer to as the ``pre near-horizon limit''. 
In fact, as we will show below, we can construct \emph{two distinct} such solutions, both flowing to the same near-horizon geometry. We wish to consider $N$ D3-branes wrapping $\mathbb{R}^{1,1}\times C$ where $C$ is the curve in the base of $Y_3$, Poincar\'e dual to the K\"ahler form of the base. We shall first consider a solution in the background of $M$ KK-monopoles and later in the background $\R^4$. To realise the D3-branes extended along the curve Poincar\'e dual to $J= e^{12}+e^{34}$, with $\dd s^{2}(B)= e_{1}^{2}+\dots +e_{4}^{2}$ we shall formally view this as two stacks of D3-branes  \cite{Bena:2006qm}. The first stack will extend along $\mathbb{R}^{1,2}\times C_{12}$, where $C_{12}$ is the curve dual to $e^{12}$, 
and the second stack along $\mathbb{R}^{1,2}\times C_{34}$ each with the same number of branes, $N$. 

We begin by briefly recalling the metric for $M$ KK-monopoles and give a few comments that will be useful for later discussion. The metric  is
\be
\dd s^{2}= -\dd t^{2}+ \dd x_1^2 + \cdots+\dd x_{5}^{2} +\dd s^{2}_{TN_M} \,,
\ee
where $\dd s^{2}_{TN_M} $ is the Taub-NUT metric\footnote{Strictly speaking, the Taub-NUT metric has $M=1$ and this is non singular near to $r\to 0$. The metric with $M>1$ has an $\R^4/\Z_{M}$ singularity in the interior, and this can be resolved by replacing the single center metric with a Gibbons-Hawking multi-center metric, where near to each center the metric looks like $\R^4$. This metric develops $M-1$ two-cycles, that collapse to zero size in the single center singular metric.}
\be
{\mb^{2}}\dd s^{2}_{TN_M}=\lb 1+\frac{M}{r}\rb \lb \dd r^{2}+r^{2}(\dd \theta^{2}+\sin^2 \theta\dd \varphi^2)\rb+\left(1+\frac{M}{r}\right)^{-1}\lb \dd \psi+ M \cos \theta \dd \varphi\rb^{2}\,.
\ee
This metric is well-known to be hyper-K\"ahler and hence Ricci-flat.

This metric approaches the singular (for $M>1$)  metric on
$\R^4/\Z_{M}$ as $r\to 0$, whilst asymptotically, as $r\to \infty$, it approaches the cylinder $\R^3\times S^1$. One can set $M=0$ in the metric, obtaining exactly the flat metric on $\R^3\times S^1$. Moreover, choosing as harmonic function  $H(r)=1+\frac{M}{r}\to \frac{M}{r}$, a simple change of coordinates shows that this is exactly the metric on $\R^4/\Z_M$. This can be interpreted as saying that in the ``near-horizon'' limit the Taub-NUT metric approaches the latter. 

Let us first write the Type IIB solution corresponding to $N$ D3-branes wrapping $\mathbb{R}^{1,2}\times C_{12}$, 
\be\ba
{\mb^{2}}\dd s^{2}(D3,12)&=H^{-1/2}(r)(-\dd t^{2}+\dd x^{2}+e_{1}^{2}+e_{2}^{2})+H^{1/2}(r)(\dd s_{TN}^{2}+e_{3}^{2}+e_{4}^{2})\,,\cr
C^{(4)}&=\frac{1}{{\mb^{2}}}\lb 1-\frac{1}{H(r)}\rb \dd t\wedge \dd x\wedge e^{12}\,.
\ea\ee
To wrap $\mathbb{R}^{1,2}\times C_{34}$ we simply relabel $12\leftrightarrow 34$. We have inserted the D3-branes into the background of $M$ KK-monopoles. In particular, as remarked above, we shall \emph{smear} the D3-branes completely along the 34 directions in the manifold $B$, this has the affect of making the function $H(r)$ harmonic on Taub-NUT and not the overall transverse space to the stack of D3s. If we now use the harmonic function rule on these two configurations we obtain the solution
\bea
{\mb^{2}}\dd s^{2}&=& H(r)^{-1}(-\dd t^{2}+\dd x^{2})+H(r)\dd s_{TN}^{2}+\dd s^{2}(B)\nonumber\\
C^{(4)}&=&\frac{1}{{\mb^{2}}}\lb 1-\frac{1}{H(r)}\rb \dd t\wedge \dd x\wedge J_{B}
\eea
As commented above $H$ must be harmonic on Taub-NUT,  as such we may take 
\be
H(r)=1+\frac{\n}{r}\,, \qquad  \hbox{with}\qquad  \n \equiv \frac{(2 \pi \ls)^4  N{\mb^{4}}}{16 \pi^{2}} \,,\label{TNharmonic}
\ee
and $N$ the number of D3-branes.
The metric takes the form 
\be
{\mb^{2}}\dd s^{2}=\frac{r}{r+\n} (-\dd t^{2}+\dd x^{2})+ \frac{r+\n}{r}\dd s_{TN}^{2}+\dd s^{2}(B)\,.\label{UV1}
\ee
We recall that $B$ is the base of an elliptically fibered Calabi--Yau threefold and as such this necessarily requires $\tau$ to vary in the solution. This is an Einstein-frame solution to Type IIB supergravity with D3-branes and varying $\tau$.

Let us now take the near-horizon limit, $r\to 0$. We have
\bea
{\mb^{2}}\dd s^{2}_{r\rightarrow 0}&=&\frac{r}{\n}(-\dd t^{2}+\dd x^{2})+\dd s^{2}(B)+\frac{\n M}{r^{2}}\lb \dd r^{2}+r^{2}\dd s^{2}(S^{2})\rb +\frac{\n}{M}(\dd \psi +M \cos \theta \dd \phi)^{2}\nonumber\\
&=& \frac{r}{\n}(-\dd t^{2}+\dd x^{2})+\n M \frac{\dd r^{2}}{r^{2}}+\dd s^{2}(B)+\n M (\dd s^{2}(S^{3}/\mathbb{Z}_{M}))\,.
\eea
If we make the redefinition $M\n ={\mb^{2}}(4 m^{2})^{-1}$ and the change of coordinate $r= 4 \n^{2} M \rho^{2}$ we obtain
\bea
\dd s^{2}&=&\frac{1}{m^{2}}\lb \rho^{2}(-\dd t^{2}+\dd x^{2})+\frac{\dd \rho^{2}}{\rho^{2}}\rb+\frac{1}{4 m^{2}}\dd s^{2}(S^{3}/\mathbb{Z}_{M})+{\frac{1}{\mb^{2}}}\dd s^{2}(B)\,,
\eea
whilst the five-form becomes
\be
F=(1+*){\frac{2m}{\mb^{2}}} \dd\vol(\mathrm{AdS}_{3})\wedge J_{B}\,,
\ee
which  recovers exactly the AdS$_3$ solution. We have done this by inserting $M$ KK-monopoles into the background of $N$ D3-branes wrapping a curve, $C$ dual to $J_B$, on the base of an elliptically fibered Calabi--Yau threefold.

Let us now consider a different pre near-horizon limit of the AdS$_3$ solution. This will be obtained by replacing the Taub-NUT metric in the Type IIB solution by the flat space quotient $\R^4/\Z_M$. We shall see that the near-horizon solution agrees with the Taub-NUT solution. We may use the previous results to immediately write down the 
metric\footnote{Of course here we can simply take $\psi$ to have period $4\pi/M$.}
\be\ba
{\mb^{2}}\dd s^{2}=& H(R)^{-1}(-\dd t^{2}+\dd x^{2})+H(R)\lb \dd R^2+\frac{R^2}{4}((\dd \psi +\cos \theta \dd \varphi)^2+\dd \theta^2 +\sin^2 \theta \dd \varphi^2)\rb\cr 
&+\dd s^{2}(B)\label{UV2}
\ea\ee
where now $H(R)$ is a harmonic function on $\R^{4}$ and we take
\be
H(R)=1+\frac{\widetilde{\n} }{R^{2}}\, \qquad  \hbox{with}\qquad   \widetilde{\n}\equiv \frac{(2 \pi \ls)^{4}M N\mb^{4} }{4 \pi^{2}} \,.
\ee
This harmonic function should be contrasted  with (\ref{TNharmonic}) in the Taub-NUT case. The self-dual five-form flux takes the form
\be 
F=(1+*){\frac{1}{\mb^{4}}} \dd H(R)^{-1}\wedge \dd t \wedge \dd x \wedge J_B\,.
\ee
Taking the near-horizon limit, $R\rightarrow 0$ we obtain
\be
{\mb^{2}}\dd s^{2}=\frac{R^{2}}{\widetilde{\n}}(-\dd t^{2}+\dd x^{2})+\frac{\widetilde{\n}}{R^{2}}\dd R^{2}+\frac{\widetilde{\n}}{4} \dd s^{2}(S^{3}/\Z_M)+\dd s^{2}(B)\,.
\ee
After rescaling $R$ as $R\rightarrow \widetilde{\n} R $ and identifying the inverse radius of AdS$_3$ to be $\widetilde{\n}=\frac{{\mb^{2}}}{ m^{2}}$ one recovers precisely the AdS$_3\times S^{3}/\Z_M$ solution
\be
{\mb^{2}}\dd s^{2}= \frac{R^{2}}{m^{2}}(-\dd t^{2}+\dd x^{2})+\frac{1}{m^{2}R^{2}}\dd R^{2}+\frac{1}{4 m^{2}}\dd s^{2}(S^3/\Z_M)+\dd s^{2}(B)
\ee
in perfect agreement with (\ref{IIBmetric}). The flux becomes
\be
F=(1+*){\frac{2 m}{\mb^{2}}} \dd\vol(\text{AdS}_3)\wedge J_B\,,
\ee
in agreement with (\ref{IIBF5}). 

 We have constructed two different UV completions of the Type IIB AdS$_3$ solution that is our main interest. 
To do so, we needed to make some technical simplifications, regarding smearing of the branes and the application of the harmonic sum rule. The resulting solutions are therefore not the  fully localized brane solutions, before taking the near-horizon limit, which are  typically very difficult to construct. However these solutions will still be useful in our discussion. Moreover, we should also keep in mind that the metric on $B$ and $\tau $ were singular in the near-horizon limit and this feature will remain.

Notice that for any $N$ and any $M$, asymptotically the metric (\ref{UV1}) goes to $\R^{1,1}\times \R^3 \times S^1\times B$. This is the metric far away from the $N$ D3-branes. 
On the other hand, the metric (\ref{UV2}) asymptotically goes to $\R^{1,1}\times \R^4/\Z_{M} \times B$. So these are clearly two different UV completions of the near-horizon geometry. 
This becomes particularly instructive in the case of $M=1$: in this case both asymptotic spaces are smooth, however the solution in the presence of 1 KK-monopole comprises an asymptotic 
 $\R^3\times S^1$ geometry, whilst the solution with \emph{no} KK-monopoles comprises an asymptotic $\R^4$ geometry. However, they flow to exactly the same AdS$_3\times S^3$ solution in the IR.

The interpretation of this fact is that in the IR the field theories constructed from the two different UV setups, flow to the ``same'' SCFT in the large $N$ limit. This means that in this limit for example the two theories must have the same central 
charges, in the large $N$ limit. However, sub-leading corrections to the central charges may be possible.

Notice that one may set $M=0$ without any immediate problem in (\ref{UV1}), obtaining the metric
\be
{\mb^{2}}\dd s^{2}= \frac{r}{r+\n}(-\dd t^{2}+\dd x^{2})+\dd s^{2}(B)+\frac{r+\n}{r}\lb \dd r^{2}+r^{2}(\dd \theta^{2}+\sin^{2} \theta \dd \varphi^{2})+\dd \psi^{2}\rb\,.
\ee
Notice that the Calabi--Yau base is a direct product with the remaining six-dimensional metric. Computing the curvature invariants of the six-dimensional metric, we find
\bea
\text{R}&=&0\,,\nonumber\\
\text{R}_{AB}\text{R}^{AB}&=&\frac{3 \n^{4}}{2 r^{2}(\n+r)^6}\\
\text{R}_{ABCD}\text{R}^{ABCD}&=&\frac{\n^2(11 \n^2+32 \n r+48 r^2)}{2 r^2(\n+r)^6}\nonumber
\eea
and therefore the metric is singular at $r=0$. In fact, upon taking the near-horizon limit there is no longer an AdS$_3$ factor. In other words, putting the D3-branes transverse to the space $\R^3\times S^1$ gives rise to a solution that does not contain an AdS$_3$ factor in the IR, and in fact has a curvature singularity  as $r\to 0$.


\section{F-theory Holographic Central Charges}
\label{sec:FTheoryCC}

In this section we compute the central charges for the solution derived in section \ref{Fads3:section}. As was noted previously, the metric on the base $B$, which is induced from the Calabi--Yau metric is singular. We shall circumvent potential problems arising with singular metrics, by carrying out our computations in the smooth Calabi--Yau threefold.


\subsection{Leading Order Central Charges}
\label{sec:IIBsugra04}

The leading order term for the central charges is given by the Brown-Henneaux formula \cite{Brown:1986nw} as summarised in appendix \ref{CentralChargeFormula}. Evaluating (\ref{cformula}) for the solution we find the leading order central charges to be
\be\ba
(\cIIBL)^{(2)}=(\cIIBR)^{(2)}=\cIIBsugra&=\frac{3 \me^{\Del}\vol(M_{7})}{2m G_{N}^{(10)}} =N^2 \frac{3 \vol(S^{3}/\Z_M)\vol(B) 32 \pi^{2}}{\vol(S^{3}/\Z_M)^{2} } \cr 
&=6N^2M\vol(B)  \,.
\ea\ee
We denote by $c^{(a)}$ the $\mathcal{O}(N^a)$ contribution to the central charge. 

In a smooth geometry we would compute  the volume of the base $B$ using the metric. However, as we emphasised repeatedly, the metric of this space is singular. There is a smooth Ricci-flat metric on the putative elliptically fibered Calabi--Yau $Y_3$. The way we will work around the absence of a smooth metric on $B$ is to compute the volume in the elliptic Calabi--Yau as follows. The $(1,1)$-form dual to $B$ is $\omega_0$, and the volume of the divisor can be computed by 
\be
\vol (B) ={1\over 2}  \int_{Y_3} \omega_0  \wedge \pi^* J_B \wedge \pi^*J_B= {1\over 2} \int_B J_B \wedge J_B \,.
\ee
Furthermore the latter integral can be evaluated by first using the fact that the curve wrapped by the D3-branes, $C$, is Poincar\'e dual to the K\"ahler form $J_B$ and then using intersection theory to write 
\be
\vol (B)  = {1\over 2} \int_C J = {1\over 2} C\cdot C \,.
\ee
Using this identification we can rewrite the central charge in terms of the self-intersection of the curve $C$ in $B$ as 
\be
(\cIIBL)^{(2)} =(\cIIBR)^{(2)}= \cIIBsugra= 3 N^2 M \, C \cdot C    \label{IIB(0,4)c}\,.
\ee 
Since $\vol(B) >0$ the curve wrapped by the D3-branes must have positive self-intersection in $B$. Using the adjunction formula \eqref{adjunction} one can express the constraint $C \cdot C > 0$ as  
\be 
C\cdot C=2(g-1)+ c_1(B)\cdot C >0 \,.
\ee

At this point we should comment about the relation of our setup to the strings in minimal 6d SCFTs, also known as non-Higgsable clusters (NHCs) \cite{Morrison:2012np}, whose central charges were computed in \cite{DelZotto:2016pvm}. The geometric condition for the NHCs is that the base of the Calabi--Yau threefold is locally $\mathcal{O}(-n) \rightarrow \mathbb{P}^1$. The curve that is wrapped by the D3-brane is the base $C^{\rm NHC}=\mathbb{P}^1$, which has self-intersection 
\be
C^{\rm NHC}\cdot C^{\rm NHC} = -n < 0 \,,\qquad n= 3,4,6,8,12 \,,
\ee
and can be collapsed. This singular limit corresponds to the conformal point. 
In appendix \ref{app:ample} the geometry of these NHCs is briefly discussed. 
The negative self-intersection implies that $C^{\rm NHC}$ is not ample, and consequently that these 2d NHC strings do not directly fit into the framework discussed in this paper.

    \subsection{$\cIIBL-\cIIBR$ at Sub-leading Order from Anomaly Inflow}
\label{sec:D7inflow}

The sub-leading contribution is obtained using anomaly inflow \cite{Aharony:2007dj}. The  difference of the left and right central charges appears as the coefficient in front of the gravitational Chern-Simons term in the bulk action \cite{Kraus:2005zm} 
\bea \label{ChernSimons}
S_{CS}(\Gamma_{\mathrm{AdS}_3})= \frac{\cIIBL-\cIIBR}{96\pi} \int_{\mathrm{AdS}_3} \omega_{CS}(\Gamma_{\mathrm{AdS}_3})\,.
\eea
To determine this coefficient we consider the three dimensional terms which arise from the dimensional reduction of the Chern-Simons terms in the worldvolume action of 7-branes. 
The Chern-Simons terms for a D7-brane were computed in \cite{Green:1996dd} and are given in terms of the curvature two-forms of the tangent and normal bundles of the brane worldvolume, 
$\mathcal{R}_T$ and $\mathcal{R}_N$, respectively,
\be\label{blablabla}
 \mu_{7}\int_{\mathcal{W}_8} C^{(4)}\wedge \sqrt{\frac{\hat{A}(4 \pi^{2}\ls^{2}\mathcal{R}_T)}{\hat{A}(4 \pi^{2}\ls^{2}\mathcal{R}_N)}}\tr \lb \me^{2 \pi \ls^{2}\mathcal{F} }\rb \subset S_\mathrm{D7}\,,
\ee
where 
\be
\mu_{7}=\frac{1}{(2 \pi)^{7}\ls^{8}} \,,
\ee
is the charge of a single D7-brane, $C_{4}$ is the potential of the five-form flux and $\mathcal{F}$ is the gauge invariant field strength of the gauge fields on the D7-brane. The trace is performed in the fundamental representation of the gauge group. 
For the computation of the $\mathcal{O}(N)$ corrections to the central charges we will only be interested in the terms coming from the tangent bundle of the D7-brane. Thus below we simply write $\mathcal{R} \equiv \mathcal{R}_T$. Up to the required order the A-roof genus $\hat{A}$ is given by
\bea
\hat{A}(\mathcal{R})&=&1+\frac{1}{12(4 \pi)^{2}}\tr (\mathcal{R}\wedge \mathcal{R})\,.
\eea

As we consider only $I_1$ singularities our set-up consists of \emph{single}
7-branes wrapped on curves $C_x$ in the base\footnote{This can be easily
  generalised to other 7-brane singularities, by including suitable
  normalisations to the trace appearing in (\ref{blablabla}).}. Note that not
  all of these 7-branes can be transformed into D7-branes under an
  $SL(2, \mathbb{Z})$ transformation. Imposing that the elliptic fibration is
  Calabi--Yau results in the constraint
\be 
[\Delta]= 12 c_1(B) = \sum_{x} \omega_x \,,
\ee
where $\omega_x$ are the two-forms dual to the curves $C_x$ wrapped by the 7-branes.

Consider a single D7-brane whose world-volume extends along $\mathcal{W}_8 = \mathrm{AdS}_3 \times S^3/\mathbb{Z}_M \times C_x$. From the D7-brane Wess-Zumino term we obtain the 3d Chern-Simons term
\bea \label{SingleD7ChernSimons}
 S_{CS}(\Gamma_{\mathrm{AdS}_3})&=&\frac{\mu_{7}\pi^{2}\ls^{4}}{24 \gs}\int_{\mathcal{W}_8} C^{(4)}\wedge \tr(\mathcal{R}\wedge \mathcal{R})\nonumber\\
&=&-\frac{\mu_{7}\pi^{2}\ls^{4}}{24 \gs}\int_{\mathcal{W}_8} F\wedge \omega_{CS}\nonumber\\
&=&\frac{\mu_{7}\me^{4 \Del}\pi^{2}\ls^{4}\vol(S^{3}/\Z_M)}{24\gs(2m)^2{\mb^{2}}}\int_{C_x}J_B  \int_{\mathrm{AdS}_{3}}\omega_{CS}(\Gamma_{\mathrm{AdS}_3})\nonumber\\
&=&\frac{N}{192 \pi}\int_{B}J_B \wedge \omega_x \int_{\mathrm{AdS}_{3}}\omega_{CS}(\Gamma_{\mathrm{AdS}_3}) \,,
\eea
where we have used the fact the trace over the fundamental representation of the gauge group is 1 as only one D7-brane is wrapped on $C_x$.

As $C^{(4)}$ is invariant under $SL(2, \mathbb{Z})$ transformations, each 7-brane gives rise to the same contribution to the 3d Chern-Simons term \cite{Aharony:2007dj}. To obtain the total contribution we therefore sum the terms arising from each 7-brane
\be \label{D7ChernSimons}\ba 
S_{CS}(\Gamma_{\mathrm{AdS}_3})=&\frac{N }{192 \pi} \sum_x \int_B J_B \wedge \omega_x \int_{\mathrm{AdS}_{3}}\omega_{CS}(\Gamma_{\mathrm{AdS}_3})  \\
=&\frac{N}{16 \pi}\int_B J_B \wedge c_1(B)  \int_{\mathrm{AdS}_{3}}\omega_{CS}(\Gamma_{\mathrm{AdS}_3}) \,.
\ea \ee
We evaluate the integral over the base by pulling back to the smooth Calabi--Yau
\be
\int_{Y_3} \omega_0 \wedge \pi^*J_B \wedge \pi^*c_1(B) =\int_{B}J_B \wedge c_1(B)= c_{1}(B)\cdot C\,.
\ee
Using this relation we determine from the coefficient of \eqref{D7ChernSimons} the difference of the left and right central charges to be 
\be
(\cIIBL)^{(1)}-(\cIIBR)^{(1)} =6 N c_{1}(B)\cdot C\,.
\ee

\subsection{Level of the Superconformal R-symmetry}
\label{sec:IIBLevel}

In this section we compute the level $k_r$ of the superconformal R-symmetry. The relation $c_R = 6 k_r$ and \eqref{IIB(0,4)c} imply that the leading order contribution to the level is given by 
\be 
k_r^{(2)} = \half N^2 M C \cdot C\,.
\ee
To compute the sub-leading order term we restrict to the case of $M = 1$ and proceed by gauging the $SO(4)_T$ isometry of the $S^3$ in the supergravity solution. 
The procedure for computing the level follows \cite{Hansen:2006wu,Dabholkar:2010rm}, where one first deforms the metric on the $S^3$ to contain connections, which depend on $\AdS_3$
\be 
\diff s^2_{S^3} \rightarrow (\diff x^p - A^{pq} x^q)(\diff x^p - A^{pr} x^r)\,,
\ee 
where $\sum_{p = 1}^4 (x^p)^2 = 1$. These connections $A^{pq} = - A^{qp}$ are one-forms on $\AdS_3$ and are identified with the $SO(4)_T \simeq  SU(2)_L \times SU(2)_R$ gauge fields for the superconformal R-symmetry $SU(2)_R$ and the flavour symmetry $SU(2)_L$.  The deformed five-form flux is \cite{Hansen:2006wu}
\be \label{DeformedF5M1}
F_5'|_{M=1}= -\frac{4\pi^2\me^{4 \Del}}{m^2{\mb^{2}}}(1 + \ast)\left( (e_3 - \chi_3) \wedge J_B\right)\,,
\ee
where $e_3$ is the volume form on the sphere bundle satisfying $\int_{S^3} e_3 = 1$ and $\dd e_3  = \chi_4 $, $\chi_4$ being the Euler class of the sphere bundle. The additional term $\chi_3$, a three-form on $\AdS_3$ satisfying $\dd \chi_3 = \chi_4$, is required for $\dd F_5 '|_{M=1} = 0$.

The reduction of the Chern-Simons term for D7-branes wrapped on this deformed
metric gives rise to Chern-Simons terms for the $SO(4)_T$ gauge fields. Upon
inserting the deformed flux (\ref{DeformedF5M1}) into the D7-brane Chern-Simons term and summing over all 7-branes as above one finds, in addition to the gravitational Chern-Simons term, 
\be  \ba 
S_{CS}(A_T)|_{M=1} 
&=\frac{N }{8 \pi}c_1(B)\cdot C  \int_{\mathrm{AdS}_{3}} \left(\omega_{CS}(A_R) + \omega_{CS}(A_L)\right)\,,
\ea \ee
where the additional factor of 2 arises from expressing the trace over the fundamental representation of $SU(2)_R$ and $SU(2)_L$ instead of the vector representation of $SO(4)_T$. 
The level of the superconformal $SU(2)_r$ R-symmetry can be extracted from the coefficient of Chern-Simons term after multiplication by $4 \pi$, namely 
\be \label{CSlevel}
S_{CS}(A) = \frac{k_r}{4\pi}\int_{\mathrm{AdS}_3} \omega_{CS}(A)\,,
\ee
where $A = \ii A^a \sigma_a/2$. From the coefficient of $\omega_{CS}(A_R)$ the sub-leading order term in the level of the superconformal R-symmetry can be extracted and found to be
\be 
k_r^{(1)}|_{M=1} = \half N c_1(B)\cdot C\,.
\label{lunchtime}
\ee

For the cases with $M > 1$, the isometry group of the solution is broken to
$SU(2)_R \times U(1)_L$. Naively, to compute the level of the superconformal R-symmetry one should still  gauge the $SU(2)_R$  by introducing gauge fields for this isometry, analogous to the $M = 1$ case. Formally, this gives exactly the same result as (\ref{lunchtime}); however this is not the complete contribution, as one would have to take into account the effects of the $M$ KK-monopoles. As we shall see in section 
 \ref{trump}, on the 11d supergravity side this will be captured by gauging the $SU(2)_\mathrm{11d}$ isometry of an $S^2$, which arises from  the base of the $S^3/\Z_{M}$ Hopf fibration. However, it should be noted that 
 $SU(2)_R$ is different from   $SU(2)_\mathrm{11d}$, and one can check explicitly that in fact the latter is \emph{not} an isometry of the Type IIB solution.


\subsection{Summary: Central Charges from F-theory}
\label{sec:sumccF}

From the computations carried out in this section the central charges in Type IIB supergravity for $M = 1$ are given by 
\be\label{moana} \ba 
\cIIBR|_{M=1} &= 3   N^2 \, C  \cdot C+3 N c_1(B) \cdot C\, , \\
\cIIBL|_{M=1} &= 3   N^2 \, C \cdot C + 9N c_1(B) \cdot C \, .
\ea \ee
In this section we have only computed these central charges to sub-leading
order in $N$. We expect $\mathcal{O}(1)$ corrections to arise from one loop
computations and will comment on these in section
\ref{sec:CentralChargeSummary}, where we compare the central charges computed
via anomaly inflow and supergravity solutions. We further point out that
  the superconformal algebra mandates that the right-moving central charge
  belongs to $6\mathbb{Z}$. To see this explicitly we make use of the adjunction formula
  (\ref{adjunction}) and rewrite it as
  \begin{equation}\label{multipleofsixN}
    \cIIBR|_{M=1} = 6N^2(g - 1) + 3N(N+1)c_1(B) \cdot C \,, 
  \end{equation}
which {exhibits manifestly that the expression is a multiple of} six, generalising to any $N$ the property of the $N=1$ right central  charge, observed in (\ref{spectrumtimes6}).

For $M >1$ we obtain 
\be \label{fairenough}
\ba 
\cIIBR|_{M>1} &= 3 M  N^2 \, C  \cdot C + \delta c^{IIB}_R  \\
\cIIBL|_{M>1} &= 3 M  N^2 \, C \cdot C +\delta c^{IIB}_L + 6N c_1(B) \cdot C \,.
\ea \ee
As explained in the previous section, the computation of the level of the superconformal R-symmetry for $M>1$ is troublesome. Instead, we uplift our Type IIB solution to 11d supergravity in the next section. In doing so we will be able to compute the $\mathcal{O}(N)$ contributions to the $M>1$ central charges, as well as $\mathcal{O}(1)$ corrections.


\section{M/F-Duality and AdS$_3$ Solutions in M-theory}
\label{sec:FMduality}


The solution found above in Type IIB supergravity is singular at the loci above which $\tau$ degenerates. We circumnavigated this problem by computing the central charges of the solutions in terms of the volume of the base $B$ in the smooth Calabi--Yau, where it is well-defined. 
To substantiate this we can utilize M/F-duality: by T-dualizing and uplifting to M-theory, the elliptically fibered Calabi--Yau threefold becomes manifest in the geometry. Assuming that there are only $I_1$ fibers, the elliptic Calabi--Yau threefold is smooth, as can be seen by direct computation. There exists a smooth Ricci-flat metric on this space by Yau's theorem \cite{MR0451180} and we may use this metric to compute the central charge.

\subsection{Dual 11d Supergravity Solution}

In this subsection we shall perform a T-duality along the Hopf fiber of the $S^{3}/\Z_{M}$ to Type IIA and then perform the uplift to 11d supergravity.  As noted in section \ref{Lens} this will preserve all supersymmetries of the original solution.

Recall that the Type IIB solution in string frame takes the form
\bea
\dd s^{2}(\M_{IIB})&=& \frac{\me^{2 \Del}}{\sqrt{\tau_{2}}}\left[ \dd s^{2}(\mathrm{AdS}_{3})+\frac{1}{{\mb^{2}}} \dd s^{2}(B)+\frac{1}{4 m^{2}}(\sigma_{1,L}^{2}+\sigma_{2,L}^{2}+\sigma_{3,L}^{2})\right ]\,,\\
F&=&-\frac{2 m}{{\mb^{2}}}  \me^{4\Del} J_{B}\wedge \text{dvol}(\mathrm{AdS}_{3})-\frac{\me^{4\Del}}{4 m^{2}{\mb^{2}}}J_{B} \wedge\sigma_{1,L}\wedge \sigma_{2,L}\wedge \sigma_{3,L}\,,\\
\tau&=&\tau_{1}+\ii \tau_{2}=C^{(0)}+\ii \me^{-\phi}\,.
\eea
The metric defined by $\frac{1}{4}(\sigma_{1,L}^{2} +\sigma_{2,L}^{2}+\sigma_{3,L}^{2})$ is that of the round, unit radius Lens space, $S^{3}/\Z_{M}$. This is obtained by quotienting the Hopf fiber, $\sigma_{3,L}$ in our conventions, by the discrete group $\Z_{M}$ which has the effect of reducing the period of $\psi$ from $4\pi$ to $4 \pi/M$. Recall that $M$ corresponds to the number of KK-monopoles in the solution before going to the near-horizon limit, as was discussed in section \ref{sec:KKM}.

Before performing the T-duality along the Killing vector $\partial_{\psi}$ we shall absorb the factor of $4 m^{2}$ into the definition of $\psi$ by making the change of coordinates
\be
y=\frac{{\mb}}{2m}\psi\,,\label{ycoorddef}
\ee
where $y$ now has period $\frac{2 \pi{\mb}}{Mm}$.
If we now perform the T-duality along  $\partial_y$ we obtain the Type IIA solution
\be\ba
\dd s^{2}(\M_{IIA})&=\frac{\me^{2\Del}}{\sqrt{\tau_{2}}}\left[ \dd s^{2}(\mathrm{AdS}_{3})+\frac{1}{{\mb^{2}}}\dd s^{2}(B)+\frac{1}{4 m^{2}}(\dd \theta^{2}+\sin^{2}\theta \dd \varphi^{2})\right]+{\frac{\me^{-2\Del}\sqrt{\tau_{2}}}{\mb^{2}}}\dd y^{2}\,,\\
C^{(1)}&=\tau_{1}\dd y\,,\\
\me^{-\hat{\phi}}&=\tau_{2}^{3/4}\me^{\Del}\,, \\
B^{IIA}&=-\frac{\cos \theta}{2m{\mb}}\dd y\wedge \dd \varphi\,,\\
F_{4}^{IIA}&=\frac{\me^{4 \Del}}{2m{\mb^{2}}} \dd \vol(S^{2})\wedge J_{B}\,.
\ea\ee
Uplifting to 11d supergravity and performing a redefinition of the torus coordinates we have 
\bea
\dd s^{2}(\M_{11})&=&\me^{\frac{8\Del}{3}}\left(\dd s^{2}(\mathrm{AdS}_{3})+\frac{\dd s^{2}(S^{2})}{4 m^{2}} +\frac{1}{{\mb^{2}}}\left[\dd s^{2}(B)+\frac{1}{\tau_{2}}(\dd \tilde{x} +\tau_{1}\dd \tilde{y})^{2}+\tau_{2}\dd \tilde{y}^{2}\right]\right)\label{Mmetric}\\
G_{4}&=&\frac{\me^{4 \Del}}{2m{\mb^{2}}}\dd \vol(S^{2})\wedge ( J_B+\dd \tilde{x} \wedge \dd \tilde{y})\,,\label{Mflux}
\eea
where $J_{B}$ is the K\"ahler form on the base. We have redefined the torus coordinates to be
\be
\tilde{x}=\me^{-2 \Del} x\,,~~\tilde{y}=\me^{-2 \Del} y\,.
\ee
The periods of the two coordinates are given by
\be
R_{\tilde{y}}=\frac{\me^{-2\Del}l_{s}^{2}}{R_{IIB}}\,,~~R_{\tilde{x}}=\frac{\me^{-2 \Del} l_{s}^{2}}{R_{IIB}}\,,\label{Mtheoryperiod}
\ee
where $R_{IIB}=\frac{1}{Mm}$ is the radius of the $S^{1}$ in Type IIB which we have T-dualised along, whose coordinate has been normalised to give the canonical $2 \pi$ period.

As remarked earlier, the Type IIB solution is singular over the discriminant
locus where the fiber degenerates. As such the 11d supergravity metric we
obtain from the explicit T-duality and uplift is only valid away from the
singular loci. To make progress, we exploit the fact that the algebraic
variety $ \mathbb{E}_\tau \hookrightarrow Y_3 \to  B$, with only $I_1$
singular fibers\footnote{As has been previously
  stated if $Y_3$ contains singularities then one can construct the smooth and
  compact resolution of the singularities of $Y_3$ and the following analysis
generalises.}, is smooth and compact, and has $c_1(Y_3)=0$, thus, by Yau's theorem, there exists a global non-singular  Ricci-flat metric, of which (\ref{singularcy}) is an approximation valid only away from the singularities. The 11d supergravity solution is therefore given by
\bea
\dd s^{2}(\M_{11})&=&\me^{\frac{8\Del}{3}}\lb\dd s^{2}(\mathrm{AdS}_{3})+\frac{1}{4 m^{2}}(\dd \theta^{2}+\sin ^{2}\theta \dd \varphi^2)+\frac{1}{{\mb^{2}}}\dd s^{2}(Y_{3})\rb\\
G_{4}&=&\frac{\me^{4 \Del}}{2m{\mb^{2}}}\dd \vol(S^{2})\wedge J_{Y_{3}}\,.\label{MF4}
\eea
This solution falls within the classification of \cite{Colgain:2010wb}, specialised to elliptically fibered Calabi--Yau threefolds.
Despite the fact that we do not know this metric explicitly, we will be able to compute the central charges for this solution as we discuss in section \ref{sec:Mthccs}.

As commented in \cite{Colgain:2010wb}, this solution agrees locally with the geometry discussed in \cite{Maldacena:1997de}. The M5-branes therefore wrap the 4-cycle Poincar\'e dual to the K\"ahler form $J_{Y_3}$, which is an ample divisor in the Calabi--Yau. Using the expansion  (\ref{KahlerFormExp})  we see that this divisor is a linear combination of $B$ and $\widehat{C}_\alpha$, which are divisors arising from pullbacks of curves in the base. As we only consider $I_1$ singularities in the fiber there are no Cartan divisors $D_i$. The presence of M5s wrapping the base of the Calabi--Yau is consistent with the $M$ KK-monopoles in the Type IIB supergravity solution described in section \ref{sec:KKM}. The sequence of dualities relating these two supergravity solutions is described in detail in \cite{Bena:2006qm}. The T-duality of $M$ KK-monopoles in Type IIB gives rise to $M$ NS5-branes along $\AdS_3 \times B$, which uplift to $M$ M5-branes wrapped on the base. The D3-branes wrapped on the curve $C$ in the base are uplifted to M5-branes wrapped on the elliptic surface $\widehat{C}$ as described in section \ref{sec:FieldTheory}. As noted in \cite{Bena:2006qm}, these two stacks of M5-branes can be deformed into one stack wrapped on a linear combination of $B$ and $\widehat{C}$  provided the curve $C$ is sufficiently ample in the base.


\subsection{M5-Brane Solutions}

Analogous to the discussion conducted in subsection \ref{sec:KKM} we shall construct the explicit smeared brane solution which gives (\ref{Mmetric}) in the near-horizon limit. To construct this solution one may either T-dualize the ``pre near-horizon'' solution obtained in (\ref{UV1}) along the Hopf fiber and then uplift or use the brane smearing techniques employed previously to combine $N$ M5-branes wrapping $\R^{1,1}\times \widehat{C}$ and $M$ M5-branes wrapping $\R^{1,1}\times B$, in the background $\R^{1,1}\times \R^{3}\times Y_3$ with $M>0$. Both methods result in the same solution given by\footnote{Note that $\Ntt$ is the same as the constant appearing in (\ref{TNharmonic}) upon using the relation $\lp^{3}=\frac{\gs l_{s}^{4}}{R_{IIB}}$ and the fact that $R_{IIB}=\frac{2}{{{\mb}}}$ for this T-duality and uplift. Recall that $R_{IIB}$ is the radius of the $S^1$ in Type IIB along which we have T-dualized with the $S^1$ coordinate having the canonical $2\pi$ period.}  
\bea
{\mb^{2}}\dd s^{2}(\M_{11})&=&\lb 
\frac{r+\Ntt}{r+\Mti}\rb ^{-\frac{2}{3}}\lb \frac{1}{\tau_2}(\dd y+ \tau_{1} \dd \psi)^{2}+\tau_{2} \dd \psi^{2}\rb
 \label{pre11d}\\
&+&\lb \frac{r+\Ntt}{r+\Mti}\rb ^{\frac{1}{3}}\lb \frac{(r+\Ntt)(r+\Mti)}{r^{2}}(\dd r^{2}+r^{2}(\dd \theta^{2}+\sin^{2}\theta \dd \varphi^{2})) \rb\nonumber\\
&+&\lb \frac{r+\Ntt}{r+\Mti}\rb ^{\frac{1}{3}}\lb \frac{r}{r+\Ntt}(-\dd t^{2}+\dd x^{2}) + \dd s^{2}(B) \rb\nonumber\\
G_{4}&=&\frac{1}{{\mb^{3}}}\dd\vol(S^{2})\wedge (\Ntt J_{B}+\Mti \dd\vol(\mathbb{E}_{\tau}))\,,
\eea
where
\be
\Ntt= 2 \pi^2 \lp^3{\mb^{3}} N  
~~\text{  and  }~~
\Mti=\frac{2 \pi^2 \lp^3{\mb^{3}} M}{\vol(\mathbb{E}_{\tau})}
\,.
\ee
Of course, as already mentioned, this solution has singularities arising from
$B$ and also $\tau$. Notice that the Calabi--Yau metric is now warped and we are unable to resolve these singularities as in the previous subsection. However here we are interested in understanding the behaviour in the radial direction $r$ and so we shall not discuss this issue further.

Taking the near-horizon limit one obtains the metric
\bea
{\mb^{2}}\dd s^{2}(\M_{11})_{r\rightarrow 0}&=& \lb \frac{\Ntt}{\Mti}\rb ^{\frac{1}{3}}\left[ \frac{r}{\Ntt}(-\dd t^{2}+\dd x^{2})+\frac{\Ntt \Mti}{r^{2}}\dd r^{2}+\Ntt \Mti(\dd \theta^{2}+\sin^{2}\theta \dd \phi^{2})\right.\nonumber\\
&&\left.+\dd s^{2}(B)+\frac{\Mti}{\Ntt}\lb \frac{1}{\tau_2}(\dd y+ \tau_{1} \dd \psi)^{2}+\tau_{2} \dd \psi^{2}\rb\right]\, .
\eea
Upon identifying the warp factor to be $
  \me^{8\Del}=\frac{\Ntt}{\Mti}$, the inverse radius squared of AdS$_3$ to be
  $\frac{{\mb^{2}}}{m^2}=4 \Ntt \Mti$ and performing the change of coordinates $r= 4
  \Ntt \Mti \rho^{2}$, $y=\sqrt{\frac{\Ntt}{\Mti}}\tilde{y}$, and
  $\psi=\sqrt{\frac{\Ntt}{\Mti}}\tilde{\psi}$ one recovers  (\ref{Mmetric})
  exactly and therefore an unwarped Calabi--Yau metric which may now be resolved. One also finds that the flux matches exactly with (\ref{Mflux}).
Asymptotically, that is $r\to \infty$, the metric approaches the space $ \R^{1,1}\times \R^3\times Y_3$, this is the space far away from the M5-branes. We emphasise that this geometry arises from $N$ M5-branes wrapped on $\R^{1,1}\times \widehat C$ plus $M$ M5-branes wrapped on $\R^{1,1}\times B$, with $B$ the base of $Y_3$, the latter M5-branes can be seen to arise from the initial $M$ KK-monopoles in the Type IIB solution. 

One may also consider the case of $N$ M5-branes wrapping only $\R^{1,1}\times \widehat{C}$ in the background  $\R^{1,1}\times \R^{3}\times Y_3$. This is the formal definition of $M=0$. The solution of this setup obtained from brane smearing is
\be\ba
{\mb^{2}}\dd s^{2}(\M_{11})&=\lb \frac{r+\Ntt}{r}\rb ^{-\frac{2}{3}}\lb \frac{1}{\tau_2}(\dd y+ \tau_{1} \dd \psi)^{2}+\tau_{2} \dd \psi^{2}\rb+ \lb \frac{r+\Ntt}{r}\rb ^{\frac{1}{3}}\dd s^{2}(B)\\
&+\lb \frac{r+\Ntt}{r}\rb ^{\frac{1}{3}}\lb \frac{r}{r+\Ntt}(\dd x^{2}-\dd t^{2})+\frac{(r+\Ntt)}{r}(\dd r^{2}+r^{2}(\dd \theta^{2}+\sin^{2}\theta \dd \varphi^{2})) \rb\nonumber\,,\\
G_{4}&=\dd\vol(S^{2})\wedge \Ntt J_{B}\,,
\ea\ee
with $\Ntt$ as before. Notice that this agrees with taking the limit $M\rightarrow 0$ in (\ref{pre11d}). Recall that $\widehat{C}$ is not an ample divisor and therefore the M5-branes do \emph{not} wrap an ample divisor as in the $M\neq 0$ case. Asymptotically the metric approaches $\R^{1,1}\times \R^{3}\times Y_3$ as before, however the metric is singular at $r=0$ now. To see this one computes the Ricci scalar to be\footnote{For ease of reading we present the result of replacing $Y_3$ with $T^6$ though the singularity persists if one reinstates the $Y_3$.}
\be
\text{R}=\frac{\Ntt^2}{3 r^{2/3}(r+\Ntt)^{10/3}}\, , 
\label{flick}
\ee
which clearly diverges at $r=0$. Upon taking the near-horizon limit one does not obtain an AdS factor, this of course matches with our previous analysis that we can only get an AdS$_3$ solution if the divisor wrapped by the branes is ample. Note that this does not imply that when $N$ M5-branes wrap $\R^{1,1}\times \widehat C$, the dual 2d field theory does \emph{not} flow to a SCFT in the IR. It just means that the IR SCFT does not have an AdS$_3$ gravity dual in 11d supergravity.

Recall, as discussed in section \ref{sec:KKM}, that in Type IIB the $M=1$ case has two UV completions. One may consider either $N$ D3-branes wrapping $\R^{1,1}\times C$ in the presence of a single KK-monopole or replacing the Taub-NUT space by flat space $\R^4/\Z_M$. Applying T-duality along the Hopf fiber of (\ref{UV2}) and uplifting we obtain the 11d supergravity solution 
 \bea
\dd s^{2}&=&(R^{2}+\Ntt)^{1/3}\lb \frac{R^{2}}{R^{2}+\Ntt}(-\dd t^{2}+\dd x^{2})+\frac{R^{2}+\Ntt}{R^{2}}(\dd R^{2}+R^{2}(\dd \theta^{2}+\sin^{2}\theta \dd \varphi^{2}))+\dd s^{2}(B)\right.\nonumber\\
&&+\left.\frac{1}{R^2+\Ntt}\lb \frac{1}{\tau_{2}}(\dd y+\tau_{1} \dd \psi)^{2}+\tau_{2} \dd \psi^{2}\rb \rb\,,\\
G_{4}&=&\dd \vol(S^{2})\wedge \lb \frac{\Ntt}{4} J_B+\Mti \dd \vol(\E_{\tau})\rb\,,
\eea
with $\Ntt$ and $\Mti$ as before. Of course in the near-horizon limit we obtain (\ref{Mmetric}), however asymptotically the metric is now degenerate. This should be contrasted with the $M$ KK-monopoles solution which has a good UV completion.\footnote{ One may, as before, consider $\R^4/\Z_M$ in place of $\R^4$, however similarly one obtains a degenerate UV completion.}

To summarise, in this section we have found the ``pre near-horizon'' solution to the 11d supergravity AdS$_3$ solution (\ref{Mmetric}). One may obtain such a near-horizon solution from two 11d supergravity solutions, both can be seen as the solution arising from a T-duality along a Hopf fiber and uplift of a Type IIB solution, (\ref{UV1}) and (\ref{UV2}) respectively. The solution arising from $M$ KK-monopoles has a good UV completion whilst the solution arising from no KK-monopoles has a degenerate UV completion.


\subsection{Flux Quantisation}

For an 11d supergravity solution to be well-defined one must quantize the fluxes through all integral cycles in the geometry. Following \cite{Witten:1996md}, the correct quantization condition to 
impose is that for all $\Sigma_4 \in H_{4}(\M_{11},\Z)$,
\be
n(\Sigma_4)=\int_{\Sigma_4}\left[\frac{1}{(2 \pi \lp)^{3}}G_{4}-\frac{p_{1}}{4}\right]\in \Z\,,\label{Mquant}
\ee
where $\lp$ is the eleven-dimensional Planck length and $p_{1}$ is the first Pontryagin class of  $\M_{11}$ defined as
\be
p_{1}=-\frac{1}{8 \pi^{2}}\Tr[\mathcal{R}^{2}]\,.
\ee
There are two types of integral four-cycles in $\M_{11}$ to consider: the
divisors $D$ in the Calabi--Yau threefold $Y_3$ as summarised in section \ref{subsec:EllFib}, and the four-cycles $S^{2}\times \E_{\tau}$ and $S^{2}\times C_{\alpha}$ with $C_{\alpha}$, as before, forming a basis of $ H_{2}(B,\Z)$.

We shall first consider the contributions from the $p_{1}/4$ term and show that they are all integral.
As the metric is a product space we have
\be 
\mathcal{R}_{\mathcal{M}_{11}}=\mathcal{R}_{\mathrm{AdS}_{3}}+\mathcal{R}_{S^{2}}+\mathcal{R}_{Y_{3}}\,.
\label{Rproduct}
\ee
where $\tensor{\mathcal{R}}{^{a}_{b}}=\frac{1}{2}\tensor{R}{^{a}_{b \mu\nu}}\dd x^{\mu}\wedge \dd x^{\nu}$. In particular, $p_{1}$ is non-trivial only on the Calabi--Yau, thus $p_1(\M_{11})=-2 c_2(Y_3)$, which is given in (\ref{c2cy}). This implies that the $p_1/4$ term integrated over the four-cycles $S^{2}\times \E_{\tau}$ and $S^{2}\times C_{\alpha}$ vanishes. On the other hand, 
the integral of $c_2(Y_3)$ over every divisor $D$  is always divisible by two, as shown in (\ref{E8}), 
\be
 \int_D c_2(Y_3) = 2(h^{1,1}(D) - 4h^{0,2}(D) +
  2h^{0,1}(D) - 4) \,,
\ee
therefore the flux quantization condition reduces simply  to
\be
n(\Sigma_4)=\frac{1}{(2 \pi \lp)^{3}}\int_{\Sigma_4}G_{4}\in \Z\, .
\ee
The form of the $G_4$-flux implies that the quantization over the divisors of $Y_{3}$ is trivial, $n(D)=0$, and therefore the relevant four-cycles to perform the quantization over are $S^{2}\times \E_{\tau}$ and $S^{2}\times C_{\alpha}$. Then we have
\bea
n(S^{2}\times C_{\alpha})&=&\frac{2 \pi\me^{4\Del}}{m{\mb^{2}}(2 \pi \lp)^{3}}\int_{C_{\alpha}}J_{Y_{3}} \, ,
\eea
where $J_{Y_3}$ is given in (\ref{KahlerFormExp}). 
Recalling that $\int_{C_{\alpha}}J_{Y_{3}}=k_\alpha \in \Z^+$, we see that imposing the condition
\bea
\Z^+ \ni \widetilde{N}=\frac{2 \pi \me^{4 \Del}}{m{\mb^{2}}(2\pi \lp)^{3}}\, ,
\eea
guarantees that $n(S^{2}\times C_{\alpha})$ is correctly quantized.
For later, we shall also need the volume of the elliptic fibration. This is constant over the base. We define the integer $\widetilde{M}$ as
\be
\widetilde{M}=\widetilde{N} \int_{\mathbb{E}_{\tau}}J_{Y_{3}} = \widetilde{N} k^0\, ,
\ee
that is $\vol(\mathbb{E}_{\tau})=\frac{\widetilde{M}}{\widetilde{N}}$. We shall show that $\widetilde{M}=M$ where the latter $M$ is that arising in Type IIB from the Lens space quotient. To see this we must use the periods of the elliptic fiber coordinates arising from the Type IIB solution, (\ref{Mtheoryperiod}). As the volume is constant over the base we may compute it away from any singularity. We find
\be 
\vol(\mathbb{E}_{\tau})=(2 \pi)^{2}R_{\tilde{x}}R_{\tilde{y}}= 
(2 \pi)^{2}m{\mb^{2}} M\me^{-4 \Del}\lp^{3} =\frac{M}{\widetilde{N}} \, ,
\ee
where we have used the relation
\be
\lp^{3}=\frac{\gs l_{s}^{4}}{R_{IIB}}\, ,
\ee
which follows from the T-duality and uplift. Using this relation we may also show that the $N$ in Type IIB is the same as the $\widetilde{N}$ in 11d supergravity. Observe that 
\be
N=\frac{16 \pi^{2}\me^{4 \Del}}{4 m^2 {\mb^{2}}M (2 \pi \ls)^{4} \gs }=\frac{2 \pi\me^{4 \Del} }{ m{\mb^{2}} (2 \pi \lp)^{3} }=\widetilde{N}\,.
\ee
We conclude that the two integers appearing in Type IIB and 11d supergravity solutions  can be identified, namely $N=\widetilde{N}$ and $M=\widetilde{M}$. For notational clarity we shall drop the tildes from now on as there is no confusion. We remark that in Type IIB $M$ corresponds to the number of KK-monopoles in the geometry whilst in 11d supergravity it is proportional to the volume of the elliptic fibration.


\section{Holographic Central Charges from M-theory}
\label{sec:Mthccs}

\subsection{Leading Order Central Charges}
\label{sec:04MLeading}

The gravitational central charge for the 11d supergravity solution AdS$_3 \times S^2 \times Y_3$ was computed in \cite{Kraus:2005vz}. We reproduce it here for completeness using (\ref{cformula}) 
\be \label{MtheoryKLCharge}\ba
(c_L^{11})^{(3)} = (c_R^{11})^{(3)} = \cMsugra&= \frac{3\me^{12 \Del}}{2m{\mb^{2}}}\frac{2^{5}\pi^{2}}{(2 \pi \lp)^{9}}\int_{M_{8}}\frac{1}{4 m^{2}{\mb^{4}}}\dd\vol(S^{2})\wedge \dd\vol(Y_3)\cr 
&=\frac{3\pi^{3} 2^{4}\me^{12 \Del}}{((2 \pi \lp)^{3}m{\mb^{2}})^{3}}\int_{Y_{3}}\frac{1}{6}J_{Y_3}\wedge J_{Y_3} \wedge J_{Y_3}\cr 
&=N^3\, C_{IJK} k^I k^J k^K \,,
\ea\ee
where we have expanded the K\"ahler form in a basis of $(1,1)$-forms on the Calabi--Yau threefold as in (\ref{KahlerFormExp}) and  $C_{IJK}$ are the triple intersection numbers as given in section \ref{subsec:EllFib}, with $I=0$  included in this expansion. This result, as noted in \cite{Kraus:2005vz}, matches the original field theory computation in \cite{Maldacena:1997de} and \cite{Harvey:1998bx}.

The K\"ahler form is expanded as in (\ref{KahlerFormExp}), where the coefficient in front of the zero-section $\omega_0$ is 
\be
k_0 = \text{vol} (\mathbb{E}_{\tau}) = \frac{M}{N} \,,
\ee
the volume of the elliptic fiber. 
The central charge \eqref{MtheoryKLCharge} can then be expanded into three terms 
\be \ba 
(c_L^{11})^{(3)} = (c_R^{11})^{(3)} &= N^3 \left( 3 k_0 k_\alpha k_\beta \int_{Y_3} \omega_0 \wedge \omega_\alpha \wedge \omega_\beta + 3 k_0^2 k_\alpha \int_{Y_3} \omega_0 \wedge \omega_0 \wedge \omega_\alpha + k_0^3 \int_{Y_3} \omega_0^3\right) \\
&= N^3 \left( 3 k_0 k_\alpha k_\beta \int_{B} \omega_\alpha \wedge \omega_\beta - 3 k_0^2 k_\alpha \int_{B} c_1(B) \wedge \omega_\alpha + k_0^3 \int_{B} c_1(B)^2\right) \\
&= 3 N^2M C\cdot C - 3 N M^{2} c_1(B) \cdot C + M^{3}(10 - h^{1,1}(B))\,   ,
\ea \ee 
where we have made use of \eqref{IntersectionRel}.


\subsection{Chern-Simons Terms and $c^{11}_L-c^{11}_R$}
\label{sec:04MSubLeading}

We now calculate $c^{11}_L-c^{11}_R$ by using the eight derivative corrections as presented in \cite{Tseytlin:2000sf}. The term that will be relevant for us is the Chern-Simons term 
\cite{Duff:1995wd} 
\bea
S_{CS}=-\frac{(4 \pi \kappa_{11})^{2/3}}{2 \kappa_{11}^{2}}\int_{M_{11}}C_{3}\wedge X_{8} \,   ,
\eea
where 
\bea
X_{8}=\frac{1}{(2 \pi)^{4}2^{6}\cdot 3}\lb \tr[\mathcal{R}^4]-\frac{1}{4}(\tr[\mathcal{R}^{2}])^{2}\rb\,  .
\eea
We wish to dimensionally reduce this to obtain Chern-Simons terms in the 3d action. From the coefficient in front of the 3d Chern-Simons term one can extract 
$c^{11}_L-c^{11}_R$ by using \eqref{CSterm}. 
Using   (\ref{Rproduct}) one can see that $\tr[\mathcal{R}^{4}]=0$. We wish to find the term proportional to (\ref{CSterm}) and so we shall drop terms that do not contribute to this if necessary 
\bea
S_{CS}&=&\frac{(4 \pi \kappa_{11})^{2/3}}{2^3 \kappa_{11}^{2}}\frac{1}{(2 \pi)^{4}2^{6}\cdot 3}\int_{M_{11}}C_{3}\wedge \tr[\mathcal{R}^{2}]\wedge \tr[\mathcal{R}^{2}]\nonumber\\
&=&\frac{(4 \pi \kappa_{11})^{2/3}}{2^3 \kappa_{11}^{2}}\frac{1}{(2 \pi)^{4}2^{5}\cdot 3}\int_{M_{11}}G_{4}\wedge \tr[\mathcal{R}^{2}]\wedge\omega_{CS}\nonumber\\
&=&\frac{\pi}{(2 \pi \lp)^{3}m{\mb^{2}}}\frac{\me^{4 \Del}}{(2 \pi)^{4}2^{7}\cdot 3}\int_{M_{11}}\dd\vol(S^{2})\wedge J_{Y_3}\wedge \tr[\mathcal{R}^{2}]\wedge \omega_{CS} \nonumber\\
&=&\frac{N \pi}{2^{ 4}\cdot 3 (2 \pi)^{2}}\int_{Y_{3}}J_{Y_3}\wedge c_{2}(Y_{3})\int_{\mathrm{AdS}_{3}}\omega_{CS}(\Gamma_{\mathrm{AdS}_3}) \,,\label{SCScomp}
\eea
from which we obtain
\be\label{CY3MCLCR}
c^{11}_L-c^{11}_R ={N\over 2}\int_{Y_{3}}J_{Y_3}\wedge c_{2}(Y_{3})\,,
\ee
which is in agreement with \cite{Kraus:2005vz}.\footnote{
Note that 
$\tr[\mathcal{R}^{2}]=16 \pi^{2} c_{2}(Y_{3})$
which is valid  for a Calabi--Yau threefold whilst working in real coordinates with the normalisation as in \cite{Bonetti:2011mw}.}

To evaluate (\ref{CY3MCLCR}) we use the expansion of the K\"ahler form in \eqref{KahlerFormExp} and the form of $c_2(Y_3)$ as in (\ref{c2cy}).
With this information we reduce the integrals in  (\ref{CY3MCLCR}) to integrals over the base of the fibration, namely\footnote{Note the integral $\int_{Y_3}$ can be translated into one over $B$ by using the intersection ring relations, and extracting the coefficient of $\sigma^2$. }
\be
\int_{Y_3} c_2 (Y_3)  \wedge \omega_0   = \int_{B} c_2(B)- c_1(B)^2 = 2h^{1,1}(B) - 8\label{c2omega0} \,.
\ee
For the remaining term, the Poincar\'e dual to $\omega_\alpha$ are divisors
$D_\alpha = \widehat{C}_\alpha$ which are pull-backs of curves in the base.
Thus the integral over $Y_3$ is only non-vanishing for those terms in
$c_2(Y_3)$, which have fiber components, \emph{i.e.} the $12 \omega_0 \wedge
c_1(B) $ term, which leads to
\be
\int_{Y_3} c_2 (Y_3) \wedge \omega_\alpha= 12  c_1(B) \cdot C_\alpha \,.
\ee
Combining these terms we find 
\be
c^{11}_L-c^{11}_R =6 N c_1(B) \cdot C +M( h^{1,1}(B) - 4)\,.
\ee


\subsection{Chern-Simons Couplings from 11d Supergravity}
\label{trump}

The 11d supergravity solution AdS$_3 \times S^2 \times Y_3$ has dual SCFTs with small  $\mathcal{N} = (0,4)$ superconformal symmetry. In order to determine the left and right central charges one must also calculate the level $k_r$ of the superconformal $SU(2)_r$ R-symmetry at sub-leading order. The leading and sub-leading corrections to the level $k_r$ were computed in  \cite{Hansen:2006wu,Kraus:2005vz} to be
\be \label{04levelk}
k_r = \frac{N^3}{6} C_{IJK} k^I k^J k^K + \frac{N}{12} \int_{Y_3} J_{Y_3} \wedge c_2(Y_3) \,.
\ee
These terms are computed by deforming the metric on the two-sphere to contain connections which depend on $\AdS_3$ only
\be 
\diff s^2_{S^2} \rightarrow (\diff x^a - A^{ab} x^b)(\diff x^a - A^{ac} x^c)\,,
\ee 
where $\sum_{a = 1}^3 (x^a)^2 = 1$. These connections are identified with the $SO(3)$ gauge fields for the R-symmetry. 

The leading order term is computed from the 11d term
\be \label{11dAFF}
S_{AFF} = - \frac{1}{12 \kappa_{11}^2} \int A'_3 \wedge G'_4 \wedge G'_4 \,,
\ee
where we have used the conventions of \cite{Tseytlin:2000sf}. For the deformed metric the fluxes are corrected by terms involving the R-symmetry gauge fields and are given by 
\be \ba 
A'_3 &= \frac{2\pi \me^{4 \Del}}{m{\mb^{2}}}e_1^{(0)} \wedge J_{Y_3} \\
G'_4 &= \frac{2 \pi\me^{4 \Del}}{m {\mb^{2}}}e_2 \wedge J_{Y_3} \,,
\ea \ee
 where $e_2$ is the unique two-form for the $S^2$ bundle satisfying $ \int_{S^2} e_2 = 1$ and $\dd e_2 = 0$. The one-form $e_1^{(0)}$ is defined by $\dd e_1^{(0)} = e_2$. The overall factors in $G'_{4}$ have been fixed by requiring that the quantization pre-deformation is the same as that post-deformation. Inserting these expressions into \eqref{11dAFF} we obtain
\be 
S_{AFF} = -\frac{(2\pi)^3\me^{12 \Del}}{12 \kappa_{11}^2 m^3{\mb^{6}}} \int_{Y_3} J_{Y_3}\wedge J_{Y_3}\wedge J_{Y_3} \int_{\mathrm{AdS}_3 \times S^2} e_1^{(0)} \wedge e_2 \wedge e_2\,.
\ee
To simplify this expression we make use of the formula derived in \cite{Hansen:2006wu}
\be 
\int_{\mathrm{AdS}_3 \times S^2} e_1^{(0)} \wedge e_2 \wedge e_2 = - \frac{1}{2(2\pi)^2} \int_{\mathrm{AdS}_3} \omega_{CS}(A)\,,
\ee
which originates from \cite{MR1776090}. Recalling the expression $N = \frac{2\pi}{(2\pi l_p)^3m{\mb^{2}}}$ we obtain 
\be  \ba 
S_{AFF} &= \frac{\pi \me^{12 \Del}}{12\kappa_{11}^2m^3{\mb^{6}}} \int_{Y_3} J_{Y_3}\wedge J_{Y_3}\wedge J_{Y_3} \int_{\mathrm{AdS}_3} \omega_{CS}(A)\\
&= \frac{N^3}{24 \pi} C_{IJK} k^I k^J k^K\int_{\mathrm{AdS}_3} \omega_{CS}(A)\,,
\ea \ee
The level $k_r$ is extracted from the coefficient of the Chern-Simons term from the definition in \eqref{CSlevel}. From this we obtain the leading order term in \eqref{04levelk}.

The sub-leading order term is found by computing $S_{CS}$ for the deformed metric, which now contains a contribution from the R-symmetry gauge fields
\be \label{subleadingk}
S_{CS}=\frac{N}{192\pi}\int_{CY}J_{Y_3}\wedge c_{2}(Y_{3}) \left(\int_{\mathrm{AdS}_{3}}\omega_{CS}(\Gamma_{\mathrm{AdS}_3}) +4 \int_{\mathrm{AdS}_{3}}\omega_{CS}(A)\right)\,,
\ee
where the trace in $\omega_{CS}(A)$ is taken over the fundamental representation of $SU(2)$. The factor of 4 appearing in the gauge Chern-Simons term arises from changing the trace from over the vector representation of $SO(3)$ to $SU(2)$ fundamental. Comparing \eqref{subleadingk} to \eqref{CSlevel} the sub-leading term indeed matches that in \eqref{04levelk}.  

Using the results from section  \ref{sec:04MSubLeading} the level can be expressed as 
\be \ba 
 k_r &= \frac{N^3}{6} C_{IJK} k^I k^J k^K + \frac{N}{12} \int_{Y_3} c_2(Y_3) \wedge J_{Y_3} \\
& = \frac{1}{2} N^2 M C\cdot C+\frac{N}{2}(2-M^2) c_1(B)\cdot C +\frac{M^3}{6}(10-h^{1,1}(B))+\frac{M}{6}(h^{1,1}(B)-4) \,.
\ea \ee
The left and right central charges can now be deduced by using the relation $c_R^{11} = 6 k_r$ \cite{Strominger:1996sh}. We obtain the central charges 
\be  \label{MTheoryCentralChargeFull}\ba 
\cMR &= 3 N^2 M C\cdot C +3 N(2-M^2) c_1(B)\cdot C+ M^3(10-h^{1,1}(B))+M(h^{1,1}(B)-4) \,,\\
\cML &=  3 N^2 M C\cdot C +3 N (4- M^2) c_1(B)\cdot C+M^3(10- h^{1,1}(B))+2 M (h^{1,1}(B)-4)\,.
\ea \ee
Interestingly, we note that the right-moving central charge $c_R^{11}$ can be shown to be an integer multiple of 6 as expected \cite{Schwimmer:1986mf}. To see this we rewrite $\cMR$ as
\be 
\ba
\cMR &=   6 N^2 M(g-1) + \left( 6 N  + 3NM(N-M) \right) c_1(B)\cdot C \\
&  ~~\;\;\,  +  6M^3 + (M-1)M(M+1)(4-h^{1,1}(B)) \, . 
\ea\ee
It is an elementary exercise to show that each term in the expression above is indeed a multiple of 6, for arbitrary values of $N,M\in \Z$. We regard this as a non-trivial check on the interpretation of $\cMR$ as the right-moving central charge of a $(0,4)$ SCFT with small superconformal algebra.

In section \ref{sec:FTheoryCC} for $M>1$ we were only able to determine the leading order central charge. To the contrary here, we have the all order expression. It would be very interesting to extend the analysis in section \ref{sec:IIBLevel} to include $M>1$ and to compare with the above expression.

\section{Central Charges from Anomalies and Comparisons}
\label{peppapig}

In this section we shall determine the central charges of the 2d SCFTs
microscopically, using a UV description  in terms of world-volume theories on
wrapped branes. To determine these we will essentially need to compute only
the anomaly polynomials of the corresponding branes, although we will discuss
some subtleties involved in these computations. This complements and extends
the central charge computation in section \ref{sec:FieldTheory} from the
dimensional reduction of the abelian $\mathcal{N}=4$ SYM theory.  Below we will invert
the order of presentation with respect to the previous sections as we find it
more convenient to begin with the M5-branes in the M-theory picture and
address the D3-branes in the F-theory picture after. We also include a section
summarising the results of the computations in the different setups and their
comparison.

\subsection{Anomalies from M5-branes}
\label{sec:M5inflow}

In this section we wish to determine the anomaly polynomial
associated to the $(0,4)$ theory on the worldvolume of the string in 5d
arising from a stack of M5-branes wrapping a compact $4$-cycle in a Calabi--Yau
threefold.

A single M5-brane has an anomaly \cite{Witten:1996hc} from the chiral modes
living on the 6d worldvolume of the brane; this anomaly must be cancelled by
anomaly inflow from the M-theory bulk. In \cite{Freed:1998tg} a certain deformation of the cubic
Chern-Simons term in M-theory was found to cancel the anomaly from a single
M5-brane, and this was generalised in \cite{Harvey:1998bx} to compute the
total anomaly polynomial of the 6d worldvolume theory on a stack of $N$
M5-branes. The anomaly polynomial is
\begin{equation}
  I_8[N] = N I_8[1] + \frac{1}{24}(N^3 - N)p_2(\mathcal{N}) \,, 
\end{equation}
where
\begin{equation}
  I_8[1] = \frac{1}{48}\left[ p_2(\mathcal{N}) - p_2({W})
  + \frac{1}{4}(p_1({W}) - p_1(\mathcal{N}))^2 \right] \,,
\end{equation}
is the anomaly polynomial for the free abelian tensor multiplet that lives on
the worldvolume of a single M5-brane and $W$, $\mathcal{N}$ are respectively
the 6d submanifold the M5-brane wraps, and the normal, or $SO(5)$ R-symmetry,
bundle associated to the transverse directions of the M5-brane worldvolume in
the 11d spacetime.

The theory living on the worldvolume of $N$ M5-branes in flat space is the
interacting $(2,0)$ superconformal field theory of type $A_{N-1}$ coupled to
the free abelian tensor multiplet. We can determine the anomaly polynomial of
the $A_{N-1}$ theory by subtracting off the contribution from the latter, 
\begin{equation}
  I_8^\text{int}[N] = (N - 1)I_8[1] + \frac{1}{24}(N^3 -
  N)p_2(\mathcal{N}) \,.
\end{equation}
This agrees with \cite{Intriligator:2000eq} where the anomaly polynomial of
the 6d $(2,0)$ theories associated to ADE Lie algebras was conjectured to be
\begin{equation}
  I_8(G) = r(G) I_8[1] + \frac{1}{24} d(G) h^\vee(G)
  p_2(\mathcal{N}) \,,
\end{equation}
where $r$, $d$, and $h^\vee$ are the rank, dimension, and the dual Coxeter
number of the ADE group $G$, respectively. 

Following \cite{Harvey:1998bx} the anomaly polynomial $I_4$ for the string arising
from the M5-brane wrapping a compact surface $P$ inside a Calabi--Yau
threefold\footnote{Note that the Calabi--Yau threefold does not, at this
point, need to be elliptically fibered. Moreover, $P$ does not have to be a (very) ample divisor.}, $Y_3$, can be determined by
integrating the 6d anomaly polynomial over $P$. For such an M-theory setup the
tangent and normal bundles decompose as
\begin{equation}
  \begin{aligned}
    TW &= TP \oplus TW_2 \, , \cr 
    \mathcal{N} &= \mathcal{N}_{P/Y_3} \oplus \mathcal{N}_3 \,,
  \end{aligned}
\end{equation}
where $W_2$ is the worldvolume of the string, $\mathcal{N}_{P/Y_3}$ is the
normal bundle of $P$ inside of the Calabi--Yau, and $\mathcal{N}_3$ is the
bundle associated to the $SO(3)_T$ global symmetry from the rotations of the
$3$ transverse directions to the string in 5d. Under these bundle
decompositions the Pontryagin classes decompose, via the splitting principle,
to
\begin{equation}
  \begin{aligned}
    p_1(\mathcal{N}_{P/Y_3} \oplus \mathcal{N}_3) &= p_1(\mathcal{N}_{P/Y_3}) +
    p_1(\mathcal{N}_3) \cr
    p_2(\mathcal{N}_{P/Y_3} \oplus \mathcal{N}_3) &= p_2(\mathcal{N}_{P/Y_3}) +
    p_2(\mathcal{N}_3) + p_1(\mathcal{N}_{P/Y_3})p_1(\mathcal{N}_3) \,,
  \end{aligned}
\end{equation}
and similarly for $p_i(W)$.

First, let us consider the integration of the anomaly polynomial of a single
M5-brane:
\begin{equation}
  48 \int_P I_8[1] = 2p_1(\mathcal{N}_3)\int_P p_1(\mathcal{N}_{P/Y_3})
  - \frac{1}{2}(p_1(W_2) + p_1(\mathcal{N}_3))\int_P (p_1(P) +
  p_1(\mathcal{N}_{P/Y_3})) \,.
\end{equation}
We can use the adjunction formula
\begin{equation}
  TY_3 = TP \oplus \mathcal{N}_{P/Y_3} \,,
\end{equation}
to rewrite the last integrand as $p_1(Y_3)$. Finally we can use the
representation of the Pontryagin classes in terms of the Chern classes,
\begin{equation}
  p_1(Y_3) = - 2c_2(Y_3) + c_1(Y_3)^2 \,,
\end{equation}
and the Calabi--Yau property of $Y_3$, $c_1(Y_3)=0$, to rewrite the two integrands in terms of the
Chern classes of $P$ and $Y_3$. In conclusion, the integral over the total anomaly
polynomial $I_8$, combining both the free and interacting theories living on
the M5-brane, is  \cite{Harvey:1998bx}
\begin{equation}\label{eqn:I4N}
  I_4[N] = \int_P I_8[N] = N I_4[1]
    + \frac{1}{24} (N^3 - N)P^3 p_1(\mathcal{N}_3) \,,
\end{equation}
where we have rewritten the integrals over $P$ as integrals over $Y_3$ using
intersection notation, and
\begin{equation}\label{eqn:I41}
  I_4[1] = \int_P I_8[1] = \frac{1}{48}\left[ 2 P^3 p_1(\mathcal{N}_3) +
  c_2(Y_3) \cdot_{Y_3} P (p_1(W_2) + p_1(\mathcal{N}_3)) \right] \,.
\end{equation}

The gravitational anomaly determines the difference between the left- and
right-moving central charges of the $(0,4)$ SCFT on the string, and can be
read off from the anomaly
\begin{equation}
  I_4 \supset \frac{c_L - c_R}{24}p_1(W_2) \,.
\end{equation}
Thus we immediately determine that
\begin{equation}
  c_L - c_R = \frac{1}{2} N c_2(Y_3) \cdot_{Y_3} P \,.
\end{equation}
From the anomaly polynomial it is also possible to read off the level
associated to the $SO(3)_T$ global symmetry by studying the $k_3
p_1(\mathcal{N}_3)/4$ term. We find
\begin{equation}
  k_3 = \frac{1}{6}N^3P^3 + \frac{1}{12}Nc_2(Y_3) \cdot_{Y_3}
    P \,.
\end{equation}
For future reference we also note that $k_3$ can be expressed directly in
terms of the Hodge numbers of $P$. Using the expansion of the Chern numbers in
terms of the Hodge numbers we have
\begin{equation}\label{eqn:intsHodge}
  \begin{aligned}
    P^3 &= 10 h^{0,2}(P) - 8 h^{0,1} (P) - h^{1,1}(P) + 10 \cr
    c_2(Y_3) \cdot_{Y_3} P &= 2 h^{1,1}(P) - 8  h^{0,2}(P)  + 4 h^{0,1} (P)  - 8 \,. 
  \end{aligned}
\end{equation}

At this point we shall specialise to considering that $Y_3$ is an elliptic
fibration. From the Shioda--Tate--Wazir theorem as described in section
\ref{subsec:EllFib} we know the divisors in $Y_3$ that generate the
Neron--Severi lattice, and we would like to compute these quantities, $c_L -
c_R$ and $k_3$, for representatives of certain linear systems of these
divisors on $Y_3$. Recall that we are interested in elliptically fibered
Calabi--Yau threefolds $\pi: Y_3 \rightarrow B$, with section, and that the
two types of basis divisors of principle interest are the base, $B$, and the
pullbacks of curves in the base, $\widehat{C}_\alpha = \pi^*C_\alpha$, such
that the curve is not contained inside the discriminant locus of the elliptic
fibration.

Let us consider an M5-brane wrapping a smooth irreducible divisor in the
linear system
\begin{equation}
  D \in | M B + N \widehat{C} | \,,
\end{equation}
where $\widehat{C}$ is a linear combination of the $\widehat{C}_\alpha$, and
compute the above quantities for $P = D$. 
The cohomology class of $D$ can be written as
\begin{equation}
  [D] = M[B] + N[\widehat{C}] \,,
\end{equation}
and thus the first intersection number that must be computed is
\begin{equation}
  \begin{aligned}[ ]
    [D]^3 &= M^3 [B]^3 + N^2 [\widehat{C}]^3 + 3M^2N
    [B]\cdot_{Y_3}[B]\cdot_{Y_3}[\widehat{C}] +
    3MN^2[B]\cdot_{Y_3}[\widehat{C}]\cdot_{Y_3}[\widehat{C}] \cr
    &= M^3(10 - h^{1,1}(B)) + 3M^2N(-c_1(B) \cdot_B C) + 3MN^2C\cdot_B C \,,
  \end{aligned}
\end{equation}
where for the final two intersections we have used the triple intersection numbers for elliptic
Calabi--Yau varieties of section \ref{subsec:EllFib}. Furthermore
\begin{equation}
  \begin{aligned}
    \frac{1}{2}c_2(Y_3) \cdot_{Y_3} [D] &= \frac{1}{2} M c_2(Y_3) \cdot_{Y_3} [B] +
    \frac{1}{2} N c_2(Y_3) \cdot_{Y_3} [\widehat{C}] \cr
    &= M(h^{1,1}(B) - 4) + 6N c_1(B) \cdot_B C \,.
  \end{aligned}
\end{equation}
Therefore we have determined that for an M5-brane wrapping an arbitrary
divisor $D$ belonging to such a linear system
\begin{equation}
  \begin{aligned}
    c_L - c_R &= 6N c_1(B) \cdot C + M(h^{1,1}(B) - 4) \,,
   \end{aligned}
    \label{clcrample}
\end{equation}
and 
\begin{equation}\label{eqn:k3}
  \begin{aligned}
    k_3 &= \frac{1}{2}MN^2C\cdot C + \frac{1}{2} N (2 - M^2)c_1(B) \cdot C
    \cr &\quad +
    \frac{1}{6}\left( M^3(10 - h^{1,1}(B)) + M(h^{1,1}(B) - 4)\right) \,.
  \end{aligned}
\end{equation}
 Note that to compute these coefficients we had to use the anomaly
polynomial for a single M5-brane, $I_4[1]$, as $M$ and $N$ may be coprime,
however when either $M$ or $N$ vanishes we see the correct result for
multiple M5-branes wrapping a single divisor as in
(\ref{eqn:I4N})\footnote{For arbitrary values of $M$ and $N$ one can
consider the anomaly of a single M5-brane wrapping either the divisor $D$ as
in (\ref{eqn:I41}), or
one can factor $D$ as $D = \text{gcd}(M,N)D^\prime$, and consider
$\text{gcd}(M,N)$ M5-branes wrapping the divisor $D^\prime$ as in
(\ref{eqn:I4N}), by computing $I_4[\text{gcd}(M,N)]$ for the divisor
$D^\prime$. It is straightforward to verify that both approaches produce
the same result.}.

At this point we have determined the difference in left- and right-moving central
charges and the anomaly coefficient for the $SO(3)_T$ normal bundle anomaly
for an the 2d $(0,4)$ theory on the worldvolume of the string from an
M5-brane wrapping an arbitrary divisor $D$ in $Y_3$. From
\cite{Maldacena:1997de} it is known that if $D$ is a very ample divisor in $Y_3$
then the computation of $k_3$ is a suitable substitute for the computation
of $k_r$, the level of the superconformal $SU(2)_r$ R-symmetry in the IR, and
thus one can compute the the right-moving central charge through the
superconformal algebra relation
\begin{equation}
  c_R = 6 k_r \,     .
 \end{equation}
In fact, when $D$ is ample the existence of an 11d supergravity
 dual of the type AdS$_3\times S^2\times Y_3$ guarantees that $SO(3)_T$ can be identified
 exactly\footnote{More specifically the $SO(3)_T$ acting on the fields of
 the interacting SCFT, is then exactly the $SU(2)_r$ superconformal R-symmetry
 of that interacting SCFT. One can see directly from the spectrum that only
 after the universal centre-of-mass hypermultiplet is separated out is the
 $SO(3)_T$ consistent with the superconformal algebra.} with the $SU(2)_r$ R-symmetry rotating the $S^2$. Thus   $c_R=6k_r=6k_3$ is valid more generally for an ample divisor $D$.

From the information just described it is  possible to compute the left-
and right-moving central charges for the $(0,4)$ SCFT living on the string
from a stack of M5-branes wrapping a compact complex surface inside a
Calabi--Yau threefold, assuming that the surfaces satisfy sufficient
topological properties that the level associated to the superconformal
R-symmetry, $k_r$, is the same as $k_3$. For a divisor $D$ inside the linear
system that we are interested in, $| M B + N \widehat{C} |$, a discussion of
exactly when this divisor may be ample in $Y_3$ is contained in appendix
\ref{app:ample}.
A necessary condition for $D$ to be an ample divisor is that
  \begin{equation}
    D \cdot C = (N - M) C \cdot C + M(2 g - 2) > 0 \,,
  \end{equation}
  as pointed out in (\ref{eqn:constDdC}). It is clear that such an
  inequality cannot be satisfied for arbitrary values of $M$, $N$, and $g$,
  however in the large $N$ limit, where $N \gg M$, and when $C$ is ample in the base, this is always satisfied. 
  For any ample $D$, which then satisfies this inequality, we can use 
(\ref{clcrample}) and (\ref{eqn:k3}),  to compute the right- and left-moving central charges on the
M5-brane wrapping $D$ and we find
\begin{equation}
  \begin{aligned}
    c_R &= 3 N^2 M C \cdot C + 3N(2 - M^2)c_1(B) \cdot C + M^3(10 -
    h^{1,1}(B)) + M(h^{1,1}(B) - 4) \, ,\cr
    c_L &= 3 N^2 M C \cdot C + 3N(4 - M^2)c_1(B) \cdot C + M^3(10 -
    h^{1,1}(B)) + 2M(h^{1,1}(B) - 4) \,.
  \end{aligned}
\end{equation}

To determine these central charges we have used that the level $k_3$ of the
$SO(3)_T$ normal bundle anomaly is the same as the level of the superconformal
R-symmetry anomaly, however this only holds if $D$ is ample in $Y_3$, which  is
exactly the requirement for when a supergravity dual of this 2d theory exists.
From the field theory side we are justified in considering a setup where $M =
0$ and we just have a stack of $N$ M5-branes wrapping the elliptic surface
$\widehat{C}$.  In appendix \ref{app:ample} we show that $\widehat{C}$ is
never itself an ample divisor, but in such a situation we would like to be
able to determine a prescription for computing the central charge of the
$(0,4)$ theory for such a stack of M5-branes, applicable even when the divisor wrapped by the M5-branes is not ample.
This will correspond to the Type IIB/D3-brane setup where there are no
KK-monopoles. We postpone this discussion for M5-branes until section
\ref{nonampleD}, while  we  now turn to the F-theory picture for this setup.

\subsection{Anomalies of 6d Self-dual Strings}
\label{sec:D3I4}

A stack of $N$ M5-branes wrapping an elliptic surface $\widehat{C}$ inside an
elliptic Calabi--Yau threefold is T-dual to a stack of $N$ D3-branes wrapping
a curve in the base of the elliptic Calabi--Yau. Such D3-brane stacks give
rise to self-dual strings in 6d, and the anomaly polynomial for such strings
was determined via inflow from the 6d theory in
\cite{Berman:2004ew,Shimizu:2016lbw} and extended to include arbitrary genus
curves in \cite{Lawrie:2016axq}. We will assume that the curve, $C$, on which
the D3-branes wrap has only transversal intersections with the discriminant
locus of the elliptic fibration. The $(0,4)$ worldvolume theory on the string
has the global symmetry group
\begin{equation}\label{eqn:lamp}
  SU(2)_R \times SU(2)_L \times SU(2)_I \,,
\end{equation}
where $SO(4)_T \cong SU(2)_R \times SU(2)_L$ is the rotation group to the
non-compact directions transverse to the string and $SU(2)_I$ is the
R-symmetry group of the 6d theory. The $SO(4)_R$ UV R-symmetry group for the
$(0,4)$ theory on the worldvolume of the string is  $SU(2)_R \times SU(2)_I$.

In \cite{Berman:2004ew,Shimizu:2016lbw} the anomaly polynomial for a self-dual
string, of charges $Q_i$ with respect to the two-form potentials $B_i$, with $dB_i$ self-dual, in a
6d $\mathcal{N} = (1,0)$ theory was determined by applying a similar analysis
as that  was introduced in \cite{Harvey:1998bx}, and which was used in
section \ref{nonampleD} for the anomaly polynomial on a stack of
M5-branes.  
The translation of the charges $Q_i$ of the strings into the curve classes from the
interpretation of the strings as coming from D3-branes wrapping the curve $C$
was included in \cite{Lawrie:2016axq}. The final result for the anomaly
polynomial, $I_4$, of the string in terms of the characteristic classes of the bundles associated to
the symmetry groups (\ref{eqn:lamp}) is
\begin{equation}
  \begin{aligned}
    I_4 &= c_2(R)\left[ \frac{1}{2}N^2 C \cdot C + \frac{1}{2}Nc_1(B) \cdot
    C\right] + c_2(L)\left[-\frac{1}{2}N^2 C\cdot C + \frac{1}{2}N c_1(B)
    \cdot C\right] \cr
    &\quad + c_2(I)\left[N\right]
    - \frac{1}{24}p_1(T)\left[ 6 N c_1(B)\cdot C\right] \,,
  \end{aligned}
\end{equation}
where we have ignored contributions from any additional global (flavour)
symmetries other than those discussed above, and where we recall that the genus of
the curve is contained inside the above expressions implicitly via adjunction (\ref{adjunction}). 
First we can determine the difference between the left- and right-moving central
charges from the gravitational anomaly term
\begin{equation}
  c_L - c_R = 6 N c_1(B) \cdot C \,.
  \label{donald}
\end{equation}
One can also read off from the anomaly polynomial the levels of the
$SU(2)_{R,L,I}$ global symmetries
\begin{equation}
  \begin{aligned}
    k_R &= \frac{1}{2}N^2 C \cdot C + \frac{1}{2}Nc_1(B) \cdot C \cr
    k_L &= -\frac{1}{2}N^2 C\cdot C + \frac{1}{2}N c_1(B)\cdot C \cr
    k_I &= N \,.
  \end{aligned}
\end{equation}
Note that the $SU(2)_r$ superconformal R-symmetry can in principle be a mix
\cite{Tong:2014yna} of the two $SU(2)$ factors in the $SO(4)_R$ UV R-symmetry. 
We  observe from the spectrum for $N=1$  that the IR R-symmetry
for the SCFT must be $SU(2)_R$ as this is the only factor
under which the bosons of all the hypermultiplets constituting the
theory are uncharged. 
Moreover, in the next subsection we will argue (using only the information on $k_L$ from this section)
that the correct R-symmetry in the IR should be simply $SU(2)_R$ for any $N$. 
Thus there is no  mixing with $SU(2)_I$ and we conclude that 
\begin{equation}
  \begin{aligned}
   c_R \, = \, 6 k_R &= 3 N^2 C \cdot C + 3 Nc_1(B) \cdot C\,,
    \end{aligned}
\end{equation}
and from (\ref{donald}) we also obtain
\begin{equation}
  \begin{aligned}
   c_L  &= 3 N^2 C \cdot C + 9 Nc_1(B) \cdot C\,.
    \end{aligned}
\end{equation}

\subsection{Anomaly from M5-branes  on $\widehat{C}$}
\label{nonampleD}

Let us now return to the M5-brane anomaly inflow, in the case that the branes wrap the elliptic surface $\widehat{C}$ in $Y_3$, which is not ample.  
We can immediately see from a study of the spectrum of a single M5-brane \cite{Lawrie:2016axq}  that $k_3$ is not a
suitable substitute computation for $k_r$ when the wrapped divisor is not
ample\footnote{This puzzle was raised  in \cite{LopesCardoso:1999fsj,Lambert:2007is}.}. Let us first consider an arbitrary divisor $P$ inside an arbitrary
Calabi--Yau threefold. We can read off from the expressions in terms of Hodge
numbers in (\ref{eqn:intsHodge}) that 
\begin{equation}
  k_3 = h^{0,2}(P) - h^{0,1}(P) + 1 \,,
\end{equation}
but a direct computation of the right-moving central charge from the spectrum
reveals that
\begin{equation}
  k_r = h^{0,2}(P) + 1 \,.
\end{equation}
This is consistent, as for $P$ an ample divisor inside a Calabi--Yau
threefold then $h^{0,1}(P) = 0$ by the Lefschetz hyperplane theorem. 

Now, let us consider multiple ($N$) M5-branes wrapping the divisor
$P=\widehat{C}$; hence $M=0$ in the notation of section \ref{sec:M5inflow}. 
Using  standard mathematical results for the
cohomologies of elliptic surfaces
\begin{equation}
  \begin{aligned}
    h^{0,2}(\widehat{C}) &= \frac{1}{2} \left(C\cdot C +  c_1(B)\cdot C
    \right) \, ,\cr
    h^{0,1}(\widehat{C}) &= \frac{1}{2} \left(C\cdot C - c_1(B)\cdot C
    \right) +1 \, = \, g \, , \cr
    h^{1,1}(\widehat{C}) &= C\cdot C +  9 c_1(B)\cdot C + 2 \,,
  \end{aligned}
\end{equation}
we can see that 
\begin{equation}\label{eqn:Hodgeelliptic}
  \widehat{C}^3 = 0 \,,\qquad c_2(Y_3) \cdot \widehat{C} = 12 c_1(B)\cdot C
  \,,
\end{equation}
and thus
\begin{equation}
  k_3 = N c_{1}(B)\cdot C \,,
\end{equation}
for an M5-brane wrapping any elliptic surface embedded inside an elliptic
Calabi--Yau as discussed. Such a result of course also follows directly from
the expression (\ref{eqn:k3}) for $k_3$ when one sets $M = 0$.

When the divisor is not
ample we follow the idea in \cite{Putrov:2015jpa} that $k_3$ is really a
substitute for computing the anomaly associated with the diagonal of the
superconformal R-symmetry, $k_r$, with an additional flavour symmetry that only
emerges, from the M5-brane point of view, in the IR
\begin{equation}\label{flavour}
  k_r = k_3 - k_F \,,
\end{equation}
where $k_F$ is the level of the emergent $SU(2)_F$ flavour symmetry.

In order to make progress in determining this flavour symmetry, we pass to the Type IIB description. The reason why
this is useful is that although it is still a UV description, the Type IIB side
captures also a \emph{flavour} (\emph{i.e.} non-R) symmetry, simply because the normal
bundle is $SO(4)_T$, which is larger than $SO(3)_T$. Notice that while R-symmetries are ambiguous,
because mixing an R-symmetry with a flavour symmetry is still an R-symmetry,
flavour symmetries do not have this ambiguity.

From the self-dual string in 6d, as is discussed in section
\ref{sec:D3I4}, we know exactly one flavour symmetry, which is the $SU(2)_L$
arising from the transverse rotations to the string, and further we can
observe from the spectrum that the $SO(3)_T$ charges of the multiplets from
the M5-brane on $\widehat{C}$ are the diagonal of the $SU(2)_R$ and $SU(2)_L$
charges of the multiplets from the D3-brane on $C$ \cite{Lawrie:2016axq}. As it is the only flavour
symmetry that we know is always present, and since it combines with the
superconformal R-symmetry in the correct way to form $SO(3)_T$ we are
justified in conjecturing that the level of the flavour symmetry, $k_F$, which
we must subtract off to compute the $k_r$ is none other than $k_L$.

From the analysis of the self-dual string we have that
\begin{equation}
  k_L = - \frac{1}{2} N^2 C \cdot C + \frac{1}{2} N c_1(B) \cdot C \,,
\end{equation}
however, as discussed in \cite{Haghighat:2015ega}, this anomaly coefficient is
not quite identified with the level of the $SU(2)_L$ symmetry on the combined
theory. In the anomaly coefficient of the $SU(2)_L$ anomaly there is a
fictitious contribution from the centre-of-mass hypermultiplet. This
universal hypermultiplet is charged under the $SU(2)_L$ however there is no
$SU(2)_L$ current algebra acting on these modes. The level of the $SU(2)_L$
current algebra on the combined theory is then determined by subtracting the
contribution\footnote{We note that there is a difference of an overall minus
sign between here and \cite{Haghighat:2015ega}.} of $k_L^\text{CoM} = +1$ from
$k_L$ to find that the level is
\begin{equation}
  k_L - 1 \,.
\end{equation}
This is then the level of the flavour symmetry of the combined theory
including the centre of mass which we then subtract from $k_3$, which is the
level of the $SO(3)_T$ normal bundle anomaly of the combined theory, to
determine the level of the superconformal R-symmetry of the combined theory.

As such the right-moving central charge as determined via the M5-brane anomaly inflow when
$M = 0$ is 
\begin{equation}
  c_R = 6(k_3 - (k_L - 1)) = 3 N^2 C \cdot C + 3 N c_1(B) \cdot C + 6 \,.
\end{equation}
We emphasise again that, as expected, this is the central charge for the
  combined theory, \emph{i.e.} the interacting theory together with the centre
  of mass. Further, we can observe that this identifies the superconformal
  R-symmetry level as
\begin{equation}
  k_r = k_3 - (k_L - 1) = k_R \,,
\end{equation}
demonstrating our statement in the previous subsection that the superconformal
R-symmetry is identified with $SU(2)_R$ for all $N$.
In this analysis we are
working under the assumption that generically there is only one $SU(2)$
flavour symmetry in the IR, and that that flavour symmetry is $SU(2)_L$. If
there are additional flavour symmetries then these could in principle also mix
with the superconformal R-symmetry to form $k_3$ and these would need to be
subtracted in addition.

\subsection{Summary and Comparison}
\label{sec:CentralChargeSummary}

Let us finally summarise and compare the results of all the computations (from anomalies and holography)
  of central charges   presented in this paper.
The theory to which the worldvolume theory on the string flows in the IR
consists of a direct sum of two SCFTs; the generically non-trivial and the
centre-of-mass conformal field theories. We shall refer to the former as the
SCFT part. Depending on the method  used we either compute
properties of the SCFT, or else of the combined theory. Generally speaking we
shall be interested in comparing the central charges of the SCFT, not
including the centre of mass; these are the quantities naturally computed by
the AdS duals as the centre of mass decouples in the near-horizon geometry.

\vskip 2mm

\noindent\underline{\textbf{The Spectrum}}

\vskip 2mm 

For a single D3-brane wrapping a curve $C$ in the base of an elliptic
threefold, or equivalently for a single M5-brane wrapping the elliptic
surface $\widehat{C}$, the massless spectrum can be computed explicitly. The
central charges as computed directly from the UV spectrum are
\begin{equation}
\hbox{Spectrum } (N=1): \qquad 
  \begin{aligned}
    c_R &= 3 C \cdot C + 3 c_1(B) \cdot C + 6\, ,  \cr
    c_L &= 3 C \cdot C + 9 c_1(B) \cdot C + 6 \,.
  \end{aligned}
\end{equation}
These are the central charges for the combined theory, including the
centre-of-mass modes. The scalar fields parametrising the position of the
string in the transverse 5d or 6d space are contained inside of a single
hypermultiplet, which is then referred to as the centre-of-mass
hypermultiplet, and contributes to the central charges
\begin{equation}
  (c_L^\text{CoM}, c_R^\text{CoM}) = (4, 6) \,.
\end{equation}
Subtracting off these modes gives the central charges for the IR
SCFT on the worldvolume of the string.

\vskip 2mm

\noindent\underline{\textbf{Anomaly Polynomial of Self-dual Strings}}

\vskip 2mm

In \cite{Shimizu:2016lbw} the anomaly polynomial for the self-dual string in
6d was written down, as we discussed in section
\ref{sec:D3I4}. This is the anomaly polynomial for the combined theory
including both the centre-of-mass and SCFT sectors. The combined theory
on the string has a global symmetry group
\begin{equation}
  SU(2)_R \times SU(2)_L \times SU(2)_I \,,
\end{equation}
where $SU(2)_R \times SU(2)_L$ comes from the transverse rotations to the
string in 6d, and $SU(2)_R \times SU(2)_I$ is the UV R-symmetry of the
worldvolume theory of the string. We are interested in computing from this
anomaly polynomial the central charges of the SCFT in the IR. First one
can determine the difference of the central charges of the combined theory
from the gravitational anomaly
\begin{equation}
  c_L - c_R = 6 N c_1(B) \cdot C \,.
\end{equation}
To determine the right-moving central charge of the SCFT we need
to know the level of the superconformal $SU(2)_r$ R-symmetry, which should be
one of the $SU(2)$ factors inside the $SO(4)$ UV R-symmetry. 
Furthermore, identifying  $SU(2)_R$ with  the IR  R-symmetry, as discussed in the previous subsection, we have computed that
\begin{equation}
  c_R = 6 k_R = 3 N^2 C \cdot C + 3 N c_1(B) \cdot C \,.
  \label{brexit}
\end{equation}
This matches the right-moving central charge computed for the SCFT from the
spectrum for $N=1$, as expected. If we subtract the free hypermultiplet
constituting the centre-of-mass degree of freedom from the difference of the
right- and left-moving central charges then we can also determine the
left-moving central charge for the SCFT as
\begin{equation}
  c_L = 3 N^2 C \cdot C + 9 N c_1(B) \cdot C + 2 \,.
\end{equation}
Again this matches the spectrum when $N=1$ as expected.

\vskip 2mm

\noindent\underline{\textbf{Type IIB Supergravity}}

\vskip 2mm

As discussed in section \ref{sec:FTheoryCC} we can also compute the central
charges for the same setup from the Type IIB supergravity dual. As such a
supergravity computation is necessarily in the near-horizon limit then the
centre-of-mass modes are decoupled and we compute directly only the central
charges of the SCFT. We will first consider the case \emph{without  KK-monopoles}, where 
in (\ref{moana}) we found that
\begin{equation}
  c^{\text{IIB}}_R = 3 N^2 C \cdot C + 3 N c_1(B) \cdot C \,,
\end{equation}
which exactly matches the right-moving central charge of the
theory from the spectrum and the anomaly analyses discussed previously. This would lead us to conclude that the Type IIB supergravity computation of
$c_R$ {\it is} in fact exact, meaning that there would be no quantum
corrections, in this precise situation, as any
sub-subleading correction would ruin the precise matching with the result in (\ref{brexit}).

In  (\ref{moana}) we also determined the left-moving central charge to be
\begin{equation}
  c^{\text{IIB}}_L = 3 N^2 C \cdot C + 9 N c_1(B) \cdot C \,,
\end{equation}
where we remind the reader  that this result is only expected to be
accurate to  order in ${\cal O}(N)$, and we expect from the alternate
approaches to the computation of the same quantity that the full result, including quantum corrections, should
have an additional $+2$.

In principle, from the Type IIB supergravity one should be able to determine  the holographic 
central charges also for $M\geq 1$, where there are in addition $M$ KK-monopoles in the system. However, as we discussed in section  \ref{sec:FTheoryCC}, in this case we can compute reliably only the leading order, ${\cal O}(N^2)$, 
 coefficients. To determine the correct ${\cal O}(N)$ contributions to the anomalies we would need to incorporate the effect of the KK-monopoles.

\vskip 2mm

\noindent\underline{\textbf{11d Supergravity}}

\vskip 2mm

In order for the 11d supergravity solution to exist it is necessary that the divisor wrapped by the
M5-brane is an ample divisor in the Calabi--Yau threefold, and from appendix
\ref{app:ample} we can see that this generally requires that $M > 0$. In this
section we shall take $M = 1$ principally so as to compare with the majority
of the different approaches, and we will show a matching for  $M > 0$
result at the end. For $M=1$ (the 11d supergravity setup dual to \emph{one} KK-monopole in Type IIB)
in section \ref{sec:Mthccs} we computed the central charges
to be
\begin{equation}\label{eqn:MthSUGRAccs}
  \begin{aligned}
    c_R^{11} &= 3 N^2 C \cdot C + 3 N c_1(B) \cdot C + 6 \,,\cr
    c_L^{11} &= 3 N^2 C \cdot C + 9 N c_1(B) \cdot C + 2 + h^{1,1}(B) \,. 
  \end{aligned}
\end{equation}
These central charges are said to be exact in \cite{Kraus:2005zm} as they can
be determined from an anomaly analysis. Since the exactness follows from an
anomaly argument these central charges should be the central charges for the
full combined theory, including the centre-of-mass degrees of freedom. Given
that the centre-of-mass contribution should be universal, regardless of the
values of $M$, $N$, we can similarly subtract one universal hypermultiplet to
determine the central charges of the IR SCFT.  We notice that the leading and
sub-leading terms are consistent with all other methods of computations for one or no KK-monopole.
As discussed in section \ref{intro}, in the near-horizon limit there is 
no difference between the setup with one or no KK-monopoles, and thus 
the leading contribution to the central charges must be identical.
We find that the central charges match also at the subleading order, and 
in fact the expression for $c_R$ matches the results obtained in the case without KK-monopoles
exactly, but it is not clear to us whether this is accidental or not. On the other hand, it makes sense that
both $c_R$ and $c_L$ do not both match exactly across the configurations with one or no KK-monopole, 
as the difference $c_L - c_R$ is a quantity that can be
computed purely in the UV, and in the UV the single KK-monopole is apparent.

The general result for all $M > 0$ was given in
(\ref{MTheoryCentralChargeFull}) and reads
\begin{equation}
  \begin{aligned}
    c_R^{11} &= 3N^2 M C \cdot C + 3N(2 - M^2)c_1(B) \cdot C + M^3(10 - h^{1,1}(B))
    + M(h^{1,1}(B) - 4) \, , \cr
    c_L^{11} &= 3N^2 M C \cdot C + 3N(4 - M^2)c_1(B) \cdot C + M^3(10 - h^{1,1}(B))
    + 2M(h^{1,1}(B) - 4) \,,
  \end{aligned}
\end{equation}
but as discussed previously, we have not determined these in the Type IIB picture, beyond the leading ${\cal O}(N^2)$ order. At this order we indeed find perfect agreement for any $N$ and $M$, see (\ref{fairenough}).

\vskip 2mm

\noindent\underline{\textbf{M5-brane Anomaly Inflow}}

\vskip 2mm

Another M-theory approach that one can take to determine the central charges
involves computing the anomaly polynomial to the string via M5-brane anomaly
inflow as described in section \ref{sec:M5inflow}. When the divisor wrapped by
the M5-brane is ample in the Calabi--Yau then this approach involves
effectively the same computation as was used to determine the central charges
from 11d supergravity, and is also a computation of the central
charges of the combined theory. The results for the central charges for
$M > 0$ from the anomaly inflow are then the same as
those given in (\ref{MTheoryCentralChargeFull}) from the 11d
supergravity.

The inflow computation however is
valid for any divisor $D$ even if it is not ample in the Calabi--Yau. As such, here 
we shall be mainly interested in the central charges for
the $M = 0$ case where the M5-brane wraps simply $\widehat{C}$. 
As described in section \ref{nonampleD} this approach does not directly
compute the central charge, but instead computes the anomaly coefficient
associated to the $SO(3)_T$ normal bundle anomaly, and the gravitational
anomaly which fixes 
\begin{equation}
  c_L - c_R = 6 N c_1(B) \cdot C \,.
\end{equation}
It is known that when the divisor wrapped is ample the computation of the
anomaly coefficient $k_3$ is a suitable substitute computation for the anomaly
coefficient of the superconformal R-symmetry, $k_r$. However when the wrapped
divisor is not ample one must subtract an emergent IR flavour symmetry from
$k_3$ to determine the superconformal R-symmetry. 
As discussed in section \ref{nonampleD} we can determine the flavour
symmetry which mixes with the superconformal R-symmetry and we can then compute
\begin{equation}
  c_R = 6(k_3 - (k_L - 1)) = 3 N^2 C \cdot C + 3 N c_1(B) \cdot C + 6 \,,
\end{equation}
which is the central charge of the combined theory. Further one can determine
the left-moving central charge of the combined theory as
\begin{equation}
  c_L = 3 N^2 C \cdot C + 9 N c_1(B) \cdot C + 6 \,.
\end{equation}


\section{Conclusions and Outlook}
\label{writesomethingsensible}

New holographic setups which allow for a controlled computational framework for both the perturbative gauge theory as well as the dual gravitational/string theory, are difficult to come by. In this paper we studied a new class of solutions of Type IIB supergravity, which allow for a varying axio-dilaton $\tau$, that is consistent with the $SL(2, \mathbb{Z})$ duality, \emph{i.e.} F-theory solutions. 
In particular, we classified the $\AdS_3$ solutions in F-theory dual to 2d SCFTs with $(0,4)$ supersymmetry, in the absence  of three-form fluxes. 
The field theory duals arise from D3-branes wrapped on curves in the base of
elliptic Calabi--Yau threefold compactifications studied in \cite{Haghighat:2015ega, Lawrie:2016axq}. 
The solutions that we have found to be the most general of this kind are of
the type $\AdS_3 \times S_3/\Gamma \times B$, where $B$ is the base of
an elliptic Calabi--Yau threefold, and the profile of the axio-dilaton is determined in terms of the complex structure of the elliptic fiber. 

Conceptually there are various points that make this duality more subtle than those involving Type IIB solutions with constant $\tau$. First of all the profile of the axio-dilaton has to be such that $\tau$ is singular along curves in the base $B$. This in turn implies that the metric on the base cannot be smooth everywhere, and thus some care needs to be taken in order to reliably apply a supergravity analysis. This is in particular subtle in Type IIB as the compactification manifold does not include the elliptic fiber, but only the base. 
Key to corroborating the consistency of this solution is the duality to 11d
supergravity, that we can perform for the solutions with $\Gamma=\Z_M$. We showed that in 11d supergravity these solutions are of the form
$\AdS_3 \times S^2 \times Y_3$, where the elliptic Calabi--Yau threefold $Y_3$ can be resolved and has a smooth Ricci-flat K\"ahler metric.

Following up on this paper, there are numerous immediate questions of interest to pursue:
An obvious extension of the present results is to include three-form fluxes and to potentially classify all $(0,4)$ AdS$_3$ solutions in Type IIB supergravity. We presented an example of such solution in Appendix \ref{app:GfluxSol}. Based on this, and other examples in the literature 
\cite{Lozano:2015bra,Wulff:2016vqy}, we expect that this class of solutions may be quite rich. Furthermore, the classification obtained here can be applied to $\mathcal{N}=(2,2)$ supersymmetry in 2d, and it  will be interesting to explore these and construct examples 
of  dual gauge theories.

{Another  class of $(0,4)$ strings in F-theory compactifications to 6d are the so-called non-Higgsable cluster strings.  As we recalled earlier, these are obtained from  D3-branes wrapped on 
 collapsed curves in Calabi--Yau threefolds, which have singular algebraic varieties as base manifolds.  In particular, these singularities can be thought of as arising from the collapse of a curve $C^{\rm NHC}\simeq \P^1$ in the local geometry of 
 $\mathcal{O}(-n) \rightarrow \mathbb{P}^1$, where the curve  has self-intersection $C^{\rm NHC}\cdot C^{\rm NHC} = -n < 0$. These can be embedded in a compact geometry by projectivizing, which results in the Hirzebruch surfaces 
 $\F_n $. It is then tantalizing to speculate that  our solutions might capture some features of the NHC strings by choosing the K\"ahler base to be $B=\F_n$, or their singular limits, {\it i.e.} the weighted projective spaces  $\P^{(1,1,n)}$. On the other hand, since $C^{\rm NHC}$ is not ample, this simple setup 
 cannot be found within the class of solutions discussed in this paper. Our
 attempts to reproduce features of the NHC strings in this holographic setup have not been successful, and it remains an open problem to 
 determine  what the appropriate holographic duals of these SCFTs, if these exist, are.}

In \cite{Lawrie:2016axq} a class of 2d $(0,2)$ theories were obtained, from
D3-branes wrapped in the base of elliptic Calabi--Yau four- and fivefolds. These are very closely related setups to the ones studied here, and naturally finding $\AdS_3$ duals to these  2d SCFTs would be very interesting. {In relation to the solutions found here, the case of Calabi--Yau fivefolds is closely related to our F-theory solutions with KK-monopole. The F-theory compactification space is $Y_3 \times TN_M$, which is a special Calabi--Yau fivefold. F-theory on elliptic Calabi--Yau fivefolds has only recently been investigated in \cite{Schafer-Nameki:2016cfr, Apruzzi:2016iac} and result in 2d  (0,2) theories for generic Calabi--Yau fivefolds. In view of this, it would be interesting to study our $\AdS_3$ solutions with $\Gamma = \mathbb{Z}_M$ in relation to the near horizon limits of D3-branes in Calabi--Yau fivefold compactifications of the type $Y_3 \times TN_M$ and determine the spectrum for general $M$ as in \cite{Lawrie:2016axq}.}

Finally,  the question of $\AdS_5$ solutions in F-theory arises, which would generalise the solutions AdS$_5 \times S^5/\Gamma$ of \cite{Fayyazuddin:1998fb, Aharony:1998xz, Aharony:2007dj} to F-theory 
solutions\footnote{Such solutions were briefly alluded to in
  \cite{Kehagias:1998gn}.} with non-trivially varying $\tau$. In particular,
  this would be interesting in relation to dimensional reductions of the
  recently obtained classification of 6d $(1,0)$ SCFTs in F-theory
  \cite{Heckman:2013pva}, which upon compactification on curves yield 4d
  $\mathcal{N}=1$ SCFTs. These theories could arise also in terms of F-theory
  on Calabi--Yau fourfolds and may have F-theoretic $\AdS_5$ duals.

We hope to return to these interesting questions in the near future \cite{toappear}.


\subsection*{Acknowledgments}

We thank Costas Bachas, Nikolay Bobev, Andreas Braun, Alejandra Castro, Michele del Zotto, Thomas Grimm, Per Kraus, Yolanda Lozano, Daniel Mayerson, Jes\'us Montero, Itamar Shamir, Stefan Vandoren and Timo Weigand for discussions. 
C.C., D.M. and J.-M.W. thank the Galileo Galilei Institute for Theoretical Physics (GGI) for their hospitality and INFN for partial support during the completion of this work, within the program ``New Developments in $\AdS_3/\CFT_2$ Holography''.  C.C. thanks the INFN and the ACRI for their sponsorship of the Young Investigator Training Program fellowship hosted at the theoretical physics group at Milano-Bicocca as well as for their hospitality during the final stages of this work. 
SSN thanks the Aspen Physics Center for hospitality during the workshop ``SCFTs in $d\geq 4$". 
C.C. and J.-M.W. are supported by
STFC studentships under the STFC rolling grant ST/N504361/1.
C.L. is supported by DFG under Grant TR33 'The Dark Universe' and under GK 1940 'Particle Phyiscs Beyond the Standard Model'. 
D.M. is supported by the ERC Starting Grant 304806 ``The gauge/gravity duality and
geometry in string theory''. SSN is supported by the ERC Consolidator Grant 682608 ``Higgs bundles: Supersymmetric Gauge Theories and Geometry (HIGGSBNDL)''.


\appendix


\section{Conventions for Gamma Matrices and Spinors}\label{spinconv}

We shall use the letters $M,N,..$ for the 10d indices, $a,b,..$ takes values $0,1,2$ and are used for the AdS$_{3}$ indices and $\mu,\nu,..\in\{1,..,7\}$ for the indices for $\M_{7}$. Following \cite{Lu:1998nu} we decompose the 10d Gamma matrices as
\bea
\Gamma^{a}&=&\rho^{a}\otimes 1\otimes \sigma_{2}\,,\\
\Gamma^{\mu}&=&1\otimes \gamma^{\mu}\otimes \sigma_{1}\,,
\eea
where $\rho^{a}$ generate Cliff(1,2) and $\gamma^{\mu}$ generate Cliff(7). Explicitly we shall take 
\be
\rho_{0}=\ii ~\sigma^{1}\,,\quad 
\rho_{1}=\sigma^{2}\,,\quad 
\rho_{2}=\sigma^{3}\,,
\ee
with $\rho_{012}=-1$. For the Cliff(7) gamma matrices we shall take
\bea
\gamma^{1}&=&-\sigma_{1}\otimes1\otimes 1\,,\\
\gamma^{2}&=&\sigma_{3}\otimes \sigma_2\otimes 1\,,\\
\gamma^{3}&=&-\sigma_{2}\otimes 1\otimes 1\,,\\
\gamma^{4} &=& \sigma_3 \otimes \sigma_1 \otimes 1 \,, \\
\gamma^5 &=& \sigma_3 \otimes \sigma_3 \otimes \sigma_1 \,, \\
\gamma^6 &=& - \sigma_3 \otimes \sigma_3 \otimes \sigma_2 \,, \\
\gamma^{7}&=&-\sigma_{3}\otimes\sigma_{3}\otimes \sigma_{3}\,,
\eea
and we have $\gamma_{1...7}=-\ii 1$.With these conventions we have 
\be
\Gamma_{11}=\1\otimes \1\otimes \sigma_{3}\,.
\ee
We follow the definitions in \cite{Sohnius:1985qm} for the various intertwiners. For the $A$ intertwiner we have
\bea
A_{10}\Gamma_{M}A_{10}^{-1}&=&\Gamma_{M}^{ \dag}\,,\\
A_{7}\gamma_{\mu}A_{7}^{-1}&=&\gamma_{\mu}^{\dag}\,,\\
A_{1,2}\rho_{a}A_{1,2}^{-1}&=&-\rho_{a}^{\dag}\,,\\
A_{10}&=&A_{1,2}~\otimes A_{7}\otimes \sigma_{1}\,,\\
A_{7}&=&1\,,\\
A_{1,2}&=&\sigma_{1}\,.
\eea
For the charge conjugation intertwiner $C$ we take
\bea
C_{10}^{-1}\Gamma_{M}C_{10}&=&-\Gamma_{M}^{\Tra}\,,\\
C_{7}^{-1}\gamma_{\mu}C_{7}&=&-\gamma_{\mu}^{\Tra}\,,\\
C_{1,2}^{-1}\rho_{a}C_{1,2}&=&-\rho_{a}^{\Tra}\,,\\
C_{10}&=&C_{1,2}\otimes C_{7}\otimes \sigma_{1}\,,\\
C_{7}&=&\sigma_{2}\otimes\sigma_{1}\otimes \sigma_{2}\,,\\
C_{1,2}&=&\sigma_{2}\,.
\eea
We have
\be
C_{10}^{\Tra}=-C_{10}\,,\quad 
C_{7}^{\Tra}=C_{7}\,,\quad 
C_{1,2}^{\Tra}=-C_{1,2}\,.
\ee

Finally the $D$ intertwiner satisfies
\bea
D_{10}^{-1}\Gamma_{M}D_{10}&=&\Gamma_{M}^{*}\,,\\
D_{7}^{-1}\gamma_{\mu}D_{7}&=& -\gamma_{\mu}^{*}\,,\\
D_{1,2}^{-1}\rho^{a}D_{1,2}&=&\rho_{a}^{*}\,,\\
D_{10}&=&D_{1,2}\otimes D_{7}\otimes \sigma_{3}\,,\\
D_{7}&=&\sigma_{2}\otimes \sigma_{1}\otimes \sigma_{2}\,,\\
D_{1,2}&=&-\ii~\sigma_{3}
\eea
They satisfy
\be
D_{10}^{*} =D_{10}^{-1}\,,\qquad 
D_{7}^{*} =D_{7}^{-1}\,,\qquad 
D_{1,2}^{*}=D_{1,2}^{-1} \,.
\ee
We now wish to decompose a 10d Majorana-Weyl spinor consistent with these conventions. We shall decompose the spinor, $\epsilon$ as $\epsilon=\psi\otimes \chi\otimes \theta$ where $\psi$ is a two-component spinor, $\chi$ an eight-component spinor and $\theta$ a two-component spinor. The chirality condition in 10d is
\be
\Gamma_{11}\epsilon=-\epsilon
\ee
which is solved by
\be
\sigma_{3}\theta=-\theta\,.
\ee
For the Majorana condition we impose that both $\chi$ and $\psi$ are Majorana and also that $\theta$ is purely imaginary. Type IIB supersymmetry is parametrised by two 10d Majorana-Weyl spinors. We may complexify the two Majorana-Weyl spinors into 
\bea
\epsilon=\psi_{1}\otimes \xi_{1}\otimes \theta
\eea
where $\xi=\chi_{1}+\ii\chi_{2}$ is a Dirac spinor. This will generically preserve $(0,2)$ supersymmetry however we are also interested in finding the equations for preserving $(0,4)$ explicitly and so the ansatz we use to accommodate both cases is
\bea
\epsilon=\psi_{1}\otimes \me^{\Del/2} \xi_{1}\otimes \theta+\psi_{2}\otimes \me^{\Del/2}\xi_{2}\otimes \theta\,.\label{spinoransatz}
\eea
The $(0,2)$ case is obtained by setting one of the $\xi$'s to zero. The warp factor appears here for later convenience. Here the $\psi_{i}$ are Killing spinors on AdS$_{3}$ and satisfy the most general Killing spinor equations for two Killing spinors on AdS$_{3}$
\be
\nabla_{a}\psi_{i}=\frac{ m}{2}\sum_{j=1}^{2}W_{ij}\rho_{a}\psi_{j}\,.
\ee
 In appendix \ref{Wsimp} we show that we may diagonalize $W$.

\section{ Killing Spinors of AdS$_3$}
\label{Wsimp}

In  general two Killing spinors on AdS$_3$ may satisfy an equation of the form 
\be
\nabla_{a}\psi_{i}=\frac{ m}{2}\sum_{j=1}^{2}W_{ij}\rho_{a}\psi_{j}\,,\label{nablaW}
\ee
with $W$ an arbitrary matrix with possible coordinate dependence. In this section we show that $W$ is in fact constant for our purposes and moreover may be diagonalised, allowing us without loss of generality, to take the Killing spinors on AdS$_3$ to satisfy 
\be 
\nabla_{a}\psi_{i}=\pm \frac{m}{2}\rho_{a}\psi_{i}
\ee
which we have done in the main text.

We shall first show that $W_{ij}$ is necessarily a constant matrix and finally show that it may be diagonalised. Observe that by multiplying (\ref{nablaW}) by $\bar{\psi}_{k}$ one has the bilinear
\be
\bar{\psi}_{k}\rho^{a}\nabla_{a}\psi_{i}=\frac{3  m}{2}W_{ij}\bar{\psi}_{k}\psi_{j}\,.\label{contractedpsieq}
\ee
The Majorana condition and the antisymmetry of $C_{12}$ imply $\bar{\psi}_{i}\psi_{j}=-\bar{\psi}_{j}\psi_{i}$ and in particular $\bar{\psi}_{i}\psi_{i}=0$ (no sum). Observe that \eqref{contractedpsieq} gives four equations for the four components of $W$:
\bea
\bar{\psi}_{1}\slashed{\nabla}\psi_{1}&=&\frac{3m}{2}W_{12}\bar{\psi}_{1}\psi_{2}\,,\\
\bar{\psi}_{1}\slashed{\nabla}\psi_{2}&=&\frac{3m}{2}W_{22}\bar{\psi}_{1}\psi_{2}\,,\\
\bar{\psi}_{2}\slashed{\nabla}\psi_{1}&=&\frac{3m}{2}W_{11}\bar{\psi}_{2}\psi_{1}\,,\\
\bar{\psi}_{2}\slashed{\nabla}\psi_{2}&=&\frac{3m}{2}W_{21}\bar{\psi}_{2}\psi_{1}\,.
\eea
As the left-hand side of all four equations and the spinors on AdS$_3$ are independent of the internal manifold coordinates it follows that $W$ is dependent only on the coordinates of AdS$_{3}$. As we wish to preserve the symmetry of AdS$_{3}$ in the solution this requires that the components of $W$ must in fact be independent of the AdS$_3$ coordinates and therefore constant. 

As the components of $W$ are constant we are able to compute the integrability condition for these spinors and commute the derivatives past the components of $W$. One has the identity 
\be
[\nabla_{a_{1}},\nabla_{a_{2}}]\psi_{i}=\frac{1}{4}\tensor{R}{_{a_{1}a_{2}}^{b_{1}b_{2}}}\rho_{b_{1}b_{2}}\psi_{i}\,.
\ee
Explicitly computing this using (\ref{nablaW}) we find
\bea
[\nabla_{a_{1}},\nabla_{a_{2}}]\psi_{i}=\frac{m^{2}}{2}W_{ij}W_{jk}\rho_{a_{1}a_{2}}\psi_{k}\,.
\eea
Upon equating the two expressions and contracting with $\rho^{a_{2}}$ whilst recalling our normalisation of AdS$_{3}$ we find 
\be
W_{ij}W_{jk}=\delta_{ik}\,.
\ee
It is therefore clear from this expression that the eigenvalues of $W$ are $\pm 1$, and furthermore that $W$ is diagonalizable over $\R$. Therefore there is no ambiguity in changing basis of the Killing spinors on AdS$_3$ to make $W$ diagonal. We shall interpret the eigenvalues of $W$ as determining the preserved supersymmetry of the putative dual SCFT. After diagonalizing $W$ the Killing spinor equation on AdS$_{3}$ is 
\be
\nabla_{a}\psi=\pm \frac{m}{2}\rho_{a}\psi\,.\label{AdSKSeq2}
\ee
It can be seen that the two equations give Killing spinors of opposite chirality on the boundary of AdS$_{3}$. Let us take global coordinates on AdS$_3$ in which the metric takes the form
\be
\dd s^{2}(\text{AdS}_{3})=-\frac{r^{2} m^{2}+1}{m^{2}}\dd t^{2}+ r^{2} \dd \phi^{2}+\frac{1}{m^{2}r^{2}+1} \dd r^{2}\,,
\ee
and satisfies $R_{ab}=-2 m^2 g_{ab}$. First computing the Majorana Killing spinors on AdS$_{3}$ for the positive sign in (\ref{AdSKSeq2}), and using the gamma matrix conventions of appendix \ref{spinconv}, we find
\be
\psi_{+}=
\me^{-\frac{\ii(t+\phi)}{2}}
\begin{pmatrix}
\sqrt{r+\frac{1}{m}\sqrt{1+m^{2}r^{2}}}\lb a-\ii \me^{\ii(t+\phi)} a^{*}\rb \\
-\frac{1}{m\sqrt{r+\frac{1}{m}\sqrt{1+m^{2}r^{2}}}}(a+\ii \me^{\ii (t+\phi)}a^{*})
\end{pmatrix}\,,
\ee
whilst for the negative sign in (\ref{AdSKSeq2}) we find the Killing spinor
\be
\psi_{-}=
\me^{-\frac{\ii(t+\phi)}{2}}
\begin{pmatrix}
-\frac{1}{m\sqrt{r+\frac{1}{m}\sqrt{1+m^{2}r^{2}}}}(b\me^{\ii \phi}-\ii b^{*} \me^{\ii t})\\
-\sqrt{r+\frac{1}{m}\sqrt{1+m^{2}r^{2}}}\lb b\me^{\ii \phi}+\ii b^{*} \me^{\ii t} \rb
\end{pmatrix}\,.
\ee
Following \cite{Balasubramanian:2000pq} the divergent piece of each spinor in the limit as we go to the boundary of AdS$_3$ ($r\rightarrow \infty$ in the coordinates we have chosen), becomes the 2d supersymmetry parameter. In our conventions the chirality matrix in 2d is $\sigma_{3}$ and therefore $\psi_{+}$ gives rise to a supersymmetry parameter in 2d with positive chirality, whilst $\psi_{-}$ gives rise to one with negative chirality. We conclude that to preserve $(0,2)$ supersymmetry in the boundary theory we must use a Killing spinor on AdS$_3$ which solves (\ref{AdSKSeq2}) with positive sign in our spinor ansatz (\ref{spinoransatz}), whilst to preserve $(2,0)$ we use one which solves (\ref{AdSKSeq2}) with negative sign.

\section{Torsion Conditions on Spinor Bilinears}
\label{app:TorsionConditions}

In the following subsections we compute the torsion conditions of the spinor bilinears. We shall keep the $\alphai_i$'s arbitrary in this appendix and specialise in the main text.
Our notation for the various spinor bilinears is given by
\bea
S_{ij}&\equiv&\bar{\xi}_{i}\xi_{j}\\
A_{ij}&\equiv&\bar{\xi}_{i}^{c}\xi_{j} \\
K_{ij}^{\mu}&\equiv&\bar{\xi}_{i}\gamma^{\mu}\xi_{j} \\
B_{ij}^{\mu}&\equiv&\bar{\xi}_{i}^{c}\gamma^{\mu}\xi_{j} \\
U_{ij}^{\mu_{1}\mu_{2}}&\equiv&\bar{\xi}_{i}\gamma^{\mu_{1}\mu_{2}}\xi_{j} \\
V_{ij}^{\mu_{1}\mu_{2}}&\equiv&\bar{\xi}_{i}^{c}\gamma^{\mu_{1}\mu_{2}}\xi_{j} \\
X_{ij}^{\mu_{1}\mu_{2}\mu_{3}}&\equiv&\bar{\xi}_{i}\gamma^{\mu_{1}\mu_{2}\mu_{3}}\xi_{j} \\
Y_{ij}^{\mu_{1}\mu_{2}\mu_{3}}&\equiv&\bar{\xi}_{i}^{c}\gamma^{\mu_{1}\mu_{2}\mu_{3}}\xi_{j}\,.
\eea
Higher order bilinears are related to the ones presented above by Hodge duality.

Using the representation of the Clifford algebra in appendix \ref{spinconv} we find that the spinor bilinears have following symmetries 
\be
A_{ij}=A_{ji}\,,\quad 
B_{ij}=-B_{ji}\,,\quad 
V_{ij}=-V_{ji}\,,\quad 
Y_{ij}=Y_{ji}\,.
\ee
Furthermore, the following identities hold
\bea
\bar{\xi}_{i}^{c}\xi_{j}^{c}&=&\bar{\xi}_{j}\xi_{i}\,,\\
\bar{\xi}_{i}^{c}\gamma^{\mu}\xi_{j}^{c}&=&-\bar{\xi}_{j}\gamma^{\mu}\xi_{i}\,,\\
\bar{\xi}_{i}^{c}\gamma^{\mu\nu}\xi_{j}^{c}&=&-\bar{\xi}_{j}\gamma^{\mu\nu}\xi_{i}\,,\\
\bar{\xi}_{i}^{c}\gamma^{\mu\nu\rho}\xi_{j}^{c}&=&\bar{\xi}_{j}\gamma^{\mu\nu\rho}\xi_{i}\,.
\eea

\subsection{Simplifying Relations}
\label{spinsimp}

Let $\eta$ be a spinor ($\xi_{i}$ or $\xi_{i}^{c}$) and let $\gamma$ be an arbitrary product of antisymmetrized gamma matrices, from (\ref{gravalgg=0}) we have the algebraic relations
\bea
\bar\eta \gamma \lb \frac{1}{2}\partial_{\mu}\Del \gamma^{\mu}-\frac{\ii \alphai_{j}m}{2}+\frac{\me^{-4 \Del}}{8}\sff\rb \xi_{j}&=&0\,,\label{algg=01}\\
\bar{\xi}_{i}\lb \frac{1}{2}\partial_{\mu}\Del \gamma^{\mu}+\frac{\ii \alpha_{i}m}{2}-\frac{\me^{-4 \Del}}{8}\sff \rb \gamma \eta&=&0\,,\label{algg=02}\\
\bar{\xi}_{i}^{c}\lb \frac{1}{2}\partial_{\mu}\Del \gamma^{\mu}+\frac{\ii \alpha_{i}m}{2}+\frac{\me^{-4 \Del}}{8}\sff\rb \gamma \eta&=&0\,,\label{algg=03}\\
\bar{\eta}\gamma \lb -\frac{1}{2}\partial_{\mu}\Del \gamma^{\mu}+\frac{\ii \alphai_{j}m}{2}+\frac{\me^{-4 \Del}}{8}\sff \rb \xi_{j}^{c}&=&0\,.\label{algg=04}
\eea
These relations are useful in simplifying the expressions obtained from computing the torsion conditions and have been used extensively in deriving the formulae in the following sections.

\subsection{Algebraic Equations}\label{algeqs}

We begin by computing some algebraic equations that will be useful in our analysis.
First note, by taking the difference of (\ref{algg=01}) and (\ref{algg=03}) 
one can derive
\be
(\alphai_{i}+\alphai_{j})A_{ij}=0\,.\label{xicixij=0}
\ee
Notice this condition implies that the scalars $A_{11}$ and $A_{22}$ must vanish irrespective of the value of $\alpha_i$. 
We can use the algebraic relations of section \ref{spinsimp} with $\gamma=1$ to find
\bea
\partial_{\mu}\Del \bar{\xi}_{i}\gamma^{\mu}\xi_{j} &=& \frac{\ii m}{2}(\alphai_{j}-\alphai_{i})\bar{\xi}_{i}\xi_{j}\label{warpkillg=0} \,.
\eea
Finally, we have two conditions involving the one-form $P$, which follow from \eqref{dilalgg=0}
\bea
P_{\mu}\bar{\xi}_{i}\gamma^{\mu}\xi_{j}&=&0\,,\label{Pkilled1}\\
P_{\mu}\bar{\xi}_{i}^{c}\gamma^{\mu}\xi_{j}&=&0\,.\label{Pkilled2}
\eea

\subsection{Differential Conditions: Scalars}\label{Scalar section}

The torsion conditions for the scalars, $S_{ij}$ and $A_{ij}$, take the form
\bea
\dd S_{ij}&=& \frac{\ii m}{2}(\alphai_{i}-\alphai_{j})K_{ij}\,,\label{dsijg=0}\\
\me^{-2 \Del}\D(\me^{2 \Del}A_{ij})&=& -\frac{\ii m}{2}(\alphai_{i}-\alphai_{j})B_{ij}\label{dbijg=0}\,.
\eea
From (\ref{dsijg=0}) we observe that $S_{ii}$ are constants and so we choose to normalise them to have unit norm. 

Notice that our equations are invariant under $GL(2,\mathbb{C})$ transformations of the spinors. Furthermore, when $\alphai_{1}=\alphai_{2}$ we find that $S_{12} = \bar{\xi}_{1}\xi_{2}$ is also a constant. Let us consider this constraint in more detail. By the Cauchy-Schwarz inequality we have
\be
|\bar{\xi}_{1}\xi_{2}|^{2}\leq |\bar{\xi}_{1}\xi_{1}||\bar{\xi}_{2}\xi_{2}|=1\,.
\ee
We separate the cases where this bound is saturated and when it is not and consider first when the bound is not saturated, namely $|\bar{\xi}_{1}\xi_{2}|=x<1$. We may multiply $\xi_{1}$ by a phase to make $\bar{\xi}_{1}\xi_{2}$ real. Now consider the following rotation on the spinors 
\be
\begin{pmatrix}
\xi_{1}\\
\xi_{2}
\end{pmatrix}
\rightarrow
\begin{pmatrix}
\xi_{1}'\\
\xi_{2}'
\end{pmatrix}=
\begin{pmatrix}
-1& 0\\
-\frac{x}{\sqrt{1-x^{2}}}&\frac{1}{\sqrt{1-x^{2}}}
\end{pmatrix}
\begin{pmatrix}
\xi_{1}\\
\xi_{2}
\end{pmatrix}\,.
\ee
This transformation preserves $S_{ii} = \bar{\xi}_{i}'\xi_{i}'=1$, where no sum is intended over the index $i$, and sets $|S_{12}| = |\bar{\xi}_{1}\xi_{2}|=0$. Now consider the case when the bound is saturated, one finds
\be
\xi_{1}= \lambda \xi_{2} \,,
\ee
where $\lambda$ can be shown to be just a phase. This relation, however,
reduces the amount of supersymmetry preserved. Therefore, to preserve four
supercharges, we set $S_{12} = 0$ when $\alphai_{1}=\alphai_{2}$.


\subsection{Differential Conditions: One-forms}

The covariant derivative of the one-forms $K_{ij}$ is given by 
\be
\nabla_{\mu_{1}}\bar{\xi}_{i}\gamma_{\mu_{2}}\xi_{j}=\frac{\ii m}{2}(\alphai_{i}+\alphai_{j})\bar{\xi}_{i}\gamma_{\mu_{1}\mu_{2}}\xi_{j}-\frac{\ii m}{2}(\alphai_{i}-\alphai_{j})g_{\mu_{1}\mu_{2}}\bar{\xi}_{i}\xi_{j}-\frac{\me^{-4 \Del}}{4}F_{\nu_{1}\nu_{2}}\bar{\xi}_{i}\tensor{\gamma}{_{\mu_{1}\mu_{2}}^{\nu_{1}\nu_{2}}}\xi_{j}\,.
\ee
Symmetrizing, we find
\be
\nabla_{(\mu_{1}}\bar{\xi}_{i}\gamma_{\mu_{2})}\xi_{j}=\frac{\ii m}{2}(\alphai_{j}-\alphai_{i})g_{\mu_{1}\mu_{2}}S_{ij}\,,\label{killg=0}
\ee
from which we observe that there are at least two Killing vectors when $i=j$. Notice that if $\alphai_{1}=\alphai_{2}$ there is an additional Killing vector, $\bar{\xi}_{1}\gamma^{(1)}\xi_{2}$. For  $\alphai_{1}\not=\alphai_{2}$ the existence of a third Killing vector depends on whether the scalar $S_{12}$ is vanishing or not.
Note also that (\ref{warpkillg=0}) vanishes whenever (\ref{killg=0}) vanishes, which implies that the Lie derivative of the warp factor along each Killing vector is vanishing.

The differential conditions on the one-forms read
\bea
\me^{-4 \Del}\dd \lb \me^{4 \Del}K_{ij}\rb&=& -\ii m (\alphai_{i}+\alphai_{j})U_{ij}-S_{ij}\me^{-4 \Del}\ff  \label{dKijg=0}\, ,\\
\D(\me^{2 \Del}B_{ij})&=&0\, .
\eea

\subsection{Differential Conditions: Higher Forms}

The differential conditions on higher degree forms are given by
\bea
\me^{-4 \Del}\dd (\me^{4 \Del}U_{ij})&=&-\frac{\ii m}{2}(\alphai_{i}-\alphai_{j})X_{ij}\,,\\
\me^{-6 \Del}\D \lb \me^{6 \Del}V_{ij}\rb &=&-\frac{3 \ii m}{2}(\alphai_{i}-\alphai_{j})Y_{ij}+\me^{-4 \Del}\ff \wedge B_{ij} \, ,\\
\me^{-8 \Del}\dd \lb \me^{8 \Del}X_{ij}\rb &=& 2 m (\alphai_{i}+\alphai_{j})*X_{ij}-\me^{-4 \Del}\ff\wedge U_{ij}\,,\\
\me^{-6 \Del}\D\lb \me^{6 \Del}Y_{ij}\rb &=& m (\alphai_{i}+\alphai_{j})*Y_{ij}\, , \\
\me^{-8 \Del}\dd \lb \me^{8 \Del}*X_{ij}\rb &=&-\frac{3}{2} \ii m (\alphai_{i}-\alphai_{j})*U_{ij}\, ,\\
\me^{-10 \Del}\D\lb \me^{10 \Del}*Y_{ij}\rb&=&-\frac{5\ii m}{2}(\alphai_{i}-\alphai_{j})*V_{ij}-\ii \me^{-4 \Del}\ff\wedge Y_{ij}\,,\\
\me^{-6 \Del}\D\lb \me^{6 \Del}*Y_{ij}\rb &=&-\frac{\ii m}{2}(\alphai_{i}-\alphai_{j})*V_{ij}-\ii \me^{-4 \Del}A_{ij}*\ff\, , \\
\me^{-8 \Del}\dd \lb \me^{8 \Del}*U_{ij}\rb&=& \ii m (\alphai_{i}+\alphai_{j})*K_{ij}\,,\\
\me^{-10 \Del}\D\lb \me^{10 \Del}*V_{ij}\rb &=& \ii m (\alphai_{i}+\alphai_{j})*B_{ij} \, ,\\
\me^{-12 \Del}\dd \lb \me^{12 \Del}*K_{ij}\rb &=&\ii m (\alphai_{i}-\alphai_{j})S_{ij}\text{dvol}(\M_{7})\,,\\
\me^{-10 \Del}\D\lb \me^{10\Del}*B_{ij}\rb &=&-\frac{3 \ii m}{2}(\alphai_{i}-\alphai_{j})A_{ij}\text{dvol}(\M_{7})\,.
\eea


\section{Supergravity Central Charges}

In this appendix we give details on the formulae used to compute the holographic central charges in the paper.
 
\subsection{Holographic Central Charges at Leading Order}
\label{CentralChargeFormula}

The leading order term in the central charge is given by the Brown-Henneaux formula \cite{Brown:1986nw}
\be
c_{\text{sugra}}=\frac{3}{2 m G_{N}^{(3)}} \,,\label{cformulaBH}
\ee
where $G_{N}^{(3)}$ is the three dimensional Newton constant obtained by the reduction of the Type IIB/11d supergravity action on the internal manifold. The relevant part of the action in dimension $d$ is 
\be \label{GenDAction}
S_{d}=\frac{1}{16 \pi G_{N}^{(d)}}\int_{M_{d}}*_{d}R^{(d)}\,.
\ee
We are interested in dimension $D = 10/11$ warped backgrounds of the form
\be
\dd s^{2}(M_{D})=\me^{2\Del}\dd s^{2}(\text{AdS}_{3})+\dd s^{2}(M_{D-3}) \,,
\ee
where $\Del$ a function of the internal manifold only. In this background the action in \eqref{GenDAction} can be expressed as  
\bea
S_{D}=\frac{1}{16 \pi G_{N}^{(D)}}\int_{M_{D-3}}\me^{\Del}(*_{D-3}1)\int_{\text{AdS}_{3}}*_{3}(R^{(3)}+...) \,.
\eea
This leading order piece is exactly the action \eqref{GenDAction} in three dimensions. From this we identify the $d = 3$ Newton constant to be 
\be
\frac{1}{G_{N}^{(3)}}=\frac{1}{G_{N}^{(D)}}\int_{M_{D-3}}\me^{\Del}\dd\vol(M_{D-3})\,,
\ee
and hence
\be\label{cformula}
c_{\text{sugra}}=\frac{3}{2m G_{N}^{(D)}}\int_{M_{D-3}}\me^{\Del}\dd\vol(M_{D-3}) \,.
\ee
In 10d the Newton's constant is
$G_{N}^{(10)}=2^{3}\pi^{6}\ell_{s}^{8}$, whilst in 11d it is given by
$G_{N}^{(11)}=2^{4}\pi^{7}\lp^{9}$.

\subsection{Holographic Central Charges at Sub-leading Order}

We may compute the sub-leading order terms in 11d supergravity by making use of the $X_{8}$ anomaly inflow polynomial \cite{Tseytlin:2000sf} and the relation 
\bea
c_{L}-c_{R}&=&96 \pi \beta\,,\nonumber\\
 S_{CS}&=&\beta \int_{\mathrm{AdS}_{3}}\omega_{CS}(\Gamma) \subset S_{3d}\label{CSterm}\,,
\eea
as found in \cite{Kraus:2005vz}. We reduce the 11d Chern-Simons term
\bea
S_{CS}=-\frac{(4 \pi \kappa_{11})^{2/3}}{2 \kappa_{11}^{2}}\int_{M_{11}}C_{3}\wedge X_{8}
\label{funky}
\eea
 on the internal space, with $X_{8}$ given by 
\bea
X_{8}=\frac{1}{(2 \pi)^{4}2^{6}\cdot 3}\lb \tr[\mathcal{R}^4]-\frac{1}{4}(\tr[\mathcal{R}^{2}])^{2}\rb\,,
\eea
and
\bea
\tensor{\mathcal{R}}{^{a}_{b}}&=&\frac{1}{2}\tensor{R}{^{a}_{b \mu\nu}}\dd x^{\mu}\wedge \dd x^{\nu}\,,\nonumber\\
2 \kappa_{11}^{2}&=&(2 \pi)^{8}\lp^{9}\,.
\eea
Integrating (\ref{funky}) by parts we have
\bea
S_{CS}  =  \frac{(4 \pi \kappa_{11})^{2/3}}{2 \kappa_{11}^{2}}\int_{M_{11}}G_{4}\wedge X_{7}\, ,
\eea
where $X_8\equiv \dd X_7$.   In our solution the internal eight-dimensional space is $S^2\times Y_3$, and given the form of the $G_4$ 
flux (\ref{MF4}), we have that  $X_7 =\frac{1}{3(2\pi)^4 2^7 } \omega_{CS}(\Gamma_{\mathrm{AdS}_3}) \wedge \tr[\mathcal{R}_{Y_3}^{2}]$  so that we determine
\bea
\beta  = \frac{\me^{4 \Del}}{3(2 \pi \lp)^3 m (2 \pi)^4 2^7} \int_{Y_3} J_{Y_3}\wedge  \tr[\mathcal{R}_{Y_3}^{2}]\, .
\eea


\section{Properties of K\"ahler and Calabi--Yau Varieties}
\label{app:Geo}

In this appendix we collect some essential theorems related to the
elliptically fibered Calabi--Yau threefolds that we consider as our
compactification spaces throughout the body of this paper. 

\subsection{Useful Relations}

First let $Y$ be a K\"ahler manifold with a given K\"ahler metric,
$g_{\mu\bar{\nu}}$. Then the K\"ahler form associated to this metric is
\begin{equation}
  J = \ii g_{\mu\bar{\nu}} \dd z^\mu \wedge \dd \bar{z}^{\bar{\nu}} \,,
\end{equation}
which is a closed $(1,1)$-form that is a representative of the cohomology
class known as the K\"ahler class; where it would be otherwise unambiguous we
shall abuse notation and use $J$ to refer to both the explicit representative
and the class. As $J$ is formed from the K\"ahler metric then it is real and
positive. This means that
\begin{equation}
  \int_C J > 0 \,, \quad \int_S J \wedge J > 0 \,, \quad \cdots \,,
\end{equation}
where $C$ is any curve in $Y$, $S$ any surface, and so on. One can find a
summary of this standard information in, for example, \cite{Hubsch:1992nu}.
Further, it is known that a compact complex manifold admits an holomorphic
embedding into projective space if and only if it admits a K\"ahler metric
whose associated K\"ahler form is an integral class \cite{MR2449178}.  As a
corollary to Yau's theorem, any compact strict Calabi--Yau, $Y_n$, of dimension
$n \geq 3$ can be embedded as a complex submanifold of a complex projective
space, and thus we can conclude that any Calabi--Yau threefold permits an integral
K\"ahler class.

After these introductory remarks we now collect several useful formulas. For
this we will specialise to the case of elliptically fibered Calabi--Yau
threefolds as in
section \ref{sec:Fsugra}, with base $B$, which is a K\"ahler surface. Various
properties of the base $B$ will feature in the main text, in particular
relations for topological invariants such as
\begin{equation}
  3\sigma(B) + 2 \chi(B) = \int_B c_1(B)^2 \,,
\end{equation}
where $\sigma(B)$ is the signature of the manifold and $\chi(B)$ is the Euler
number. In terms of the Hodge numbers of $B$ these can be written as
\begin{equation}
  \begin{aligned}
    \chi(B) &= 2 - 4h^{0,1}(B) + 2h^{0,2}(B) + h^{1,1}(B) \cr
    \sigma(B) &= (2h^{0,2}(B) + 1) - (h^{1,1}(B) - 1) = b_2^+ - b_2^- \,,
  \end{aligned}
\end{equation}
where $b_2^\pm$ are the number of self-dual and anti-self-dual two-forms of $B$. 
So far we have only assumed that $B$ is a compact K\"ahler surface.

Now let us further suppose that $B$ is the base of an elliptic fibration $\pi:
Y_3 \rightarrow B$ with section. As explained in the main text this does
restrict the type of K\"ahler surfaces that can function as $B$. In particular, the existence of the
section implies that
\begin{equation}\label{h01B}
  \pi_1(B) = 0 \quad \implies \quad h^{0,1}(B) = 0 \,.
\end{equation}
Furthermore the elliptic fibration must be Calabi--Yau which means that
\begin{equation}
  h^{0,2}(B) = 0 \,,
\end{equation}
as otherwise any $(0,2)$ forms on $B$ would give rise to $(0,3)$ forms on
the Calabi--Yau. Summarising, if $B$ is the base of an elliptically fibered
Calabi--Yau threefold then 
\begin{equation}
  \int_B c_1(B)^2 = 3 \sigma(B) + 2 \chi(B) = 10 - h^{1,1}(B) \,.
\end{equation}  
This agrees with the results given in \cite{Bonetti:2011mw}, and is a
general result for any base $B$ which may support a non-trivial Calabi--Yau
elliptic fibration over it.

We will require in the main text to determine the second Chern
class of the Calabi--Yau threefold when integrated over an arbitrary divisor $P$ of $Y_3$.
We have that
\begin{equation}\label{E8}
  \int_P c_2(Y_3) = \int_P (c_2(P) - c_1(P)^2) = 2(h^{1,1}(P) - 4h^{0,2}(P) +
  2h^{0,1}(P) - 4) \,,
\end{equation}
where the first equality follows via adjunction. As we can see the integral
over the second Chern class over any divisor is always an even integer.


\subsection{Ample Divisors in Elliptically Fibered Calabi--Yau Threefolds}\label{app:ample}

We shall now collect results about the ampleness properties of
divisors in an elliptically fibered Calabi--Yau threefold. An
M5-brane wrapping a divisor $D$ will only have an AdS dual when $D$ is ample,
as the divisor must be dual to a $(1,1)$-form in the K\"ahler cone of the
Calabi--Yau, following from the 11d supergravity solution in section
\ref{sec:FMduality}.  

First we shall be general and consider $Y$ any smooth algebraic variety, with
$D$ a divisor on $Y$. The Nakai--Moishezon \cite{MR0151461,MR0160782}
criterion for ampleness (see \emph{e.g.} \cite{MR2095471} for an in depth discussion)
is that
\begin{equation}
  D^{\text{dim}(X)} \cdot X > 0 \,,
\end{equation}
for every closed subvariety $X$ in $Y$. We remark that since the
Nakai--Moishezon criterion is just the intersection theory dual of the
statement that
\begin{equation}
  \int_X \omega^{\text{dim}(X)} > 0 \,,
\end{equation}
where $\omega$ is the dual $(1,1)$-form to the divisor $D$; in this way we can
see that every ample divisor is dual to a $(1,1)$-form inside of the K\"ahler
cone of $Y$.\footnote{A subset of the ample divisors consists of the
  \emph{very} ample divisors, which are those divisors which are linearly
  equivalent to the hyperplane class of a projective embedding of $Y$
  \cite{MR0463157}.}

With this in hand we shall now specifically consider a smooth elliptically
fibered Calabi--Yau threefold, $\pi: Y_3 \rightarrow B$, and the ampleness of
the divisors thereon. It was described in section \ref{subsec:EllFib} that an
elliptic fibration, with trivial Mordell--Weil group, has three distinct
classes of divisors which span the N\'eron--Severi lattice of divisors of $Y_3$.
These are the zero-section, which provides a copy of $B$ in the fiber, the
pullbacks of the curves in the base, $\widehat{C}_\alpha = \pi^*(C_\alpha)$,
and the Cartan divisors associated to the resolution of singularities, $D_i$. 
We will be interested in the triple intersection numbers of these divisors.
The triple intersection numbers that are of interest to us were determined in
\cite{Bonetti:2011mw}, and were recapped in (\ref{eqn:IntersectionNumbers}).

Let us first consider a smooth Weierstrass model $Y_3$, where we 
recall that there are no resolution divisors, we consider a divisor in the linear system
\begin{equation}
  D \in | M B + N \widehat{C}| \,.
\end{equation}
We are interested in knowing for what values of $M, N \geq 0$ is this
divisor not ample. We know from the Nakai--Moishezon criterion for ampleness
that
\begin{equation}
  D \cdot \Sigma > 0 \,,
\end{equation}
for every curve $\Sigma$ in $Y_3$, which includes the curve $C$ in $B$ which
$\widehat{C}$ is the pullback of, \emph{i.e.} $\widehat{C}$ is the elliptic surface
obtained by restricting the fibration to $C$. We can then compute
\begin{equation}
  D \cdot C = M B \cdot C + N \widehat{C} \cdot C = M B \cdot B \cdot
  \widehat{C} + N B \cdot \widehat{C} \cdot \widehat{C} \,,
\end{equation}  
where in the final equality we have used that 
\begin{equation}
  C = B \cdot \widehat{C} \,.
\end{equation}
Using the triple intersection numbers listed in (\ref{eqn:IntersectionNumbers}), along with adjunction, 
\begin{equation}
  c_1(B) \cdot C = C \cdot C + 2 - 2g \,,
\end{equation}
we can see that there is the constraint 
\begin{equation}\label{eqn:constDdC}
  D \cdot C = (N - M) C \cdot C + M(2g - 2) > 0 \,.
\end{equation}
For $N \gg M$, this is equivalent to the statement that $D$ is not ample in $Y_3$ if
$C$ is not ample in $B$. It is also clear from this formula that, for example,
when $M = N$ we need we consider an elliptic surface $\widehat{C}$, where the
base curve $C$ is such that
\begin{equation}
  g \geq 2 \,,
\end{equation}
and ampleness clearly implies a non-trivial interdependence between $M$,
$N$, and $g$.
Further one would like to determine whether there are constraints on ampleness
when $M=0$. While the constraint (\ref{eqn:constDdC}) only requires that $C$
must have a strictly positive self-intersection in the base we further note
that the Nakai--Moishezon criterion for ampleness requires also that the
triple-intersection of the divisor in $Y_3$ be strictly positive. For an
elliptic surface we observe that
\begin{equation}
  \widehat{C} \cdot \widehat{C} \cdot \widehat{C} = 0 \,,
\end{equation}
as was evidenced directly from the Hodge numbers in (\ref{eqn:Hodgeelliptic}),
and thus we determine that when $M=0$ the divisor cannot be ample.

For the case that is not a smooth Weierstrass model we can consider a divisor
in the linear system
\begin{equation}
  D \in | M B + N \widehat{C} + M_i D_i | \,,
\end{equation}
and consider again $D \cdot C$, however we should not include in this sum the
Cartan divisor associated to the affine node of the Dynkin diagram as it is
not an independent divisor inside the Neron--Severi lattice, and so the $M_0$
will not be a free parameter. We can see again from the triple intersection numbers
(\ref{eqn:IntersectionNumbers}) that
\begin{equation}
  D_i \cdot B \cdot \widehat{C} \,,
\end{equation}
is only non-zero when $D_i$ is precisely the divisor associated to the affine
node, and so the same conclusion on the constraints on $M, N$ will hold as in
the smooth Weierstrass case.

Finally one can study the case where $C$ is a smooth rational curve of
self-intersection $C \cdot C = -n$ for $n = 3,\cdots,12$, excepting $n =
9,10,11$.  These setups involve $C$ not being ample in $B$, and correspond to
the non-Higgsable clusters \cite{Morrison:2012np}. The self-intersection of the
curve in the base is severe enough that it mandates a total space singularity
above that curve in the Weierstrass model, with a specific kind of singular
fiber located above the curve $C$ depending on $n$. In such a setup one can
again compute
\begin{equation}
  D \cdot C = (n - 2)M - nN > 0 \,   	,
\end{equation}
which, given that $n$ is a positive integer generally requires that $M > N$,
if the divisor $D$ is to be ample.


\section{AdS$_3\times S^3\times S^2\times \P^1$ Solution with Three-Form Fluxes}
\label{app:GfluxSol}

In this section we present an  AdS$_3$  solution of Type IIB supergravity, preserving $(0,4)$ supersymmetry,  that includes non-zero three-form flux $G$. This is obtained by reduction and T-duality of an 
 11d supergravity solution constructed in \cite{Kelekci:2016uqv}. Of course  this does not fit into the classification of this paper as it has three-form flux,  however it can be derived from the 11d supergravity of \cite{Kelekci:2016uqv} 
 exactly as the solutions in the main body of the paper. We include it here as it may be interesting to explore this class of solutions in the future.

The 11d supergravity geometry in \cite{Kelekci:2016uqv} has the form
AdS$_3\times S^2\times S^2\times  Y_2$, where $Y_2$ is a K3 surface, with a
particular four-form flux. Here we will assume that $Y_2$ is again an
elliptically fibered Calabi--Yau and in keeping with the notation in the main
body of the paper, we shall denote the base of this fibration as $B_1\simeq
\P^1$. We shall call the vielbeins on $B_1$ $e^7$ and $e^8$. The solution of  \cite{Kelekci:2016uqv} can be shown to take the form 
\bea
\dd s^{2}&=& \dd s^{2}(\text{AdS}_{3})+R_{1}^{2}(\dd \theta_{1}^{2}+\sin^{2}\theta_{1}\dd \varphi^2_1)+R_{2}^{2}(\dd \theta_{2}^{2}+\sin^{2} \theta_{2}\dd \varphi_{2}^{2})+\dd s^{2}(Y_2)\,,\\
G_{4}&=&k_1 R_{1}R_{2} \vol(S^{2}_{1})\wedge \vol(S^{2}_{2})+R_{1}\vol(S^{2}_{1})\wedge J_{Y_2}+ R_{2}\vol(S_{2}^{2})\wedge \Re[\Omega_{Y_{2}}]\,,
\eea
where $J_{Y_2}$ is the K\"ahler form of the Calabi--Yau two-fold $Y_{2}$ and $\Omega_{Y_{2}}$ is the holomorphic two-form.  The constants $k_{1}$ and $m$,\footnote{As before $m$ is the inverse radius of AdS$_3$ and appears in the solution implicitly through $\dd s^{2}(\text{AdS}_{3})$.} are fixed in terms of the radii of the two $S^2$'s to be
\be
k_{1}= \sqrt{R_{1}^{2}+R_{2}^{2}}\,,~~m=\frac{\sqrt{R_{1}^{2}+R_{2}^{2}}}{2 R_{1}R_{2}}\,.
\ee
We then chose the Calabi--Yau two-fold to be elliptically fibered and we may then reduce on one of the cycles of the torus and then T-dualize along the other, again we take the metric ansatz for the fibration analogous to that given in (\ref{singularcy}) for the threefold. We present this duality chain below first reducing on the $x$ direction to Type IIA supergravity and then T-dualizing along the $y$ direction to Type IIB supergravity.

\noindent {\bf Type IIA Supergravity Solution from Dimensional Reduction}

Reducing the 11d supergravity solution along the $x$ direction one obtains the solution
\bea
\dd s^{2}&=& \frac{1}{\sqrt{\tau_{2}}}\lb  \dd s^{2}(\text{AdS}_{3})+R_{1}^{2}(\dd \theta_{1}^{2}+\sin^{2}\theta_{1}\dd \varphi^2_1)+R_{2}^{2}(\dd \theta_{2}^{2}+\sin^{2} \theta_{2}\dd \varphi_{2}^{2})+\dd s^{2}(B_{2})+\tau_{2} \dd y^{2}\rb\,,\nonumber\\
\dd C^{(3)}&=&k_{1} R_{1}R_{2} \vol(S_{1}^{2})\wedge \vol(S_{2}^{2})+R_{1}\vol(S_{1}^{2})\wedge J_{B}-\frac{\tau_{2}}R_{2}\vol(S_{2}^{2})\wedge e^{8}\wedge \dd y\,,\\
\dd C^{(1)}&=&\dd \tau_{1}\wedge \dd y\,,\\
B^{IIA}&=& R_{1}\cos \theta_{1} \dd \varphi_{1}\wedge \dd y-\frac{R_{2}}{\sqrt{\tau_{2}}} \cos \theta_{2}\dd \varphi_{2}\wedge e^{7}\,,\\
\me^{-2\Phi_{IIA}}&=&\tau_{2}^{\frac{3}{2}}\,.
\eea

We may now perform the T-duality along $y$ to obtain a Type IIB supergravity solution.

\noindent{\bf Type IIB Supergravity Solution from T-duality} 

The metric and fluxes in \emph{string frame} following from the T-duality are
\bea
\dd s^{2}&=& \frac{1}{\sqrt{\tau_{2}}}\lb  \dd s^{2}(\text{AdS}_{3})+R_{1}^{2}(\dd \theta_{1}^{2}+\sin^{2}\theta_{1}\dd \varphi^2_1)+R_{2}^{2}(\dd \theta_{2}^{2}+\sin^{2} \theta_{2}\dd \varphi_{2}^{2})+\dd s^{2}(B_{2})\right. \nonumber\\
&&\left.+( \dd y-R_{1}\cos \theta_{1} \dd \varphi_{1})^{2} \rb \, ,\\
F&=&(1+*)(k_{1}R_{1}R_{2}\vol(S_{1}^{2})\wedge \vol(S_{2}^{2})+R_{1}\vol(S_{1}^{2})\wedge J_{B})\wedge (\dd y-R_{1}\cos \theta_{1}\dd \varphi_{1}) \, ,\\
\dd C^{(2)}&=&-\sqrt{\tau_{2}}R_{2}\vol(S_{2}^{2})\wedge e^{8}\,,\label{dC2}\\
H&=&\frac{R_{2}}{\sqrt{\tau_{2}}}\vol(S_{2}^{2})\wedge e^{7}\,,\label{dB}\\
\me^{-\Phi}&=&\tau_{2}\,,\\
C^{(0)}&=&\tau_{1}\,.
\eea
We may now put this solution into Einstein frame and write the fluxes in the $SU(1,1)$ formalism. Inserting into the definition of $G$,
\be
G=\frac{\ii}{\sqrt{\tau_{2}}}(\tau \dd B -\dd C^{(2)})\, ,
\ee 
 equations (\ref{dC2}) and (\ref{dB}) one obtains for the three-form flux
\be
G=-R_{2} \vol(S_{2}^{2})\wedge \bar{\Omega}_{B}\,,
\ee
and the metric in Einstein frame is
\bea
\dd s^{2}&=&  \dd s^{2}(\text{AdS}_{3})+R_{1}^{2}(\dd \theta_{1}^{2}+\sin^{2}\theta_{1}\dd \varphi^2_1)+R_{2}^{2}(\dd \theta_{2}^{2}+\sin^{2} \theta_{2}\dd \varphi_{2}^{2})+\dd s^{2}(B_{1}) \nonumber\\
&&+( \dd y-R_{1}\cos \theta_{1} \dd \varphi_{1})^{2}\,,
\eea
whilst the five-form flux remains the same. If one redefines the $y$ coordinate as $y\rightarrow R_1 \psi$ then one sees that this is just a Hopf fibration over $S_{1}^{2}$ and the metric is AdS$_3\times S^3 \times S^2 \times B_1$. One finds that this solution preserves $(0,4)$ supersymmetry, after explicitly computing the Killing spinors, and that this corresponds to the \emph{large} superconformal algebra of the dual SCFT  \cite{Kelekci:2016uqv}.


\providecommand{\href}[2]{#2}\begingroup\raggedright\endgroup

\end{document}